\documentclass[pre,a4paper,twocolumn,floatfix]{revtex4-1}
\usepackage[utf8]{inputenc}
\usepackage[T1]{fontenc}
\usepackage{lmodern}
\usepackage{amssymb}
\usepackage{amsmath}
\usepackage{amsbsy}
\usepackage{bm}
\usepackage{esint}
\usepackage{braket}
\usepackage{enumitem}
\usepackage{graphicx}
\usepackage{color}
\usepackage{url}
\usepackage{siunitx}
\usepackage{listings}
\usepackage{courier}

\usepackage{hyperref}
\hypersetup{
    colorlinks=true,
    linkcolor=blue,
    filecolor=blue,      
    citecolor=blue,
    urlcolor=blue,
}
    \newcommand{\I}{\mathrm{i}}
    \newcommand{\E}{\mathrm{e}}

    \DeclareMathOperator{\CO}{C}
    \DeclareMathOperator{\GR}{G}
    \DeclareMathOperator{\FT}{F}
    \DeclareMathOperator{\MV}{M}
    \DeclareMathOperator{\SW}{S}
    \DeclareMathOperator{\CZ}{Z}

    \DeclareMathOperator{\Tr}{Tr}

\begin{document}
\title{Interacting quantum walk on a graph}
\author{Alberto D.\ Verga}\email{Alberto.Verga@univ-amu.fr}
\affiliation{Aix-Marseille Université, CPT, Campus de Luminy, case 907, 13288 Marseille, France}
\date{\today}
\begin{abstract}
  We introduce an elementary quantum system consisting of a set of spins on a graph and a particle hopping between its nodes. The quantum state is build sequentially,  applying a unitary transformation that couples neighboring spins and, at each node, couples the local spin with the particle. We observe the relaxation of the system towards a stationary paramagnetic or ferromagnetic state, and demonstrate that it is related to eigenvectors thermalization and random matrix statistics. The relation between these macroscopic properties and interaction generated entanglement is discussed.
\end{abstract}
\maketitle

\section{Introduction}

An isolated system, initially out of equilibrium, relaxes to thermal equilibrium. The understanding of the thermalization, or more generally, equilibration mechanisms behind this elementary experimental fact range from classical thermodynamics and kinetic theory, to the fundamentals of quantum statistical physics. Indeed, the very existence of a relaxation seems to be in contradiction with the unitary evolution of physical systems. Already in the twenties von Neumann \cite{Neumann-2010qf} realized that despite the global conservation of the entropy, expected values of local observables approached their microcanonical values for most quantum states \cite{Goldstein-2010rz,Tasaki-2010mz}. The aim of von Neumann was to demonstrate ergodicity in quantum systems without the assumption of disorder, or a kind of Stoßzahlansatz as Boltzmann used in his celebrated H-theorem. In the way, von Neumann assumed that macrostates must possess an enough rich structure, which now we can be related to quantum chaos \cite{Rigol-2012yq,Santos-2010qr}. In the same spirit Landau and Lifshitz stated that averaging over a Gibbs distribution is essentially equivalent to taking the expected value of an observable in a given energy state \cite{Landau-1980eu,Borgonovi-2016qe}.

In the recent literature, hypothesis of this kind are known as the ``eigenvalue thermalization'' \cite{Deutsch-1991vn,Srednicki-1994ys,Jensen-1985hb,Rigol-2008uq,Dymarsky-2018pi,Alessio-2016fj}: Deutsch \cite{Deutsch-1991vn} assumed an integrable system perturbed by a random matrix Hamiltonian, assumption that ensures Gaussian eigenstates and the validity of large deviations estimates for the fluctuations around the microcanonical values of the observables; Srednicki \cite{Srednicki-1994ys} showed that thermalization follows from the Berry's conjecture of chaotic eigenstates \cite{Berry-1977qq}, stating that in a semiclassical approximation of a classically stochastic motion, the corresponding wavefunction becomes a Gaussian random function. The eigenvalue thermalization hypothesis could be experimentally tested in an optical lattice of interacting bosons \cite{Kaufman-2016mz}, and in a small system of coupled superconducting qubits \cite{Neill-2016fr}.

Therefore, in order to explain the relaxation of an isolated system, in a pure quantum state with a well defined macroscopic energy, one must identify the basic mechanisms leading to thermal, or more generally, chaotic energy eigenstates. In this respect, it is worth mentioning Cardy's view that thermalization occurs in a pure state as a consequence of interference and entanglement with no need of ergodicity or coupling to a heat bath \cite{Cardy-2008ud}. 

In this paper we investigate what are the ingredients needed for a quantum isolated system to reach a highly entangled state amenable to thermal behavior. At variance with the usual approach in which one defines the quantum system by a Hamiltonian and solves the Schrödinger equation to obtain the energy levels and states, we are more interested in the mechanisms leading to ``typical'' quantum states \cite{Gemmer-2003yq,Popescu-2006qr,Goldstein-2006hl}; we want to explore the possibility of constructing such states from a series of local unitary operations applied to an initially particular state, which builds to a generic one possessing the properties of a thermal state. Accordingly, we adopt a quantum information approach in which the quantum state is viewed as a resource that can be manipulated by a series of one or two qubits gates \cite{Nielsen-2010fk,Raussendorf-2012xr}. As a consequence of the tensor product structure of the many-body quantum state, a network or graph of nodes and edges constitutes a natural representation of the physical system.

The present approach is reminiscent to the so called background independent theories of gravity, as quantum ``graphity'' \cite{Konopka-2006gd,Hamma-2011fk}, in which laws are formulated in terms of combinatorics; however, we do not search for an emergent theory but focus on the collective behavior of a subsystem in Hilbert space.

The study of the dynamics of a quantum systems defined on a graph is of primary interest in quantum computation \cite{Aharonov-2001ty,Kempe-2003fk}, in particular, discrete quantum walks can realize universal computations \cite{Lovett-2010rm}; their generalization to multiple particles \cite{Rohde-2011pi} was experimentally implemented \cite{Schreiber-2012fk}, showing many-body effects, such the appearance of interaction driven bound states \cite{Ahlbrecht-2012ec,Verga-2018}. In addition to these systems defined on a graph, it is also of interest to investigate quantum networks \cite{Acin-2007ij}, important in a range of domains from quantum communication to gravity \cite{Bianconi-2015uo}; a physical implementation of a spin-network was recently reported \cite{Li-2017uq}, as a first step in the direction of quantum gravity simulation.

We build our model on these two ingredients: the quantum walk of a particle on a graph, and the graph itself, which is defined as a set of interacting spins (much as a quantum network). The system's evolution will be given by the successive application of unitary operations, resulting essentially in shuffling and swapping state amplitudes. The model is presented in the next section, followed by a brief account of the phenomenology and a characterization of the observables; we find that the subsystems set in a stationary state with well defined physical properties. In addition, we compute the spectrum of the unitary operator by exact diagonalization to get full information on the structure of the eigenvectors and the statistical distribution of the eigenvalues. Finally, we discuss the role of degeneracy and entanglement in thermalization.

\section{Model}

\begin{figure}
  \centering
  \includegraphics[width=0.25\textwidth]{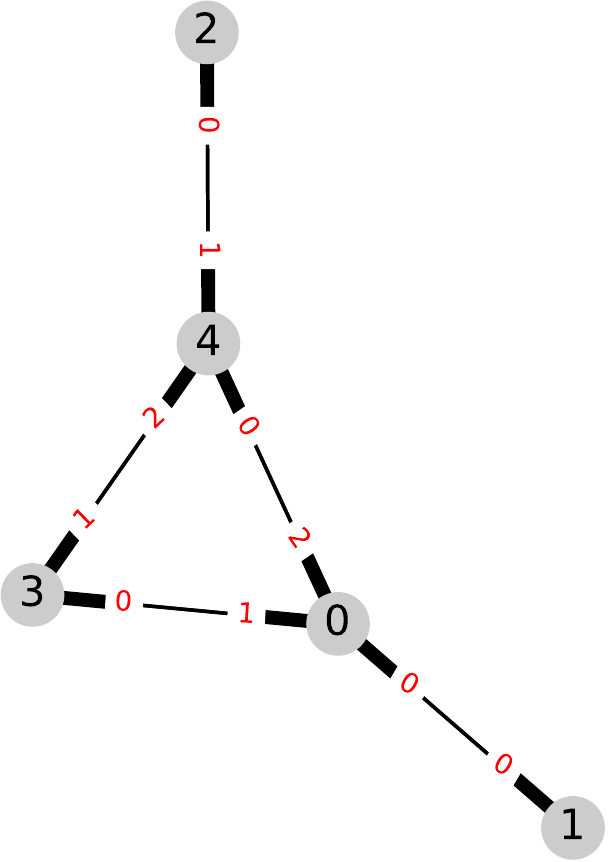}
  \caption{Graph definition. Each node is labeled by an integer $x$ (in black), which corresponds to the spin location and particle position. The subnodes of node $x$ are labeled by $c$ (in red, near the nodes, on the edges) in increasing order of the neighbors $y$ of $x$, they correspond to the particle internal degree of freedom (color).
  \label{f:b}}
\end{figure}

In this work we propose quantum systems defined on a network (a graph of nodes and links), with two kinds of degrees of freedom: an itinerant one, we call a particle (or walker) that can jump between neighboring nodes, and fixed ones, attached to each nodes, we call spins (or qubits). The ``space'' is specified by the topology of the graph; ``time'' is counted in steps, where the unitary transformation of the previous state is applied. The ``evolution'' of this state is given by the rules governing the particle jumps between the network nodes, its interaction with the graph spins, and the interactions between spins.

The model can be viewed from the point of view of quantum information, and in particular, the so-called one-way quantum computer, in which a highly entangled \emph{cluster state} is the resource on which a measurement protocol drives the computation \cite{Raussendorf-2003rm}. We replaced the projective operation (measure) on the cluster state (a set of interacting qubits) with the interaction of the graph state with a quantum walker \cite{Kempe-2003fk}. 

More formally, we consider a set of \(N\) half spins that defines the nodes \(x\) of a graph \(G\); each spin \(s_x\) interacts with a finite number of other spins \(s_y\), determining the graph connectivity. Two interacting spins belong to the same graph edge \((x,y)\). A particle can jump between nodes, along the connected edges, according to its internal degrees of freedom \(c\), we call colors (see Fig.~\ref{f:b} and Appendix \ref{s:py}). The particle interacts with the node spin it visits. The Hilbert space of the quantum system is spanned by the basis vectors
\begin{equation}
    \ket{xcs} = \ket{x} \otimes \ket{c} \otimes \ket{s_0s_1\ldots s_{N-1}}
\end{equation}
where the nodes labels \(x=0,\ldots,N-1\) correspond to the particle position, \(c=0,\ldots,d-1\) is the particle color quantum number with \(d\) the maximum graph degree, and \(s=s_0 s_1 \ldots s_{N-1}\) is a set of \(N\) binary numbers \(s_x = 0,1\), specifying the two values of the spin, `0' up, `1' down. The Hilbert's space dimension is then \(N\times d \times 2^N\). We use throughout \(\hbar=1\).

The spin-spin interaction changes the phase of the `11' two spins configuration,
\begin{equation}
  \label{e:CZ}
    \CZ\ket{xc,s=\ldots s_x \ldots s_y \ldots} = \begin{cases} - \ket{xcs} & \text{if } s_x= s_y = 1  \\  \ket{xcs} & \text{otherwise} \end{cases}
\end{equation}
where \((x,y)\) are the edges converging at \(x\); if restricted to a two spins space its matrix form would be \(\mathrm{diag}\, (1,1,1,-1)\). An important property of the \(\CZ\) gate is that it maximally entangles two \(x\) oriented spins (both spins in one of the eigenstates
\[
  \ket{\pm}=\frac{\ket{0} \pm \ket{1}}{\sqrt{2}}
\]
of the Pauli matrix \(\sigma_x\)). The spin-particle interaction,
\begin{align}
  \label{e:SW}
  \SW \ket{x,0,\ldots s_x=1 \ldots} & = \ket{x,1,\ldots 0 \ldots}\,,\\
  \SW \ket{x,1,\ldots s_x=0 \ldots} & = \ket{x,0,\ldots 1 \ldots}\,.
\end{align}
exchange the `0, 1' colors with the `1, 0' local spin, respectively, leaving the other colors unchaged; hence it can flip individual spins; restricted to a two qubit space it is the usual swap gate. Therefore, the gates \(\CZ\) and \(\SW\) can entangle the particle position and color states with the spin states, eventually favoring random spins at the nodes (maximally entangled state).

\begin{figure*}
\centering%
\includegraphics[width=0.3\textwidth]{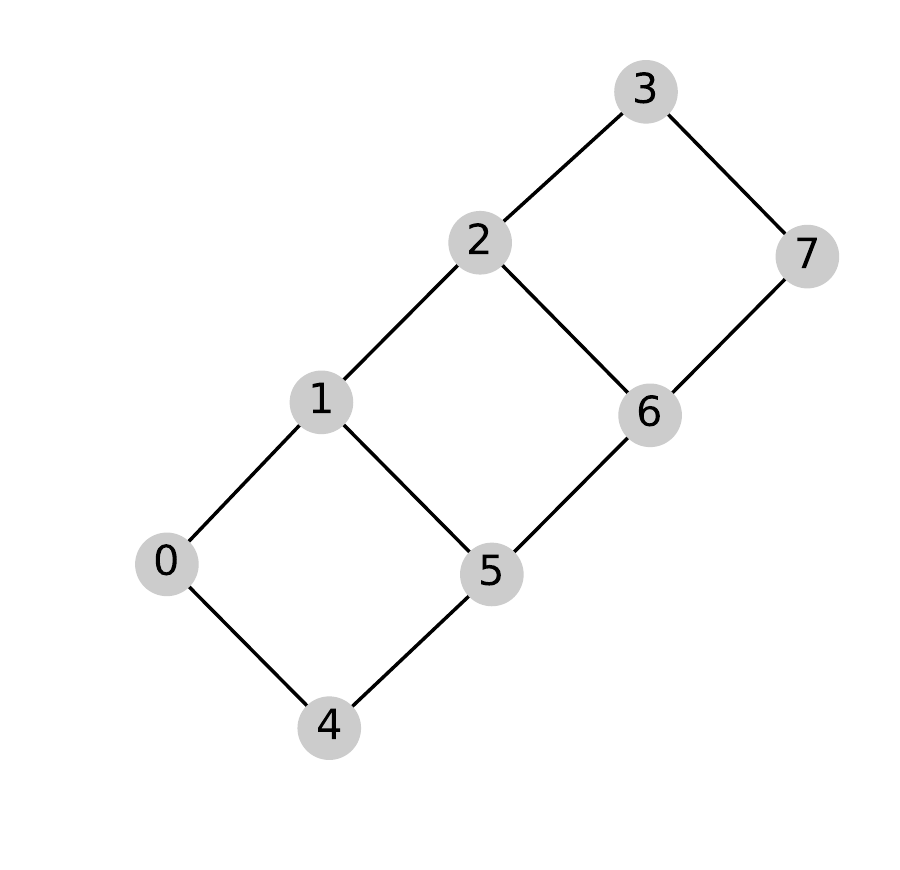}%
\includegraphics[width=0.3\textwidth]{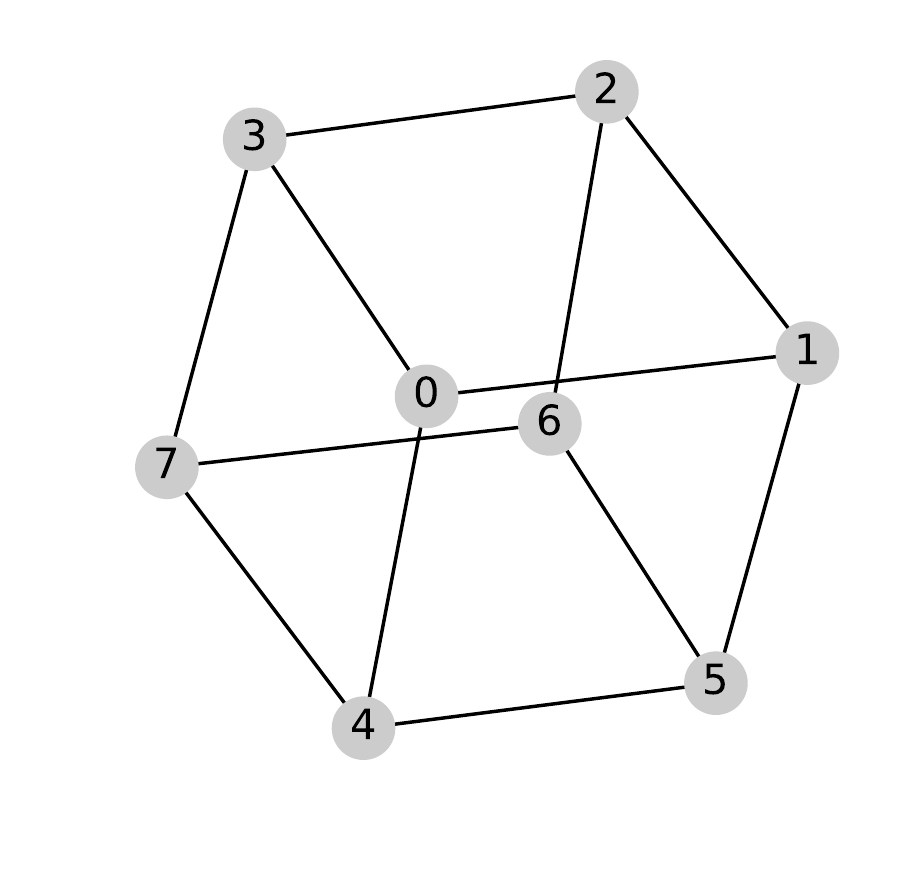}%
\includegraphics[width=0.3\textwidth]{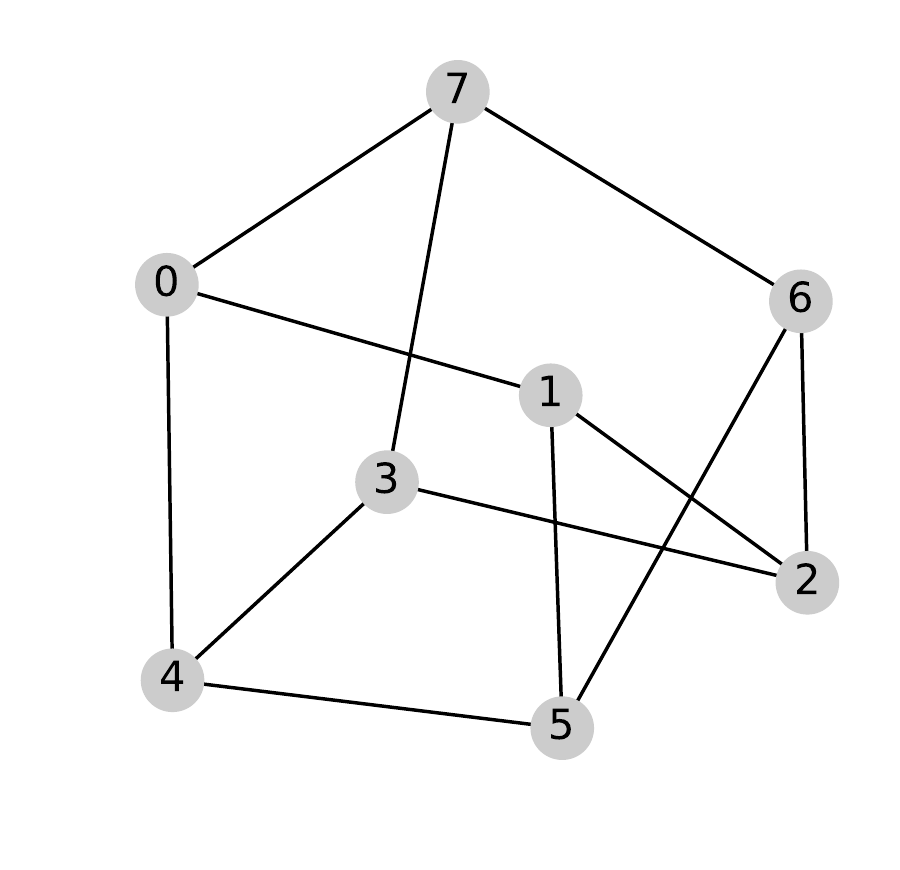}\\
~\hfill ladder `l' \hfill cube `c' \hfill möbius `m' \hfill~ \\
\includegraphics[width=0.3\textwidth]{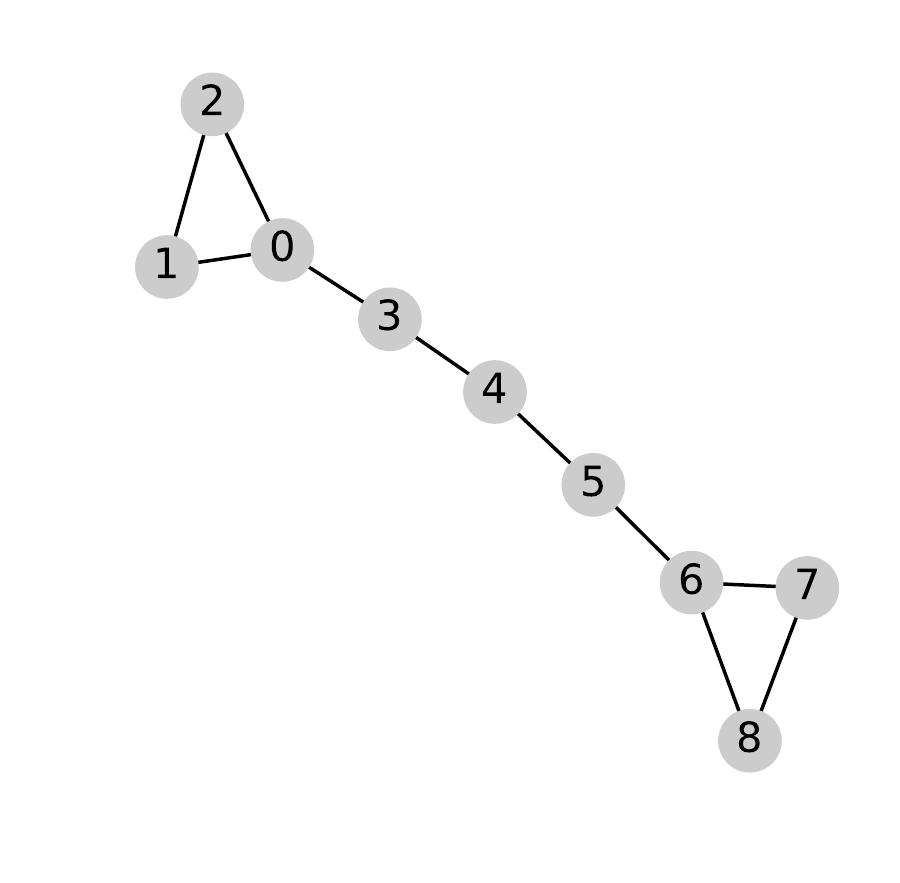}%
\includegraphics[width=0.3\textwidth]{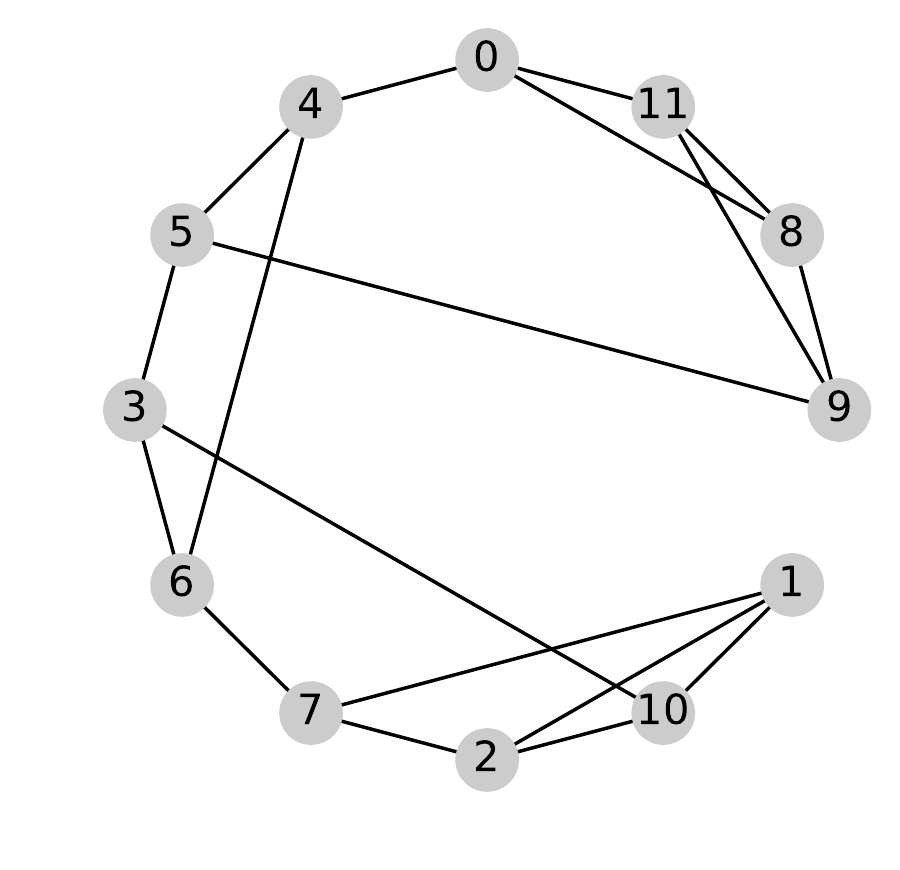}%
\includegraphics[width=0.3\textwidth]{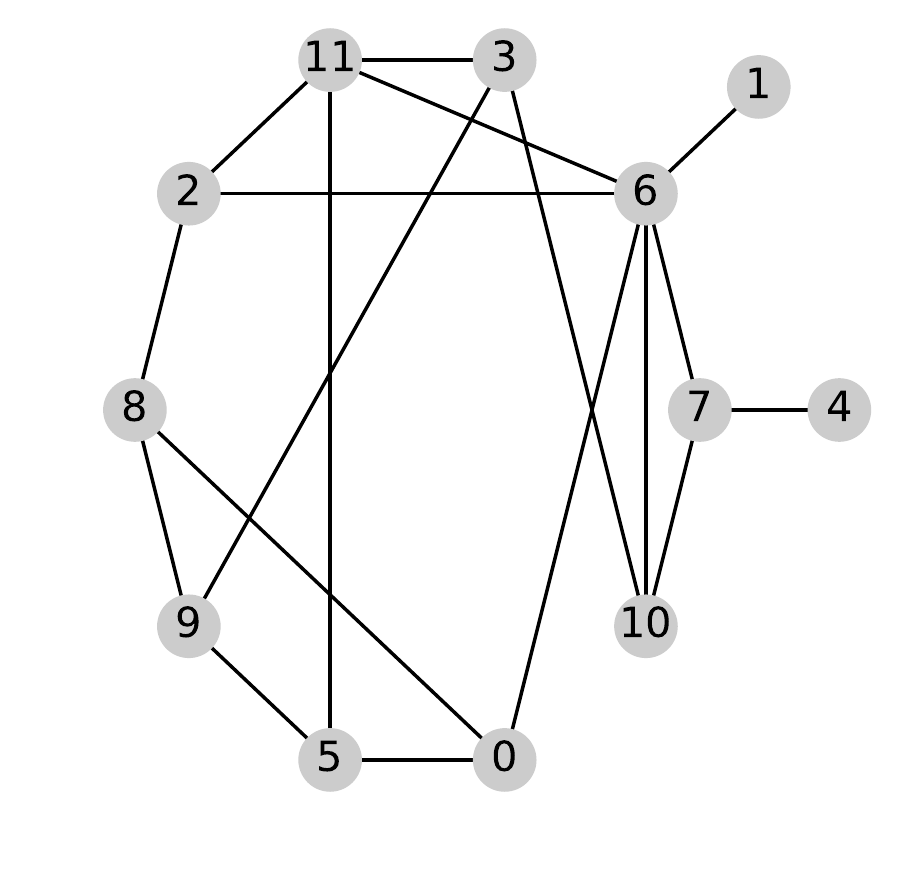}\\
~\hfill chain `X' \hfill random `R' \hfill random `G' \hfill~
    \caption{Graphs. First row, three variants of a regular $8$ vertex graph; second row, a symmetric chain `X' and two random graphs, $R(3,12)$ with regular $3$ degrees per node, and $G(12,3)$ with also $12$ nodes, and mean degree of $3$.}
    \label{f:g}
\end{figure*}

In addition to these local interactions, the particle executes a quantum walk by swapping its position \(x\) with the neighboring nodes \(y\), controlled by the different colors, one for each of the \(d_x\) edges. The motion operator \(\MV\) is defined by \cite{Berry-2011qq}, 
\begin{equation}
  \label{e:MV}
\MV \ket{x c_y s} = \ket{y c_x s}\,,
\end{equation}
where \(c_x\), \(c_y\) are the corresponding colors: the \(c_x\) amplitudes of node \(x\), as a result of the application of \(\MV\) are transferred to the nodes \(y\). Finally, the particle color at node \(x\) changes at each step of the walk, by application of the coin \(\CO = \GR\),
\begin{equation}
  \label{e:GR}
  \braket{x'c's'| \GR |x c s} = \left(\frac{2}{d_x} - \delta_{c,c'}\right) \,
    \delta_{x,x'} \delta_{s,s'}\,,
\end{equation}
the Grover coin, or \(\CO = \FT\) the Fourier coin:
\begin{equation}
  \label{e:FT}
  \braket{x'c's'| \FT |x c s} = \frac{1}{\sqrt{d_x}} \exp (\I 2 \pi c c' / d_x)\, 
    \delta_{x,x'} \delta_{s,s'}\,;
\end{equation}
where \(d_x\) is the degree of node \(x\). The Grover coin distributes the amplitudes equally between the starring edges of a given node, thus respecting the graph geometry, while the Fourier operator is an unbiased coin that equally superposes the amplitudes over the edges (much as a classical coin in a random walk). In two dimensions the Grover matrix reduces to \(\sigma_x\), and the Fourier matrix becomes the Hadamard matrix.

The evolution of the system is ensured by the repeated application of the unitary operator \(U\):
\begin{equation}
  \label{e:U}
  \ket{\psi(t+1)} = U \ket{\psi(t)} \,,\quad  U = \CZ\SW\MV\CO \,,
\end{equation}
(we use 1 as the time unit) where \(\GR\) acts on colors, \(\MV\) on nodes and colors, \(\SW\) on colors and spins, and \(\CZ\) between spins. The initial state is usually the ket \(\ket{000}\), with all spins up, the color 0 and the particle position at node 0. In Appendix~\ref{s:py} we present the numerical implementation of the model.

Typical graphs are represented in Fig.~\ref{f:g}. We consider regular graphs, like the `ladder' and the `cube'; almost linear graphs, as the `chain'; graphs differing in their topology, like the `cube', which is planar, and the `möbius', which has one crossing; or random graphs with regular degree (like `R'), or inhomogeneous degree (like `G'). 

It is worth remarking that the model thus described do not contain dimensional parameters, nor adjustable parameters: its structure is determined by the graph and its dynamics by \(U\), which is a pure numerical matrix. In addition, no obvious symmetries are present, due in particular to the subnode labeling that controls the motion of the particle, which depends on the arbitrarily denoted nodes. More importantly, although \(U\) is defined by strictly local rules, its corresponding effective Hamiltonian,
\begin{equation}
  \label{e:H}
  H \equiv \I \ln U\,,
\end{equation}
is, in general nonlocal, in the sense that it contains board range of interactions between distant nodes in the graph. 

\begin{figure*}
    \centering
    ~\hfill \includegraphics[width=0.2\textwidth]{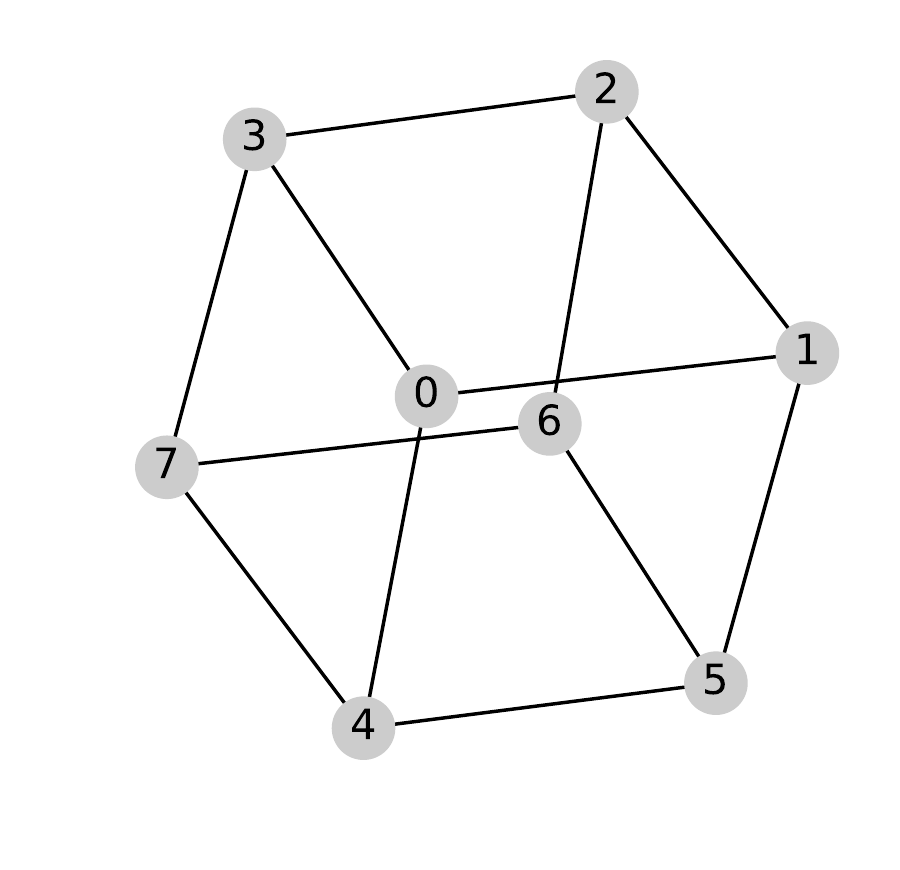} \hfill %
    \includegraphics[width=0.2\textwidth]{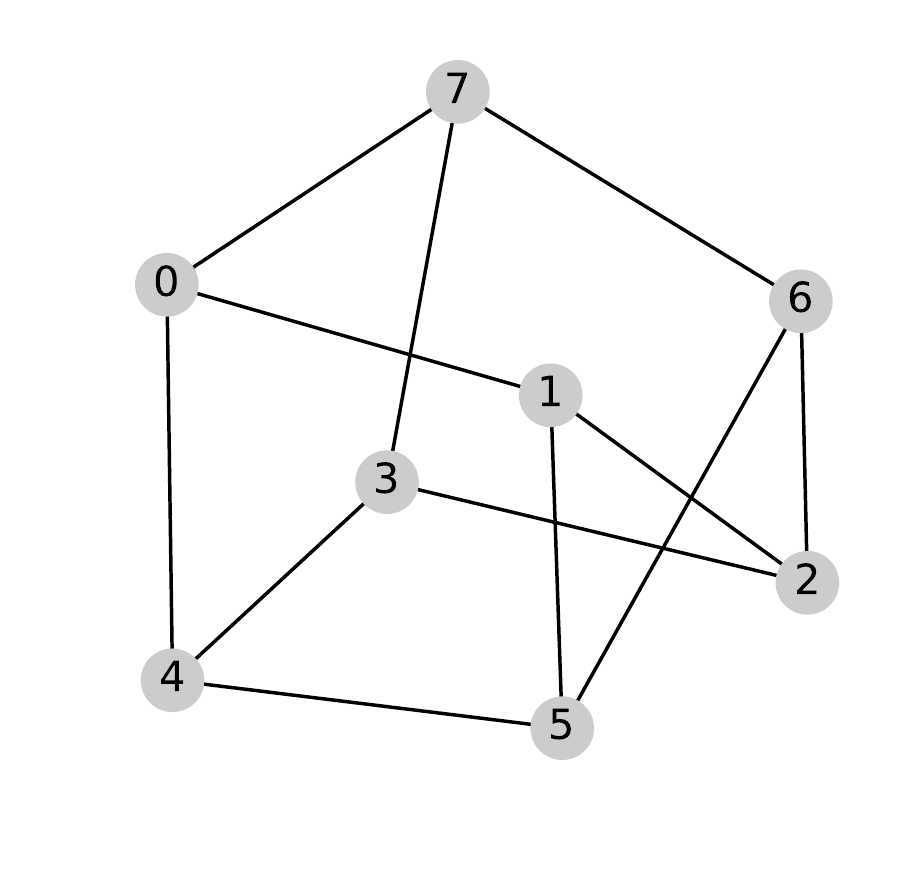} \hfill ~\\
    ~\hspace{1cm} Grover \hfill Fourier \hfill Grover \hfill Fourier \hspace{1.5cm}~\\
    \includegraphics[width=0.24\textwidth]{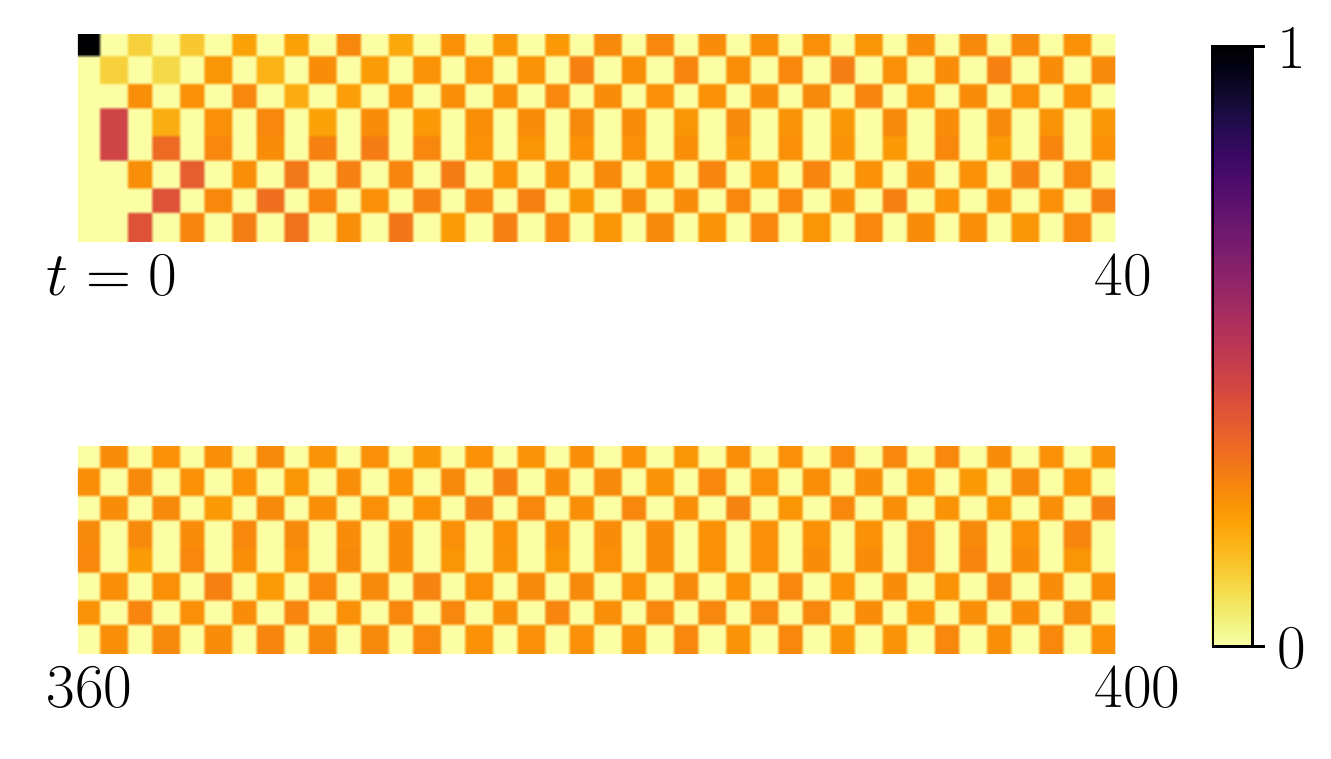} \hfill %
    \includegraphics[width=0.24\textwidth]{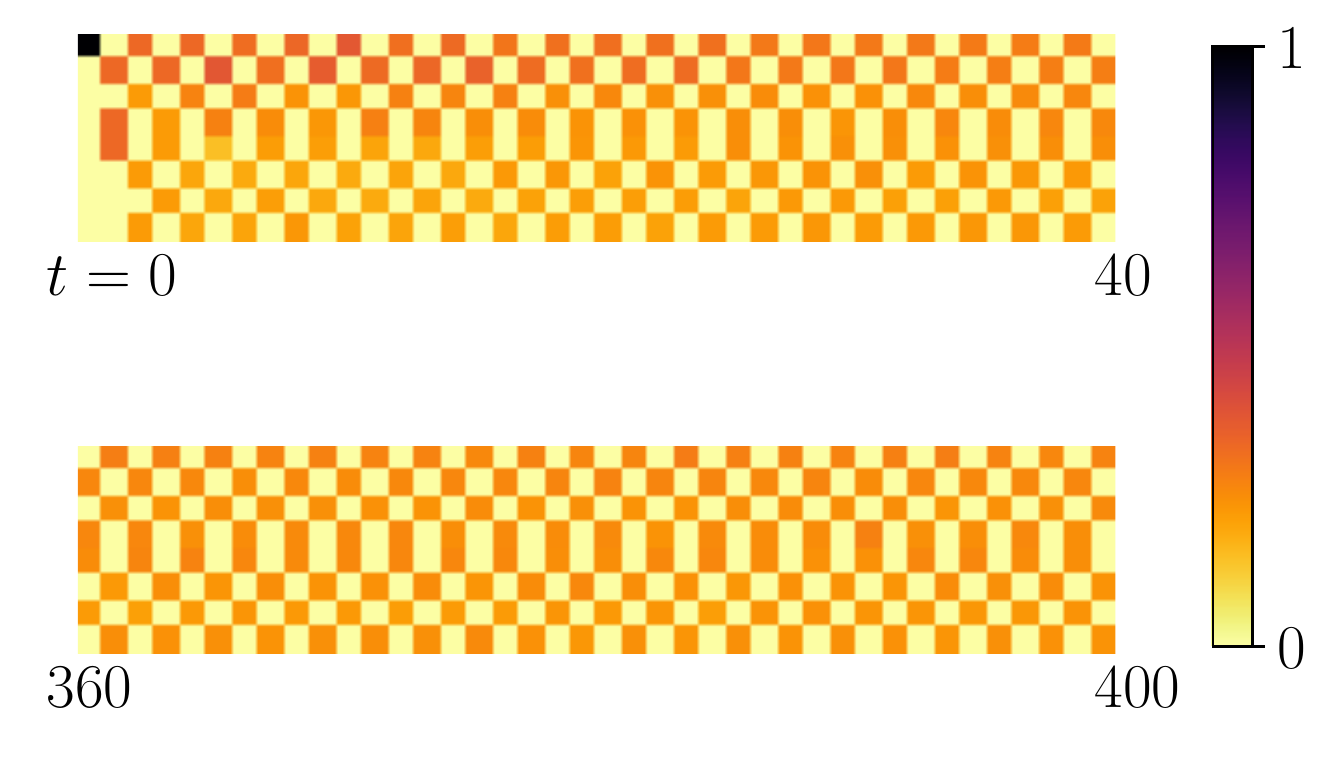} \hfill %
    \includegraphics[width=0.24\textwidth]{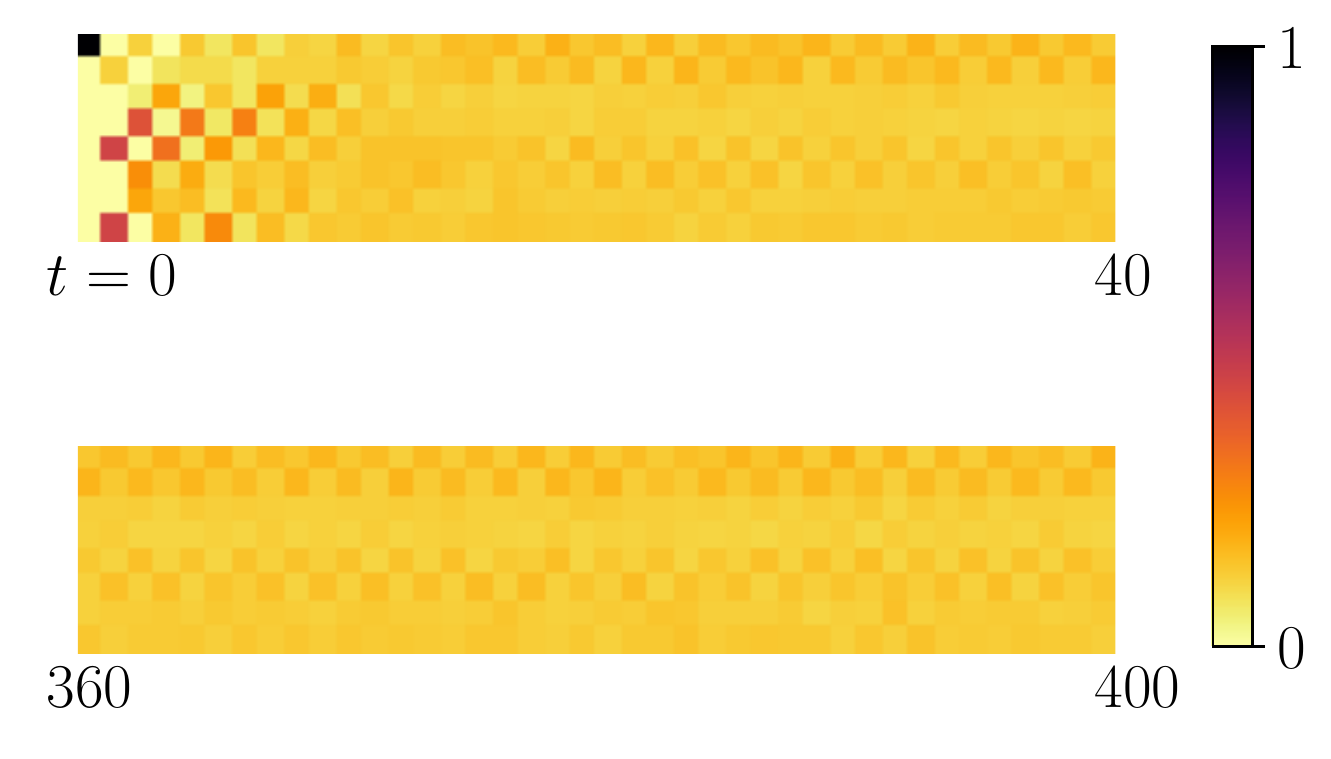} \hfill %
    \includegraphics[width=0.24\textwidth]{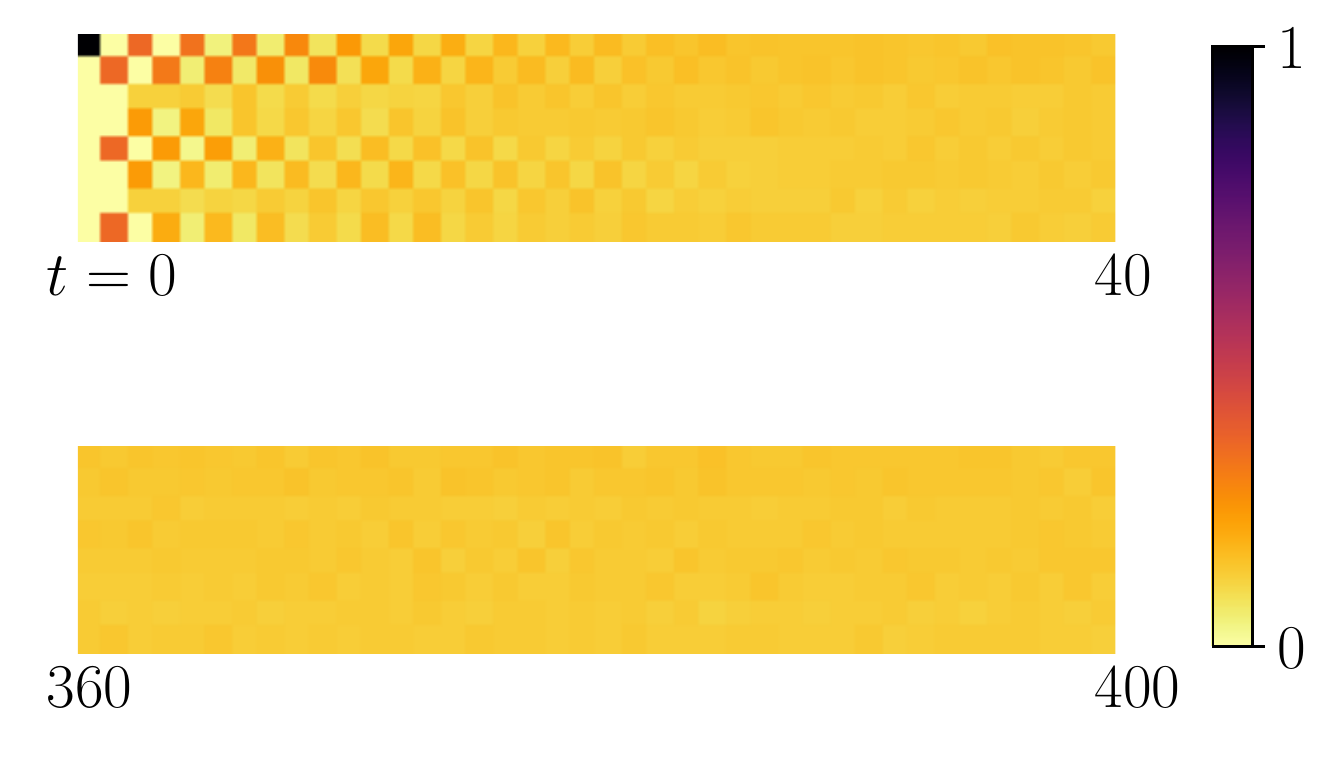} \\
    \includegraphics[width=0.24\textwidth]{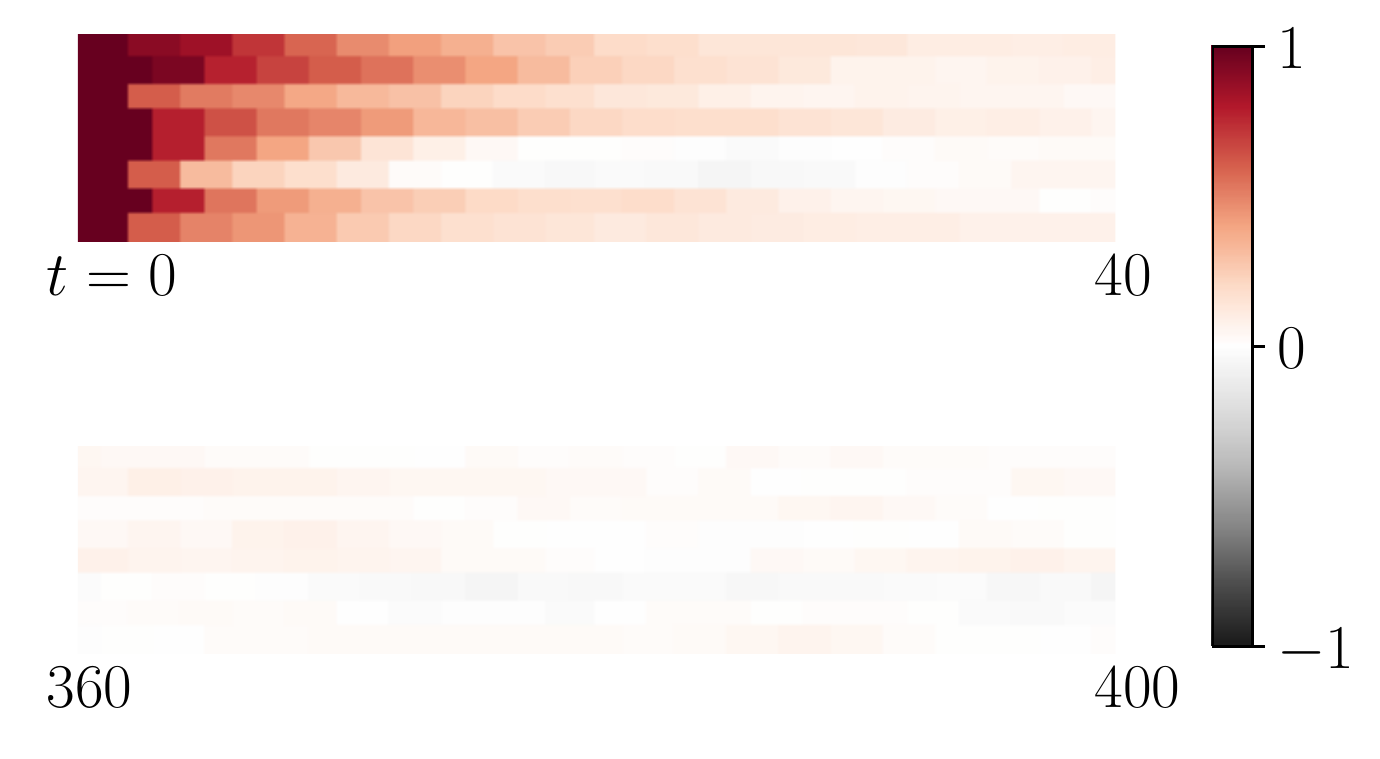} \hfill %
    \includegraphics[width=0.24\textwidth]{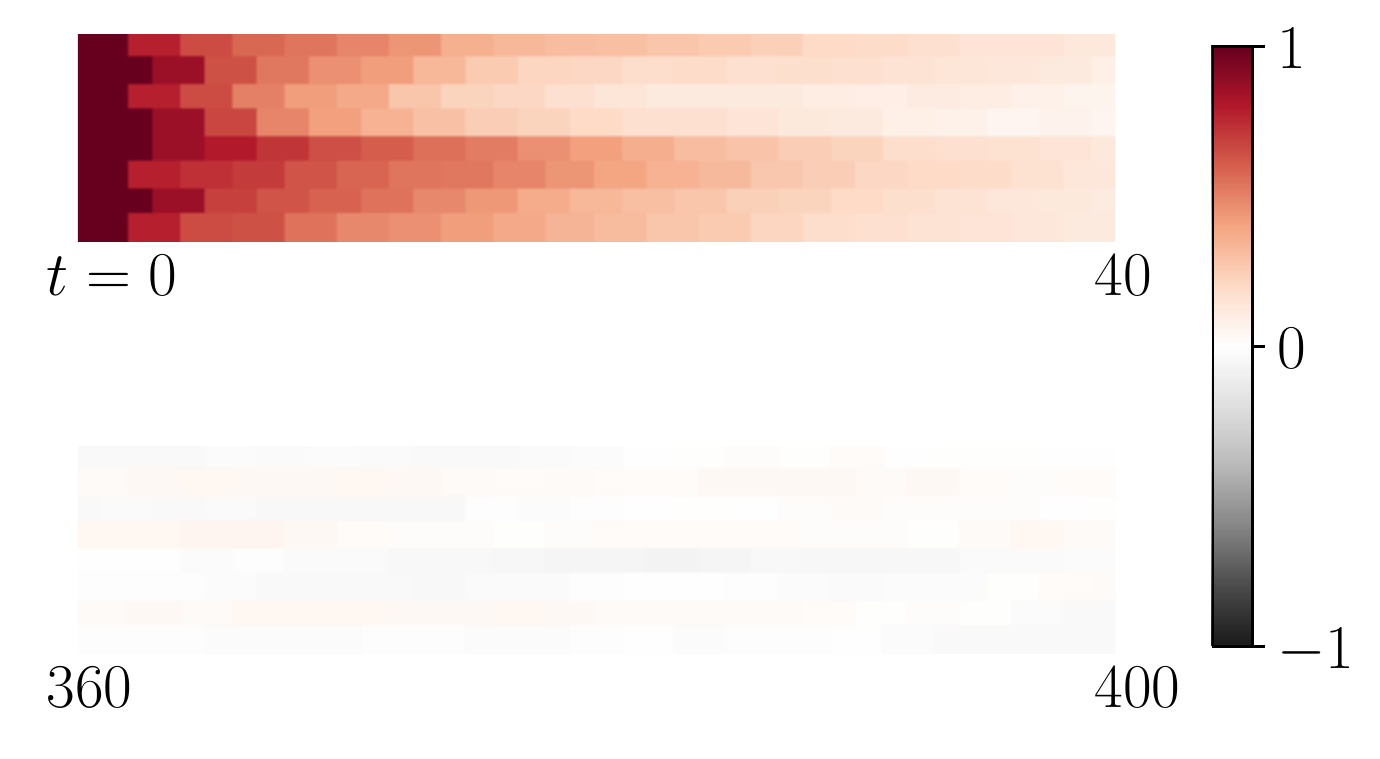} \hfill %
    \includegraphics[width=0.24\textwidth]{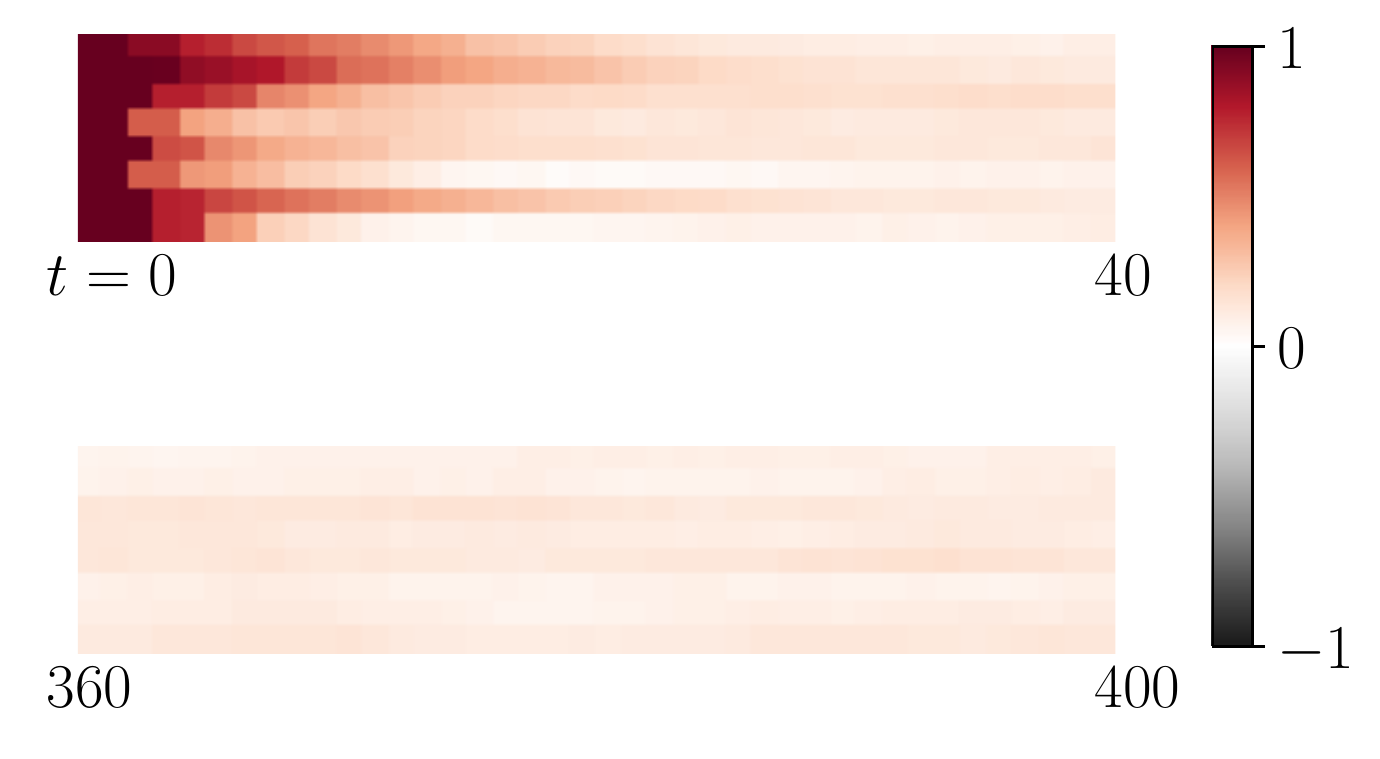} \hfill %
    \includegraphics[width=0.24\textwidth]{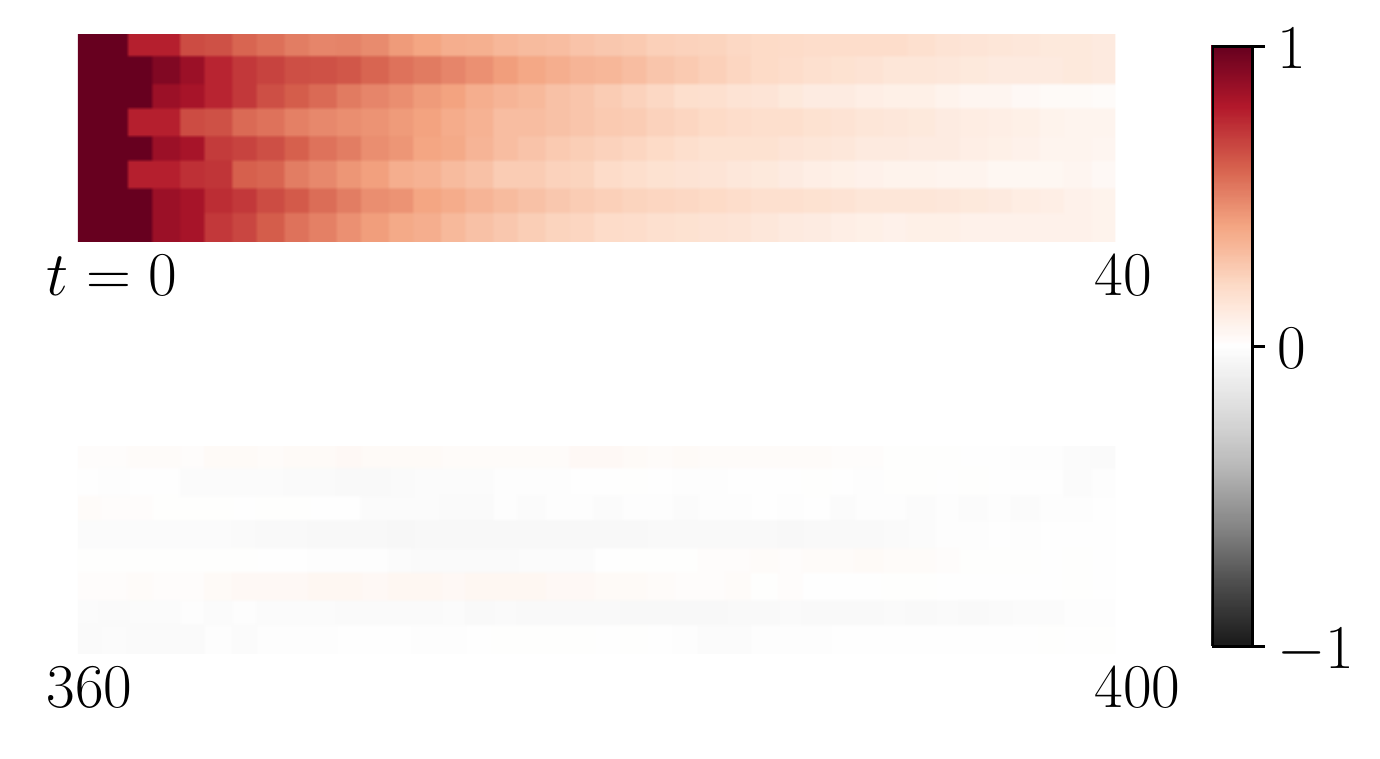} \\
    \includegraphics[width=0.24\textwidth]{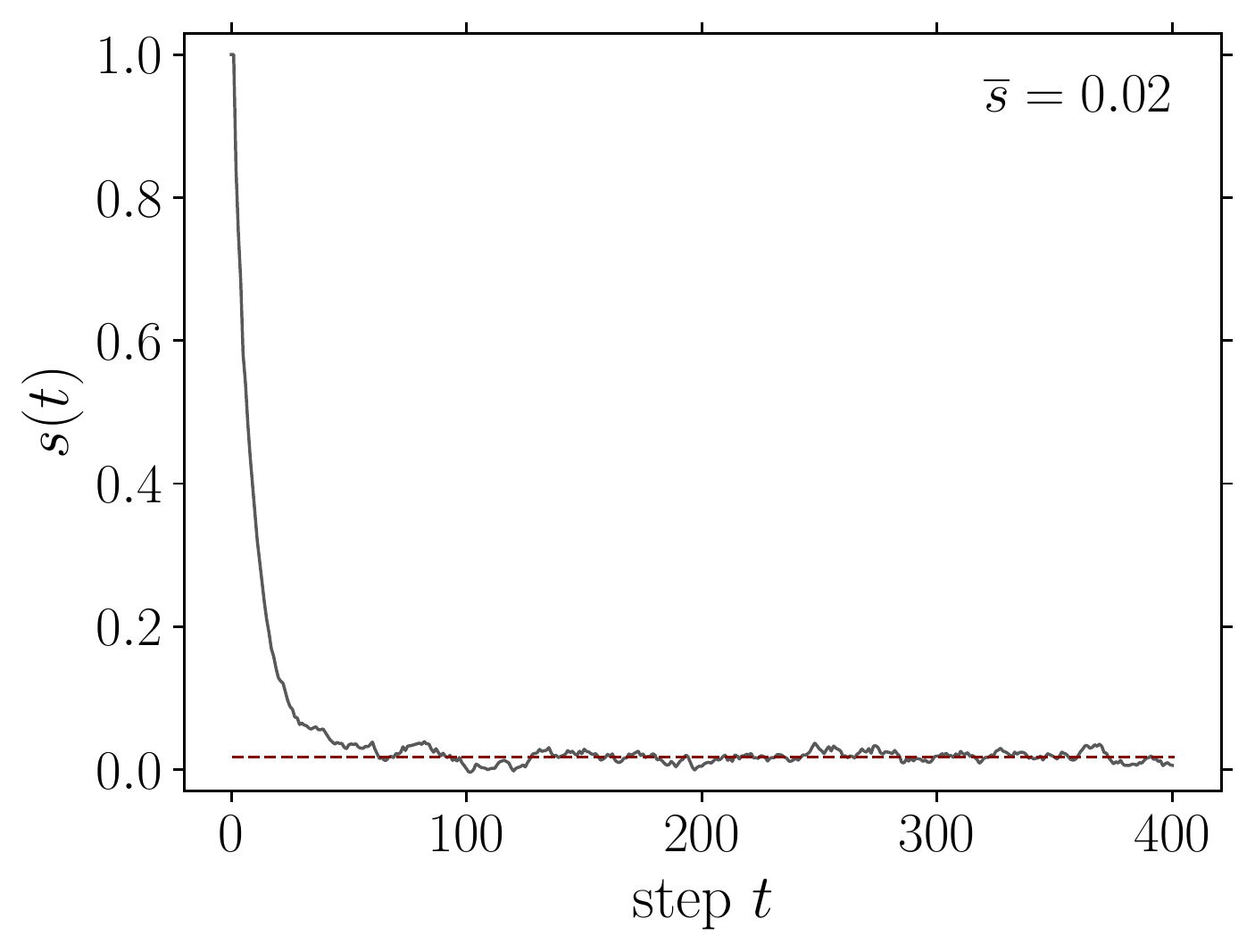} \hfill %
    \includegraphics[width=0.24\textwidth]{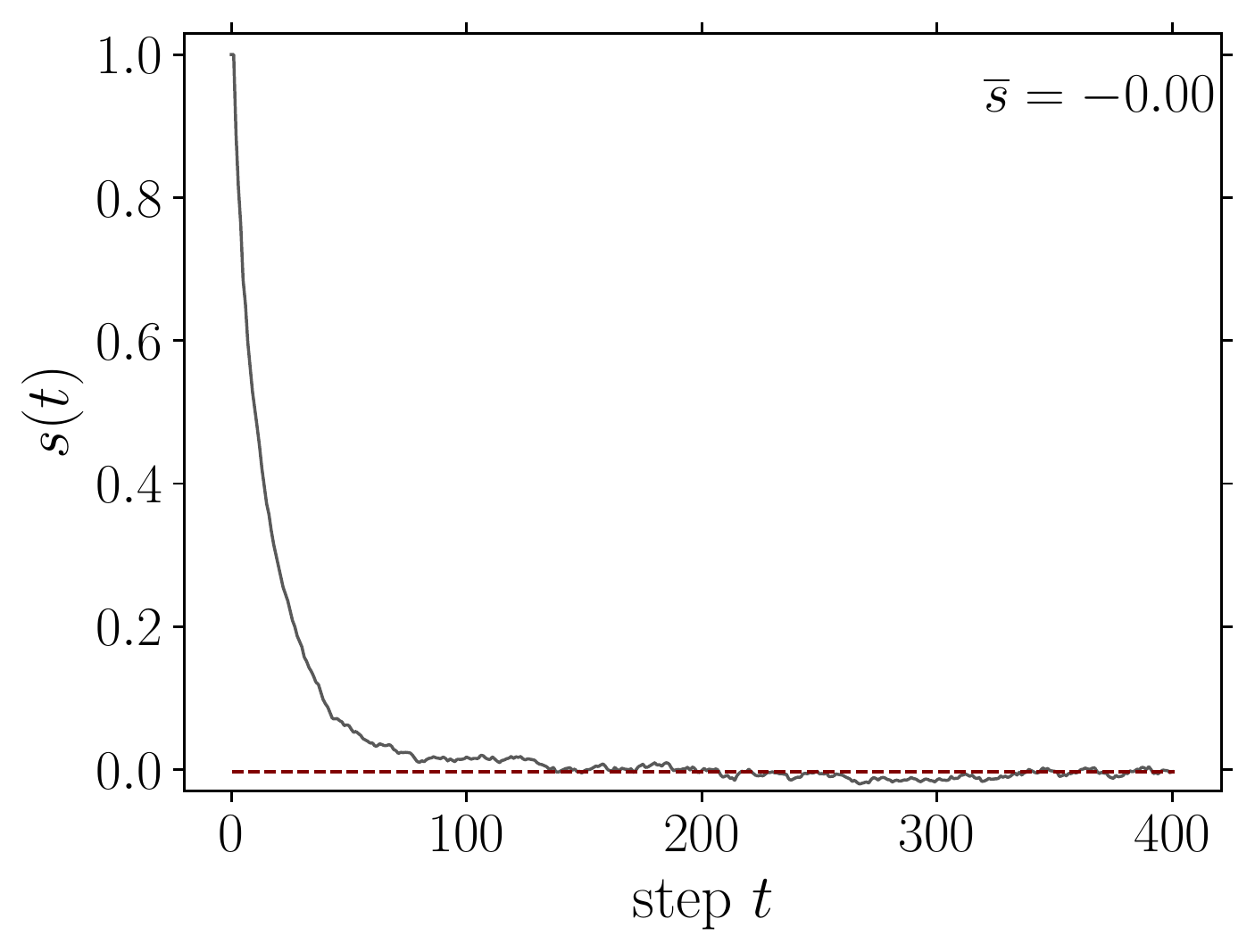} \hfill %
    \includegraphics[width=0.24\textwidth]{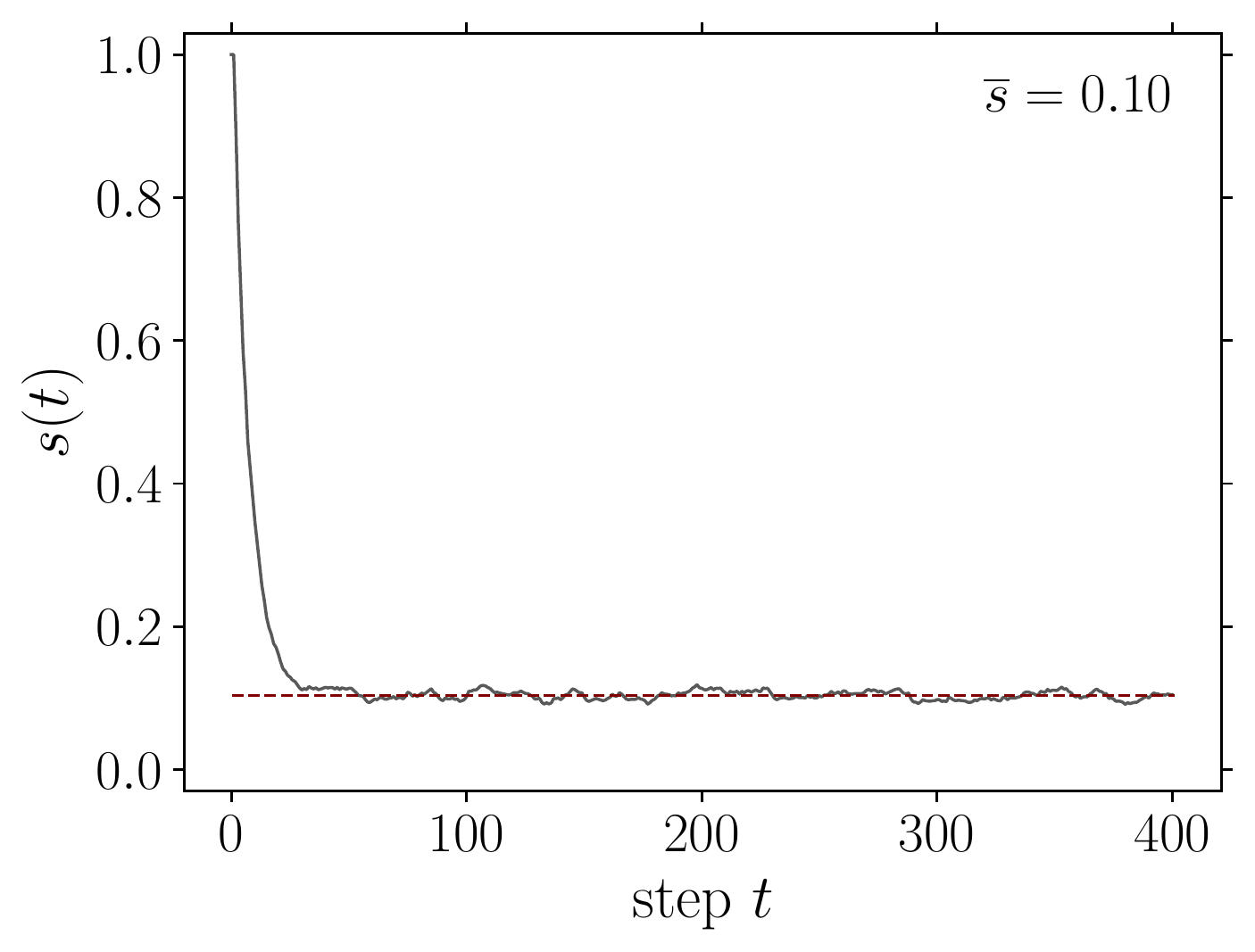} \hfill %
    \includegraphics[width=0.24\textwidth]{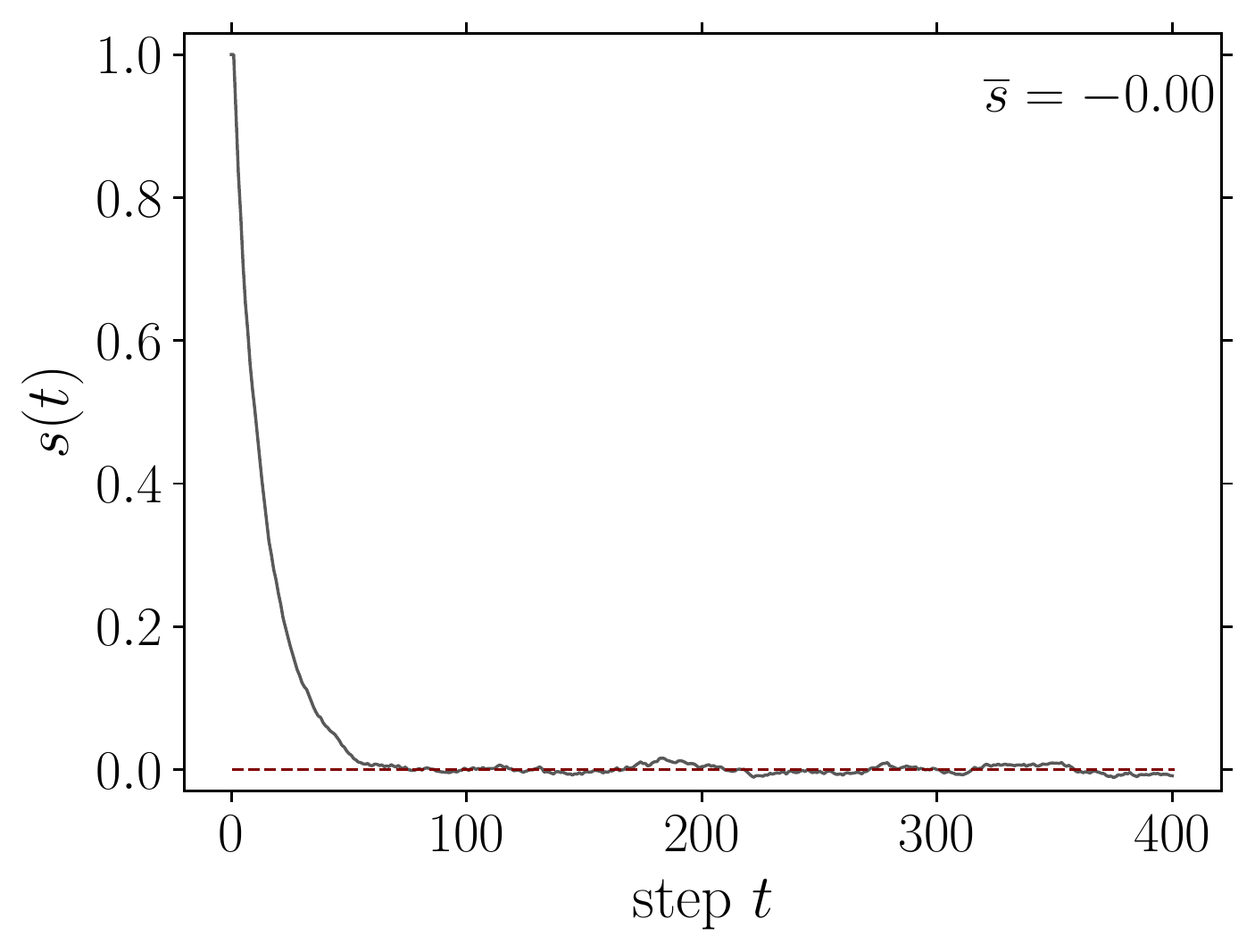} \\
    \includegraphics[width=0.24\textwidth]{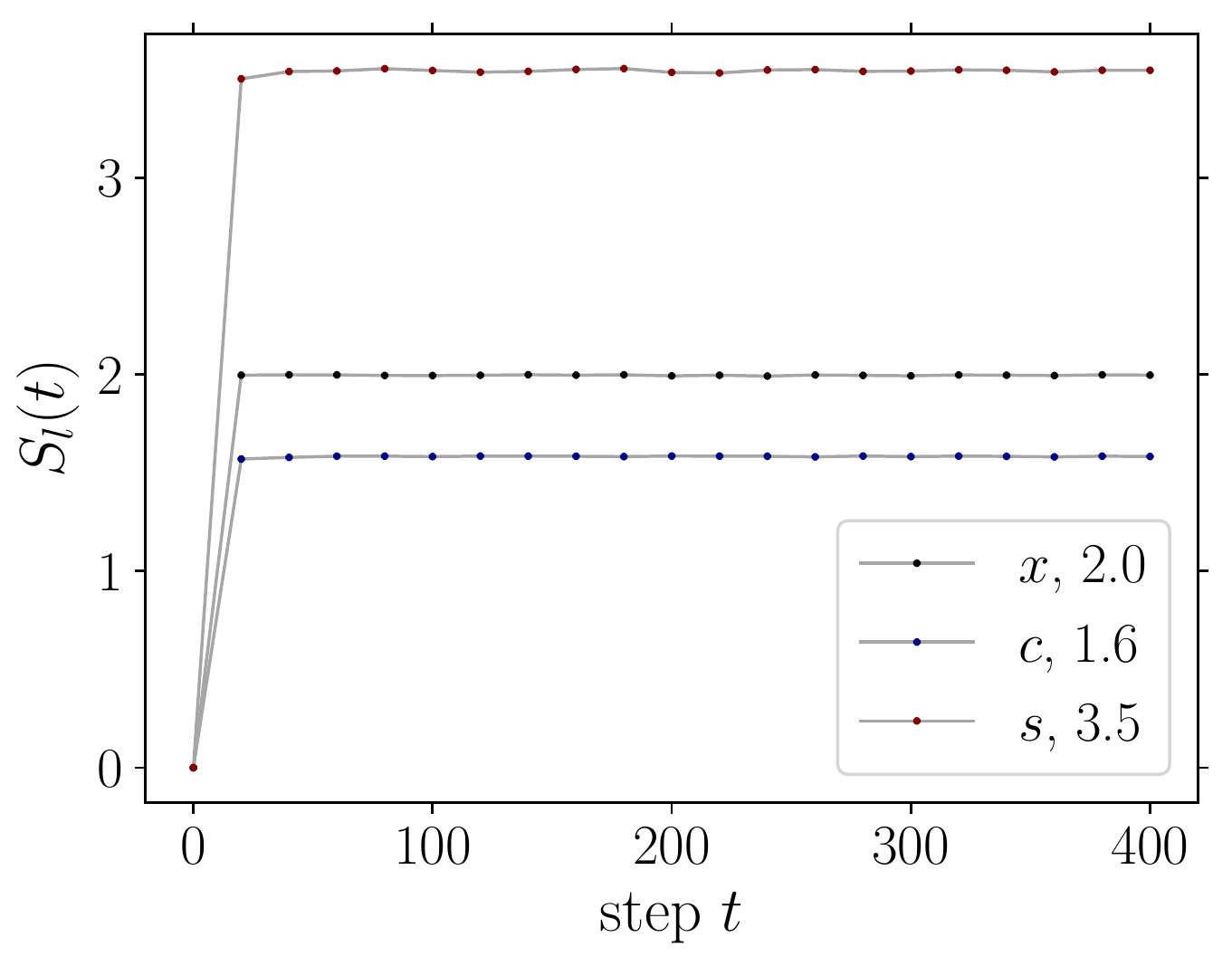} \hfill %
    \includegraphics[width=0.24\textwidth]{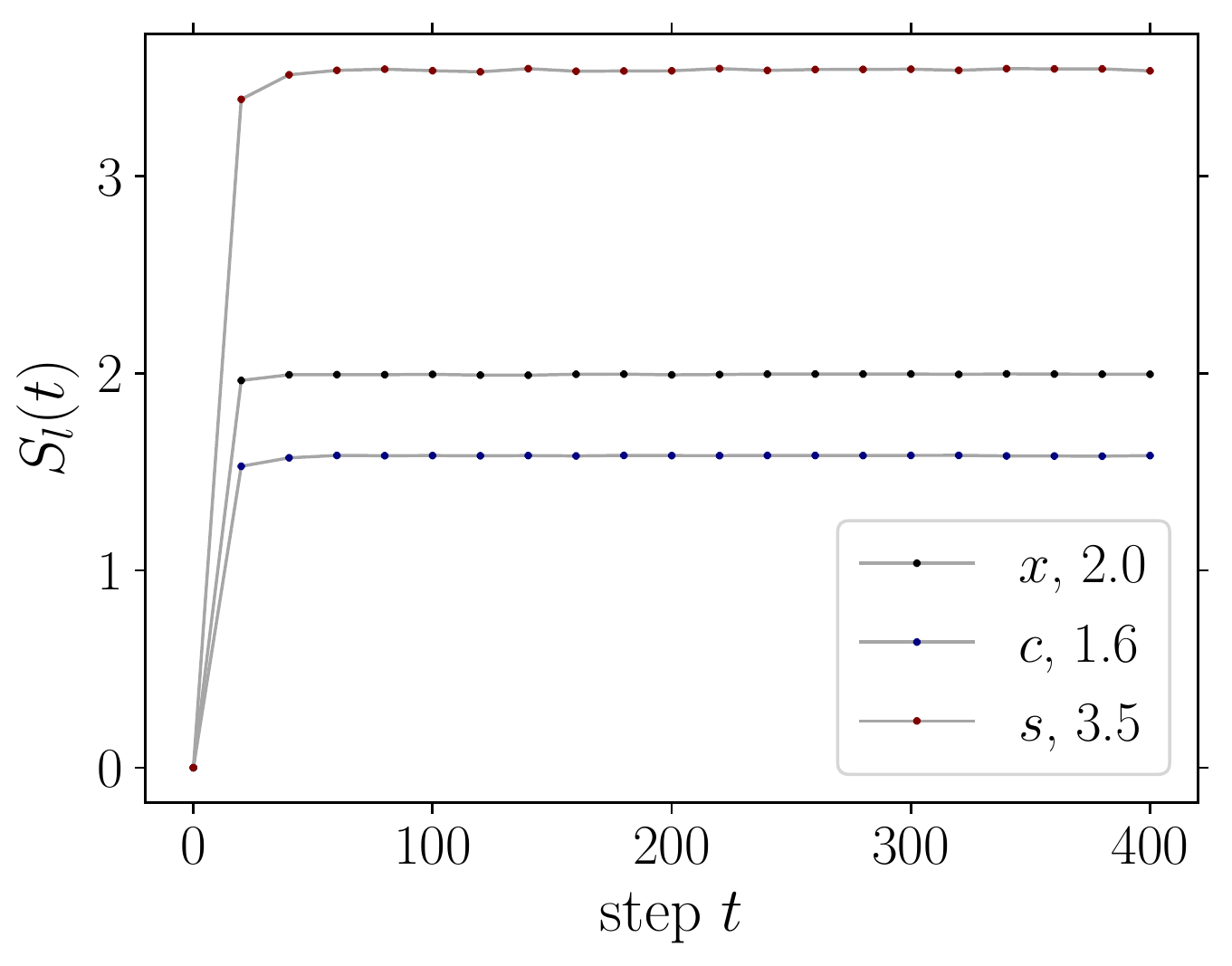} \hfill %
    \includegraphics[width=0.24\textwidth]{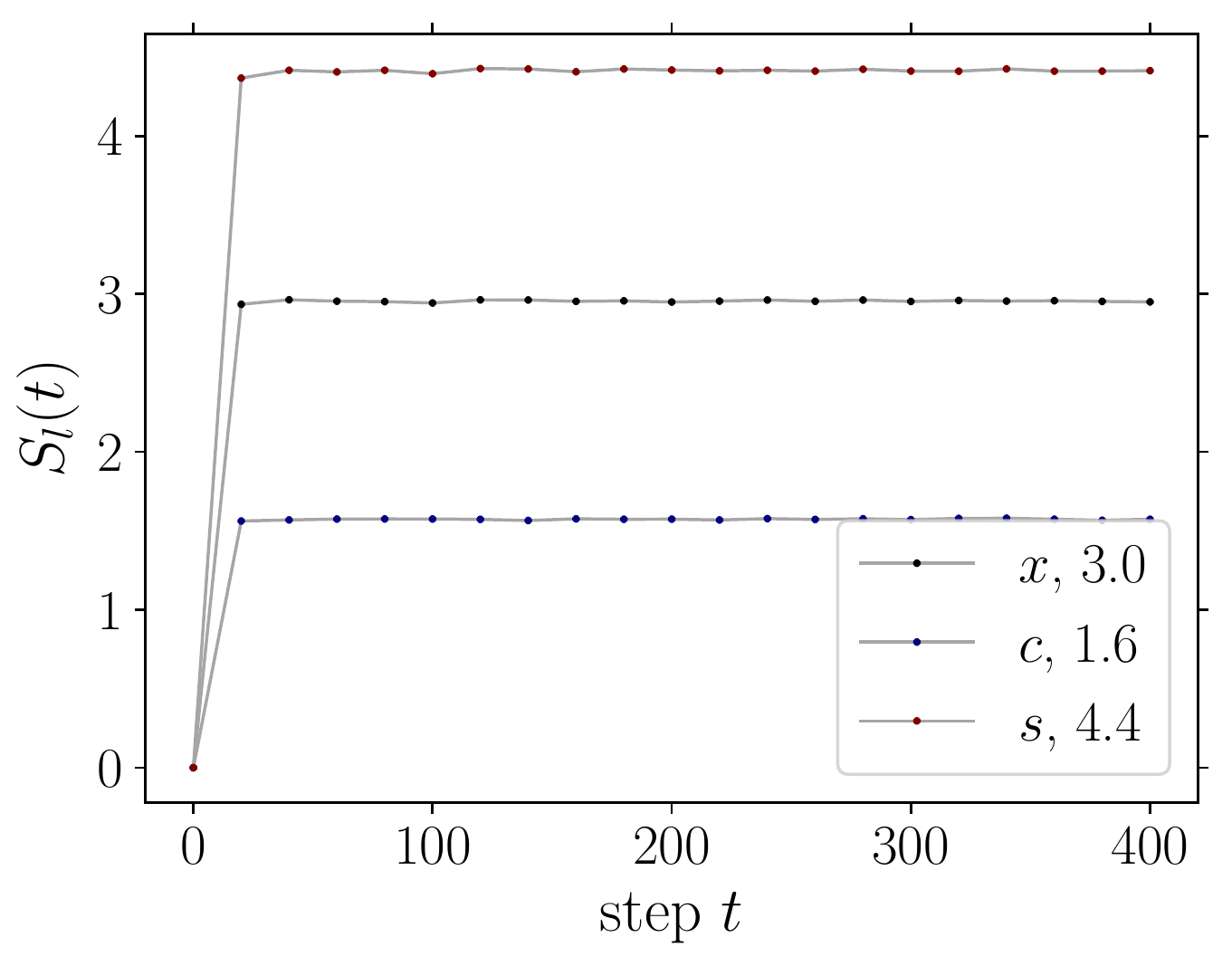} \hfill %
    \includegraphics[width=0.24\textwidth]{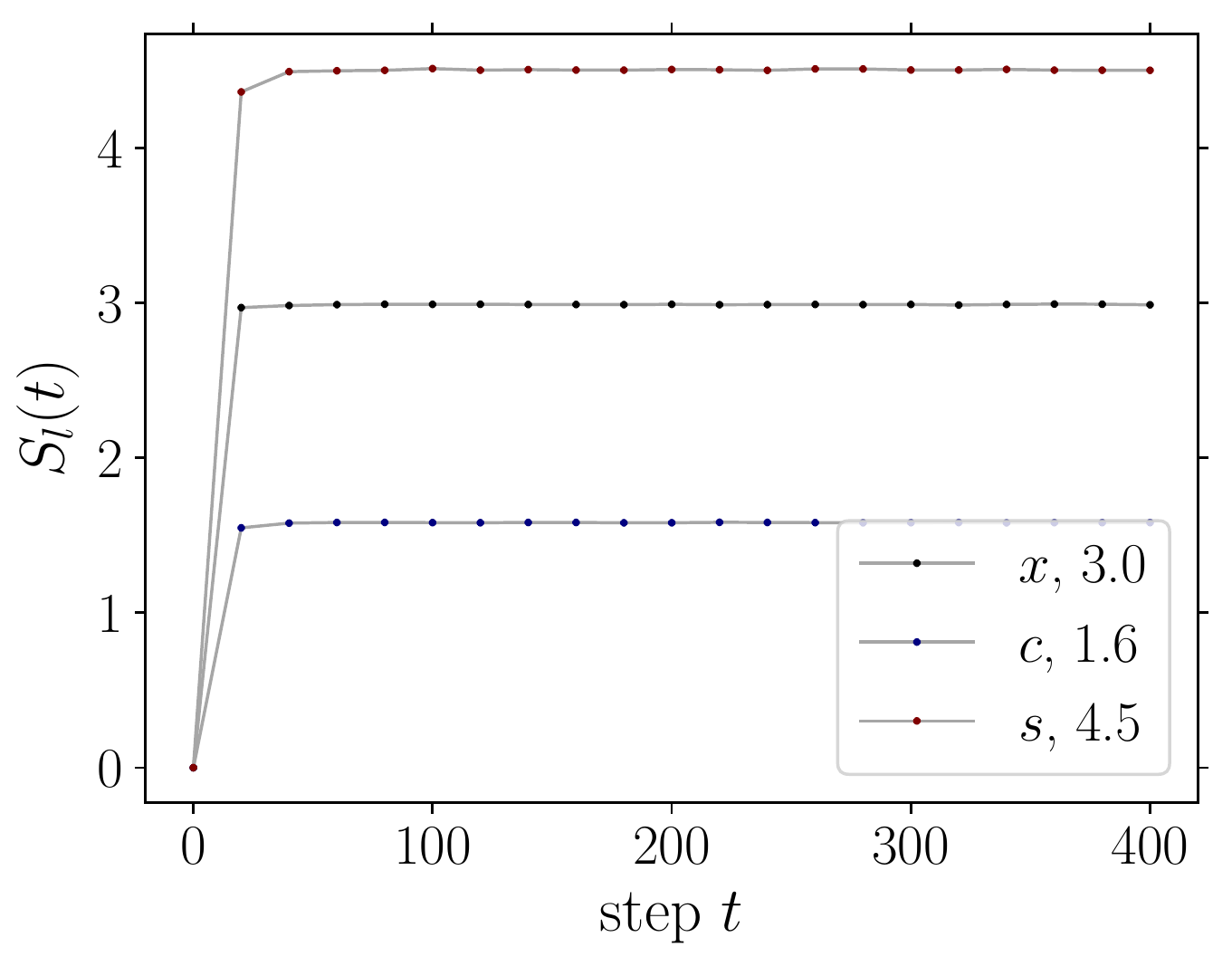}
    \caption{Lattice graphs: comparison of the cube (columns 1 and 2) and möbius (columns 3 and 4) graphs using Grover (columns 1 and 3) and Fourier (columns 2 and 4) operators. Row 1, position distribution over the 8 nodes (image pixels) as a function of the time step (top panel $t=(0,40)$, bottom panel $t=(l360,400)$); Row 2, spin distribution over the nodes; Row 3, mean spin as a function of step number (the dasched line gives the mean value); Row 4, spin (red, upper line), position (gray, middle line) and color (blue, lower line) entanglement entropy $l=x,c,s$.}
    \label{f:PH}
\end{figure*}

\section{Phenomenology}

In order to investigate the properties of the quantum state \(\ket{\psi(t)}\), build after \(t\) applications of the unitary matrix \(U\), we define macroscopic quantities characterizing the probability distribution of the particle over the nodes \(p(x,t)\) (density), 
\begin{equation}
  \label{e:p}
  p(x,t) = \Tr_{cs} \rho(xcs;t)\,,
\end{equation}
and the mean value of the spin \(s(x,t)\) in the \(z\) direction (graph magnetization):
\begin{equation}
  \label{e:sxt}
    s(x,t) = \Tr_{yc} \sigma_z(x) \rho(ycs;t)\,,
\end{equation}
where 
\begin{equation}
  \label{e:rhot}
   \rho = \rho(xcs;t) =  \ket{\psi(t)} \bra{\psi(t)}\,,
\end{equation}
is the density matrix at step \(t\) of the total system, and,
\begin{equation}
  \label{e:psi}
    \ket{\psi(t)} = \sum_{\{x,c,s\}} \psi_{xcs}(t) \ket{xcs}  \,,
\end{equation}
is the quantum state, with \(\psi_{xcs}\) its complex amplitudes in the \(\ket{xcs}\) basis. We denote \(\Tr_l\) the partial trace over the degrees of freedom corresponding to the specified label \(l = x,c,s\); \(\sigma_z(x)\) is the \(z\)-Pauli matrix of node \(x\). We also compute the mean value of the spin density over time,
\begin{equation}
  \label{e:sm}
  s(t) = \frac{1}{T-t_0+1} \sum_{t=t_0}^T s(x,t) \,,
\end{equation}
where \(t_0\) is a suitable initial time in the stationary state, chosen to avoid the relaxation transient.

\begin{figure}
  \centering
  \includegraphics[width=0.2\textwidth]{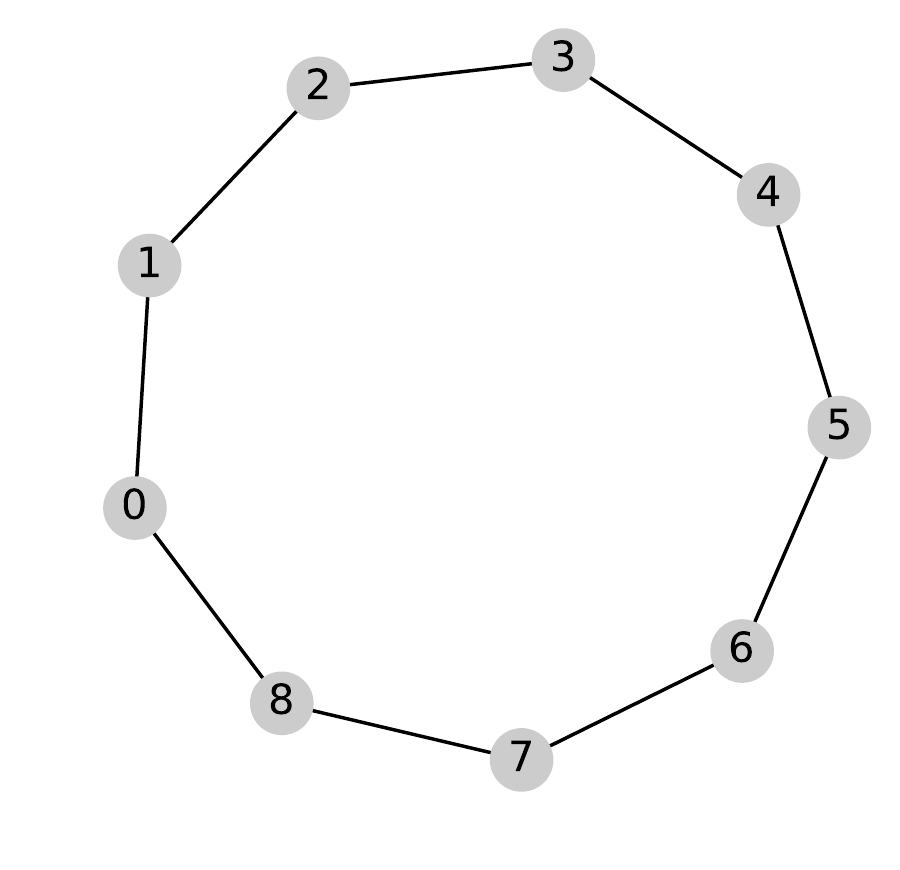}\\
  \includegraphics[width=0.48\textwidth]{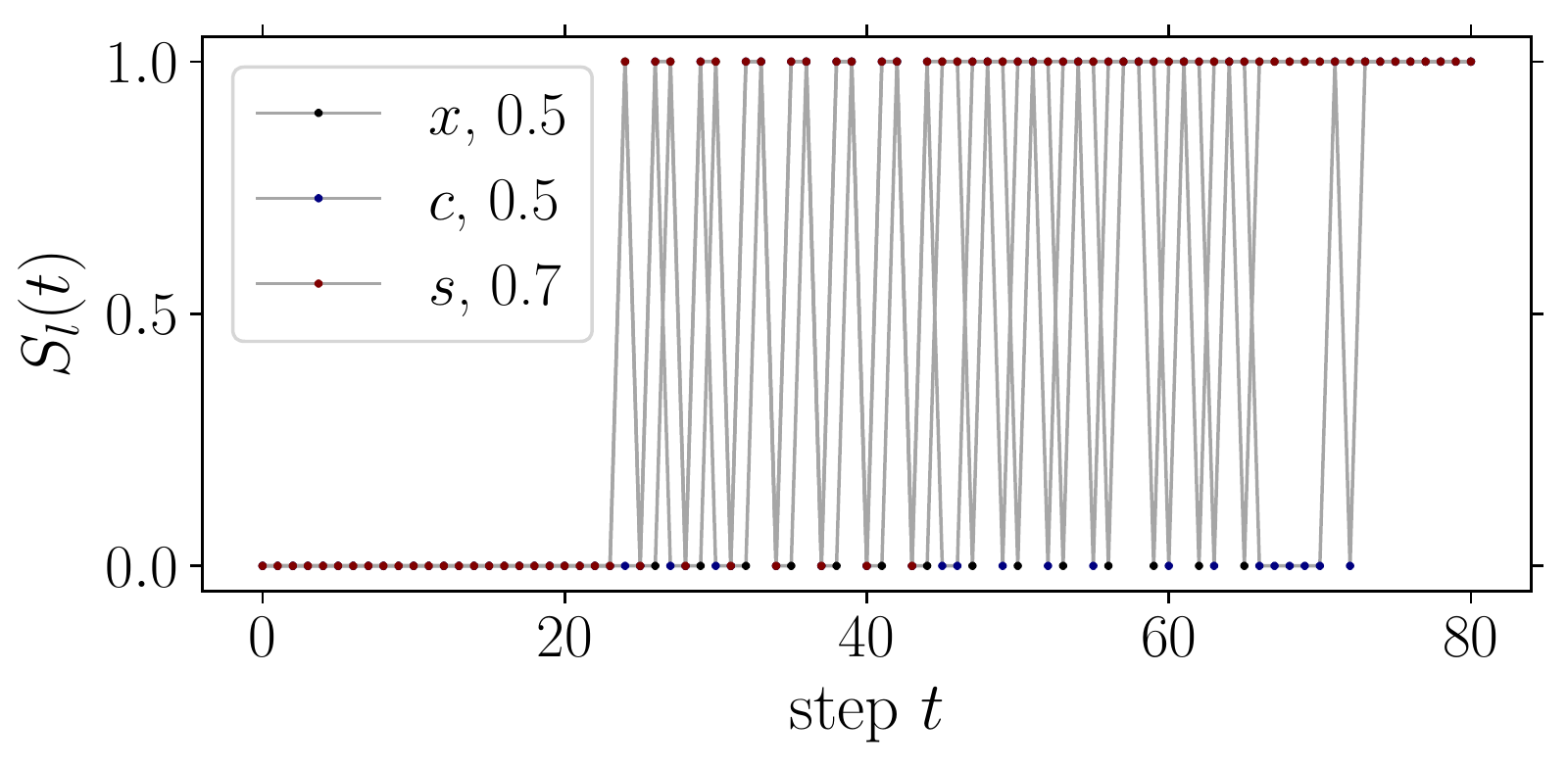}\\
  \includegraphics[width=0.49\textwidth]{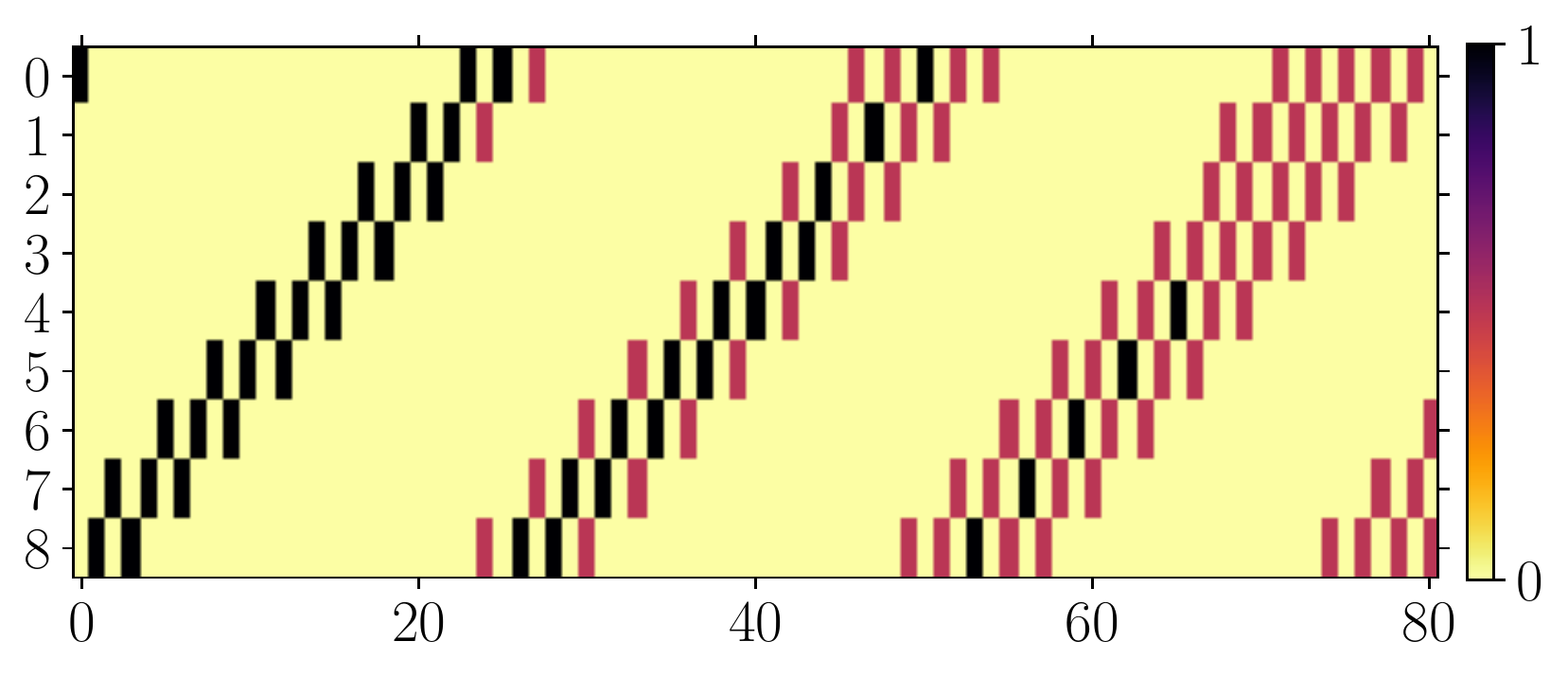}\\
  \includegraphics[width=0.49\textwidth]{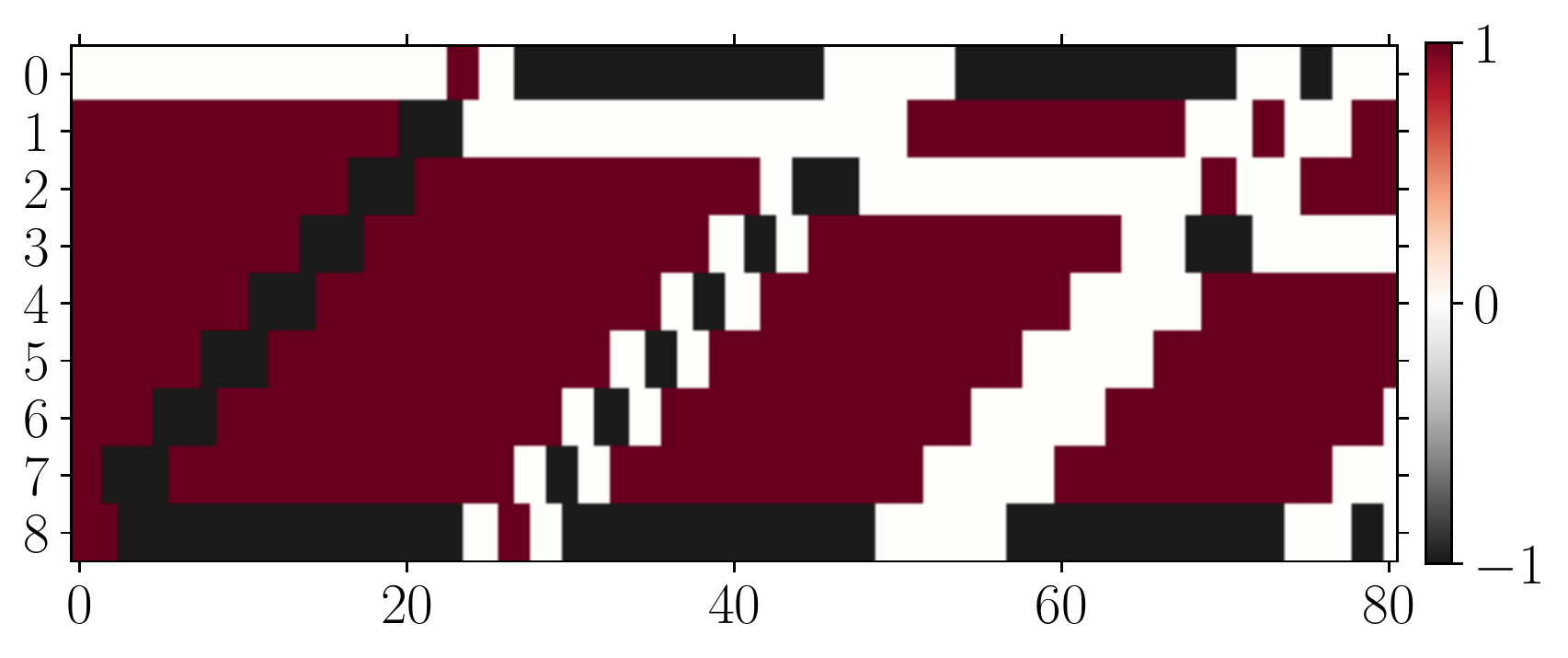}
  \caption{Cycle graph with $9$ vertices. Entanglement entropy as a function of time step, $l=x,c,s$ (top panel); position distribution probability (middle panel), and magnetization (bottom panel).
  \label{f:Y9}}
\end{figure}

In addition to these local properties, we can characterize global properties of the system's state, using the entanglement von Neumann entropy; we split the total degrees of freedom into two sets, one among position, color and spin \(l = x,c,s\), and the other including their complementary set \(\bar{l} =  \{c,s\}, \{x,s\}, \{x,c\}\):
\begin{equation}
  \label{e:ENT}
  S_l(t) = - \Tr \rho(l;t) \log \rho(l; t)\,, \quad
  \rho(l; t) = \Tr_{\bar{l}} \rho
\end{equation}
where \(\rho(l; t)\) is the partial trace of the total density matrix, over the $l$-label complementary set \(\bar{l}\). We use the notation \(\log = \log_2\) for the logarithm in base 2.

We present in Fig.~\ref{f:PH} an overview of the properties of the interacting quantum walk for two similar graphs (same number of nodes and same regular degree), `cube' (`c') and `möbius' (`m'), and two choices of the coin operator \(\GR\), and \(\FT\). In the initial state the particle is at node 0, and the spins are up. The images of the probability to find the walker at node \(x\) (represented by the eight vertical pixels) show an interesting difference between the two graphs (first row). The walker on `c' is split into two states, one on the nodes `0257', and the other on `1346' (alternating darker pixels), which correspond to paths with distance 2 steps (according to the node numbering). These two states are related by a swap symmetry, which is precisely the action of the \(\MV\) operator on the position subspace. Such alternating pattern is not possible on the `m' graph, which presents a uniform distribution of the position over the whole graph, this graph lacking the swap symmetry. These patterns are similar for the two coins.

The graph magnetization tends to be uniform (second row), and generally sets in a `paramagnetic' state (with vanishing mean spin), except for `m' with the Grover coin, for which a remaining magnetization is present at long times. The combined action of \(\SW\) and \(\CZ\) is to flip and entangle spins, which happens when node spins are in an equally probable superposition of up and down spin. In the case of the `m' graph, with the Grover coin (which respects the graph symmetry), some frustration appears due to the existence of cycles with an odd number of nodes; this feature is absent in the `c' graph. The Fourier coin breaks the distinction between cycles by superposing the amplitudes around each node, and then it can avoid this kind of frustration (interference), restoring the zero magnetization state. In this respect, the behavior of the entanglement entropy is illuminating. The coin entropy is maximal, whatever the coin; we can thus concentrate on the position and spin entropies. The position and spin entanglement entropies differ, between `c' and `m', in just one qubit: while the `cube' encode two qubits into its position (corresponding to a subset of \(4\) nodes), the `möbius' graph encodes \emph{three} qubits, the maximum possible for \(8\) nodes. This behavior can be related to the cycles of four nodes observed in the `cube', not present in the `möbius' graph. The stationary values of the entropy are independent of the type of coin used in the quantum walk. Note also that the spin entropy is slightly larger with the Fourier coin than with the Grover coin, in agreement with the difference found in the mean magnetization: larger entanglement entropy corresponding to random spins. 

The scenario just presented is substantiated by the behavior of similar graphs with a higher number of nodes. We considered ladders with even and odd number of rungs,  periodic ladders like `c' (planar graphs), or antiperiodic like `m' (graphs with one crossing). Even ladder graphs show the same magnetization and entanglement properties as `c'; odd ones (for example, a five node length ladder, forming two pentagons), exchange their entanglement properties: the planar graph can code a supplementary qubit with respect to the crossed one. Crossed graphs, independently of the number of ladder rungs, support a non vanishing magnetization (with the Grover coin). More generally, random graphs with the same number of nodes and similar regular or mean degree, behave like the `möbius' or larger equivalent graphs, suggesting that this represents the ``typical'' case: the `cube' class, is special in the sense that the position entanglement entropy do not reach its maximum value, reflecting the existence of equivalent cycles, cycles related by the swap symmetry.

To better appreciate the interplay of the interaction operators \(\SW\) and \(\CZ\), we present the dynamics of a cyclic graph (with \(9\) nodes) in Fig.~\ref{f:Y9}. The initial state is a superposition of spin up and down at node \(0\):
\[
\ket{\psi(0)} = \frac{\ket{000} + \ket{001}}{\sqrt{2}}  \,,
\]
(a basis ket is, for this graph, an eigenvector of \(U\)). After an initial cycle, the entanglement entropy start to get values of \(1\) for the three types of degrees of freedom \(l=x,c,s\); in particular, after about \(50\) steps, a spin qubit is present. This is related to the appearance of a pair of neighboring zero spin nodes (white squares), the \(\CZ\) entangles. The position entanglement entropy also codes one qubit, which can be related to the two correlated nodes (red squares), appearing for the first time at step \(24\). Note that entanglement arise just after the completion of a cycle: this is an effect of the swap motion that allows superposition of amplitudes only after a turn (that may depend on its parity, for a given subnode labeling). 

\begin{figure}
  \centering
  \includegraphics[width=0.45\textwidth]{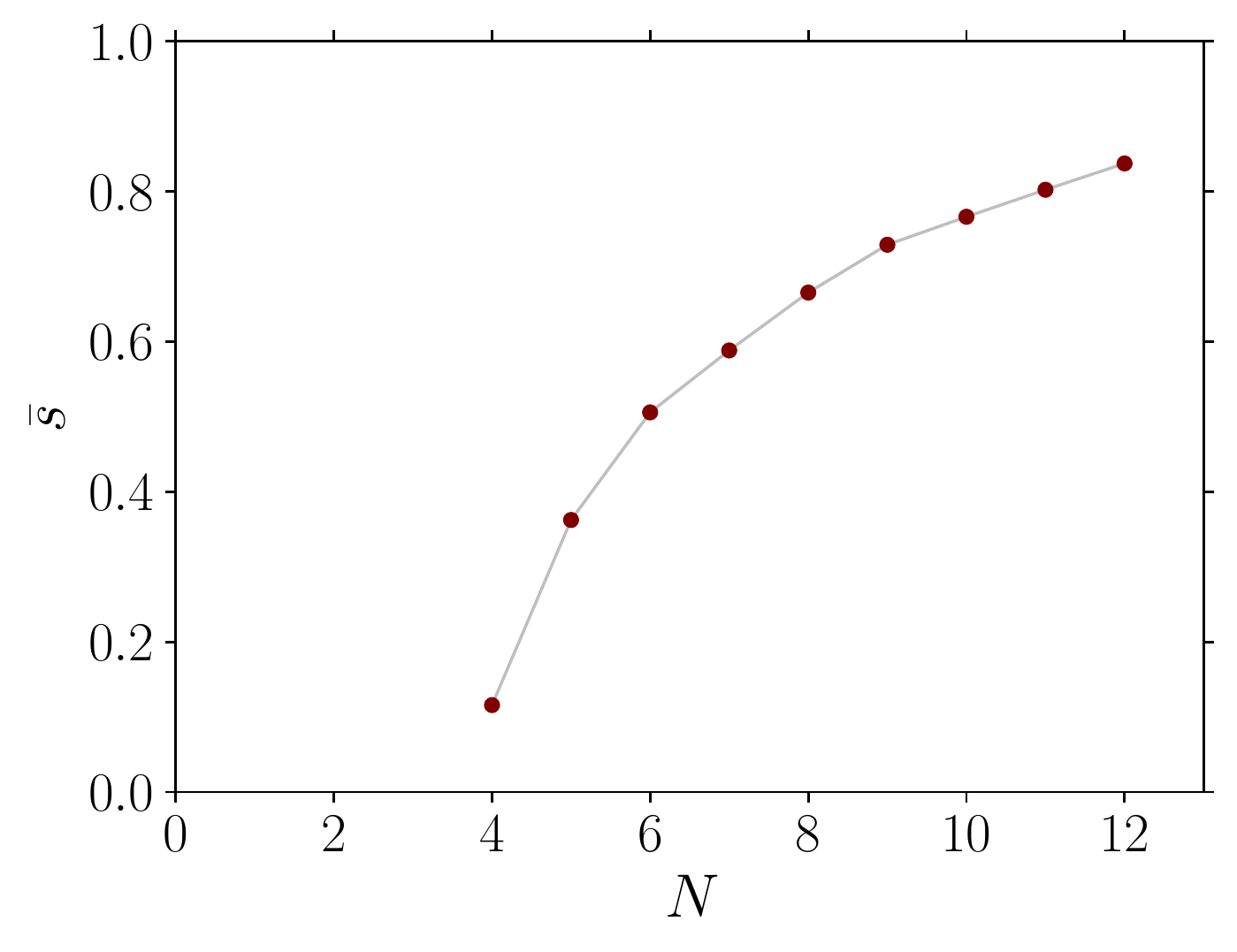}
    \caption{Complete graph of order $N$ magnetization $\bar{s}$.
  \label{f:K}}
\end{figure}

\begin{figure}
  \centering
  \includegraphics[width=0.45\textwidth]{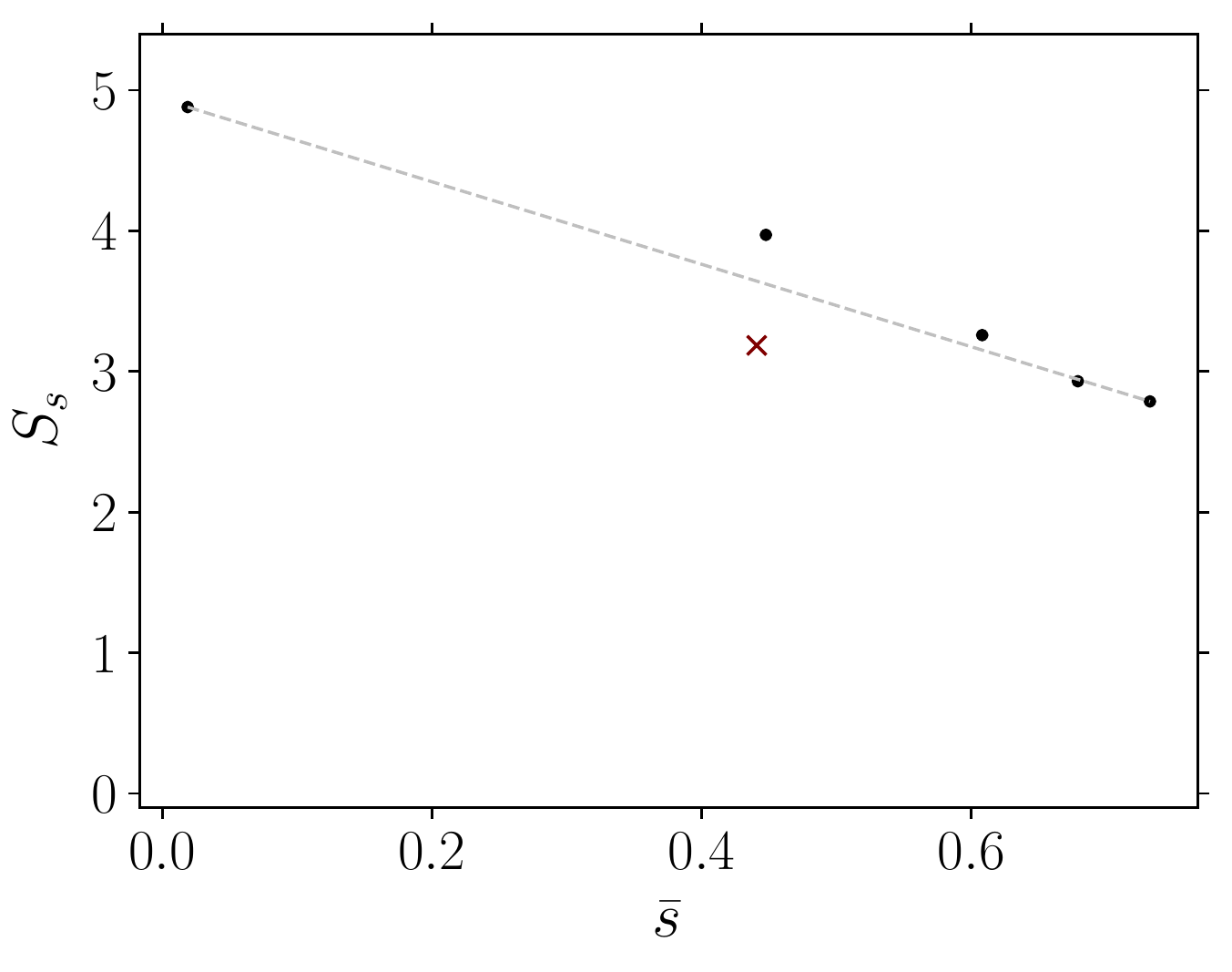}
    \caption{Spin entanglement entropy as a function of the mean magnetization (Grover coin), for random graphs of $10$ nodes, and the number of edges $m(m-1)/2$, with $m$ in the range $5$ to $10$ (the last one is the complete graph of order $10$). The dashed line is a guide to the eye. The cross mark, corresponds to a sparsea$m=5$, almost one dimensional graph, with $m=5$, do not follow the general linear pattern.
  \label{f:K10}}
\end{figure}

We also note that there is no a priori straightforward relationship between entanglement and total magnetization: taking the mean value of the spins over all nodes, implies mixing independent and correlated spins. However, we observed that the magnetization grows with the degree of the graphs, in the case of the Grover coin. This fact is well illustrated by the case of complete graphs \(K(N)\), as shown in Fig.~\ref{f:K}, where the magnetization is represented as a function of the number of nodes \(N\). In contrast, for the Fourier coin, the magnetization generally vanishes, strengthening the fact that the value of the asymptotic magnetization depends in a subtle way on the graph geometry and its entanglement properties. For instance, we find that, fixing the number of nodes, the increase of the number of edges is generally accompanied with an increase of the magnetization. This increase is in addition correlated with the spin entanglement. An illustration is provided by the plot of the spin entanglement entropy as a function of the magnetization in Fig.~\ref{f:K10}, computed using a Grover coin, for a set of order \(10\) random Erdős-Rényi graphs (`G' in Fig.~\ref{f:g}), between a sparse one with \(10\) edges (a chain ended by a cycle of four nodes), and the complete graph \(K(10)\), with \(45\) edges, showing the tendency of a linear decrease of the spin entropy with increasing magnetic order.

\begin{figure*}
    \centering
    \includegraphics[width=0.16\textwidth]{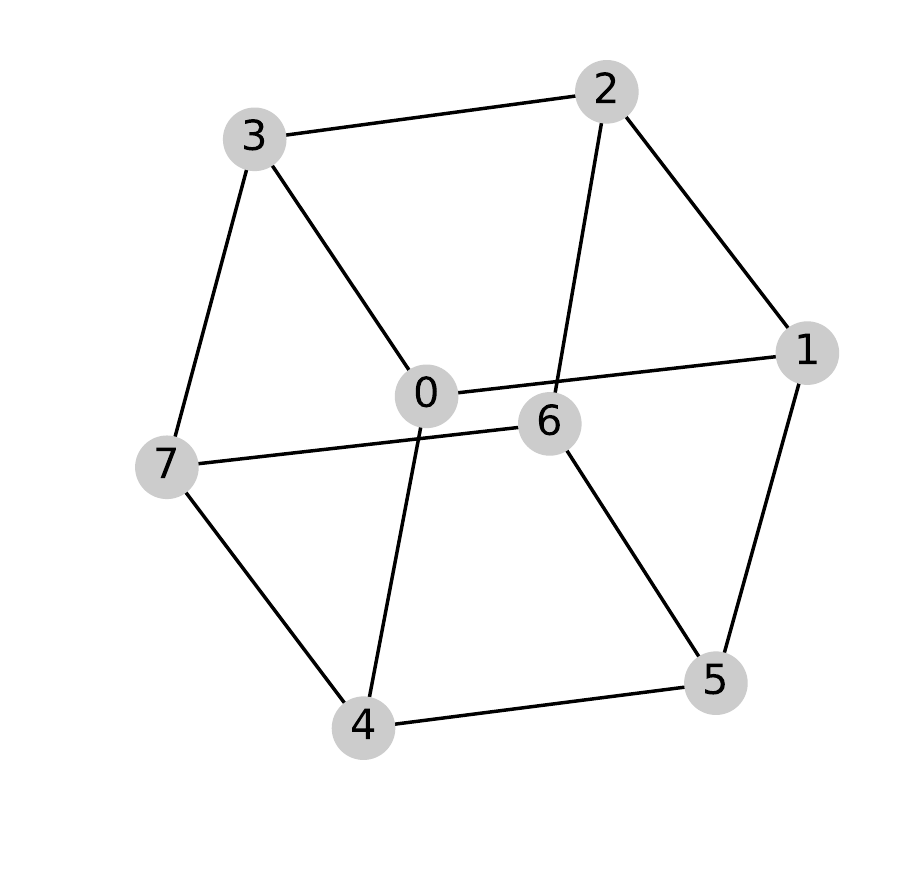} \hfill %
    \includegraphics[width=0.26\textwidth]{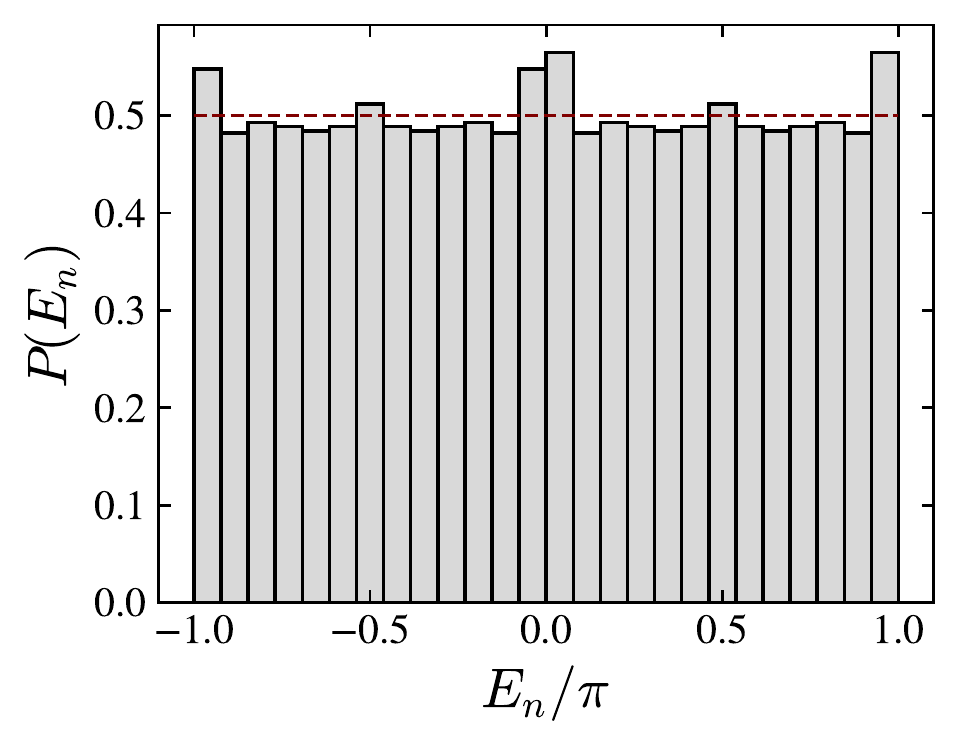} \hfill %
    \includegraphics[width=0.26\textwidth]{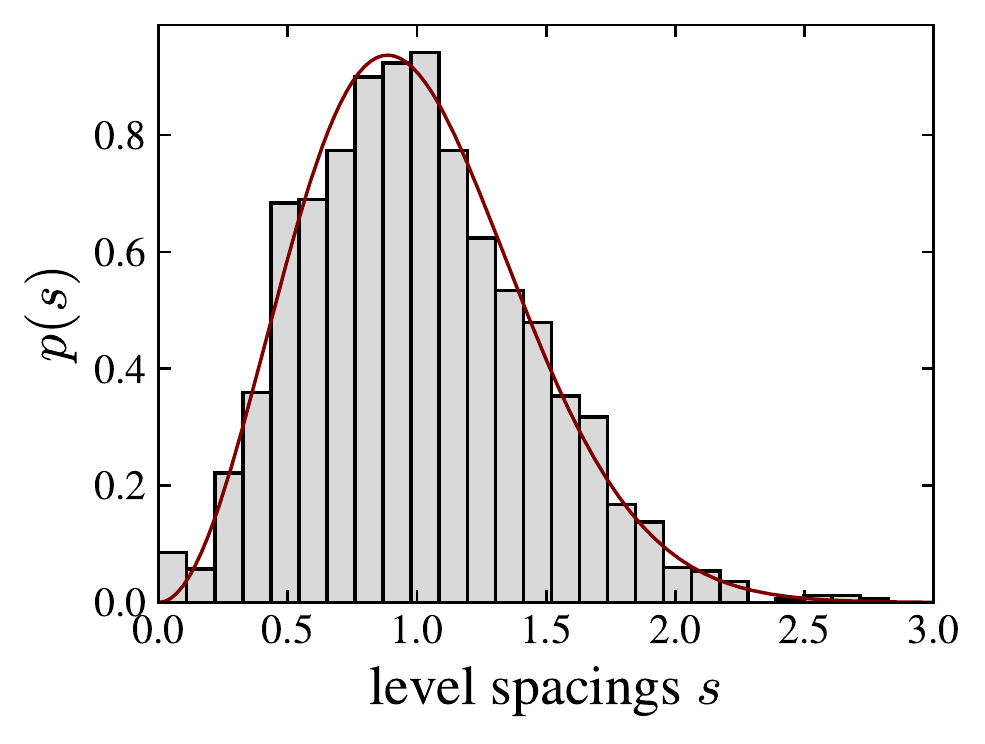} \hfill %
    \includegraphics[width=0.26\textwidth]{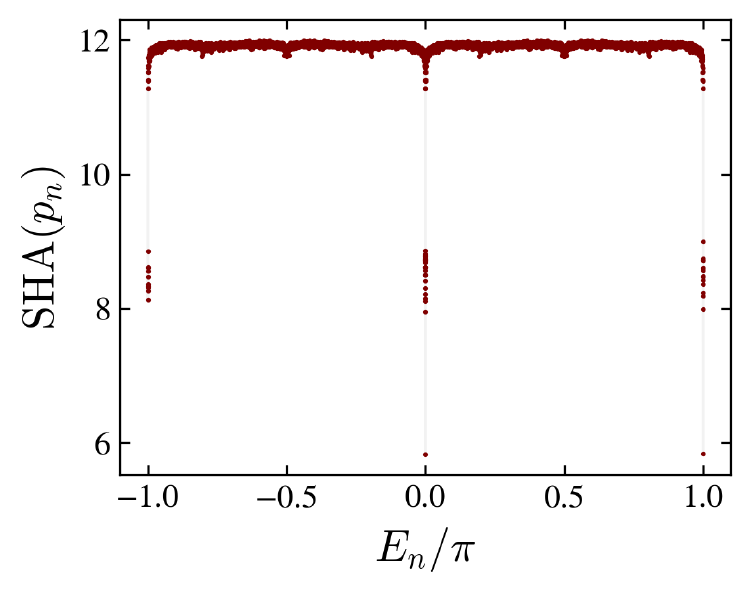} \\
    \includegraphics[width=0.16\textwidth]{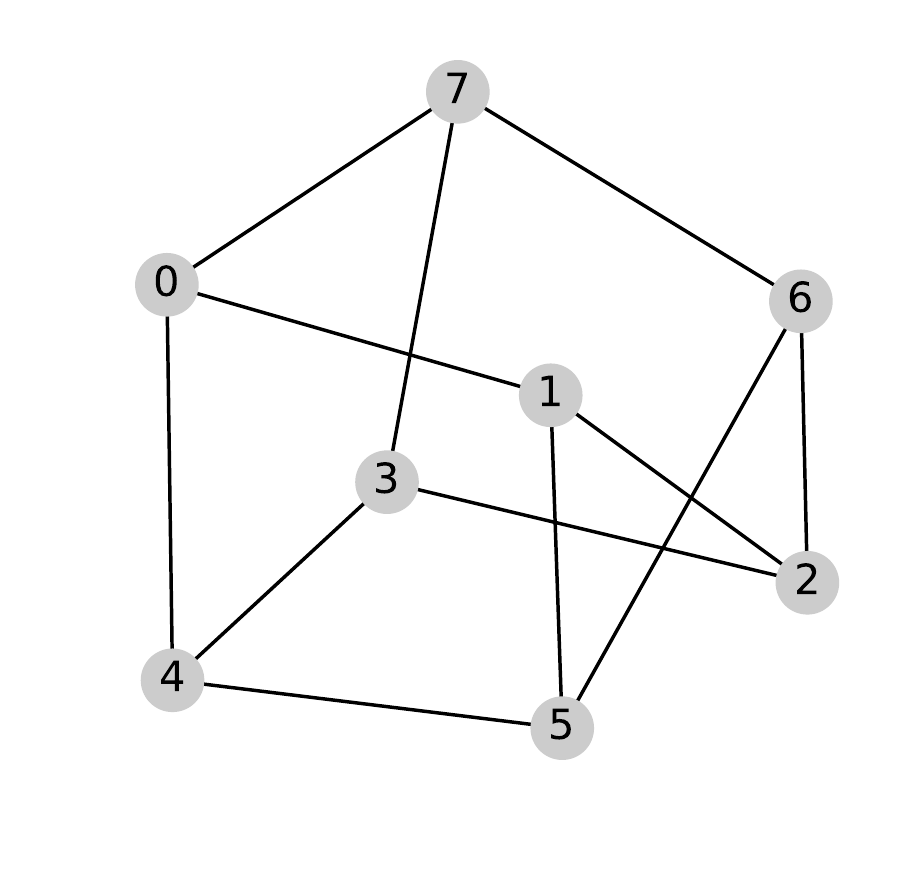} \hfill %
    \includegraphics[width=0.26\textwidth]{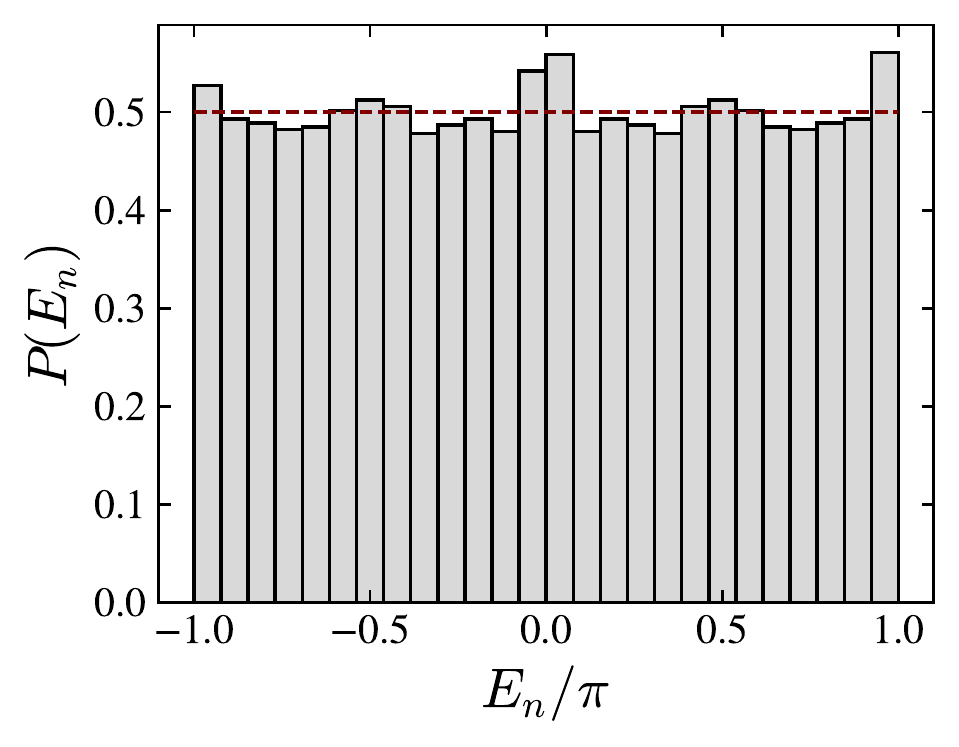} \hfill %
    \includegraphics[width=0.26\textwidth]{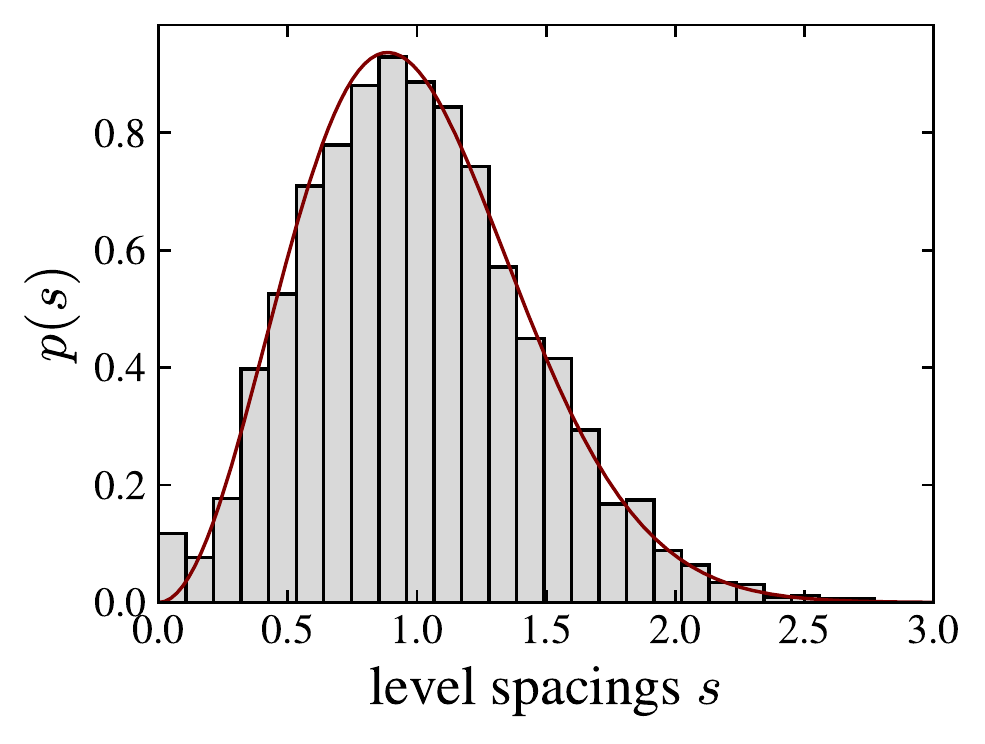} \hfill %
    \includegraphics[width=0.26\textwidth]{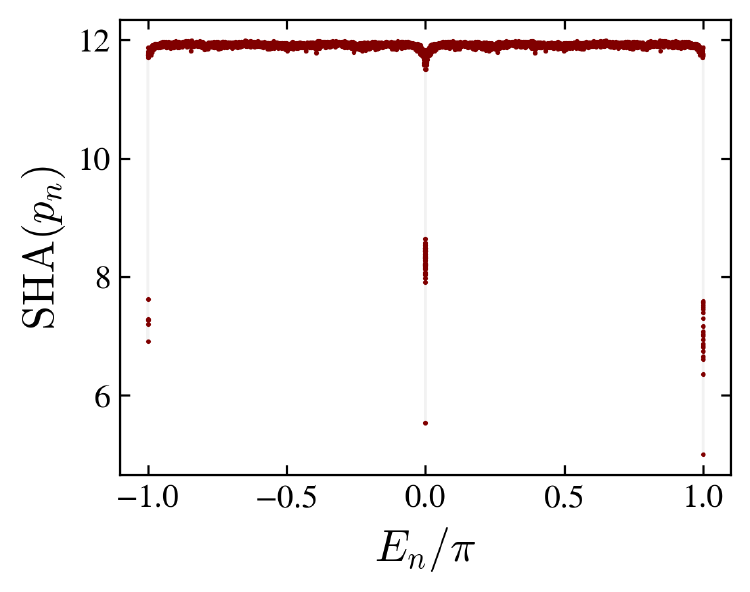} \\
    \includegraphics[width=0.16\textwidth]{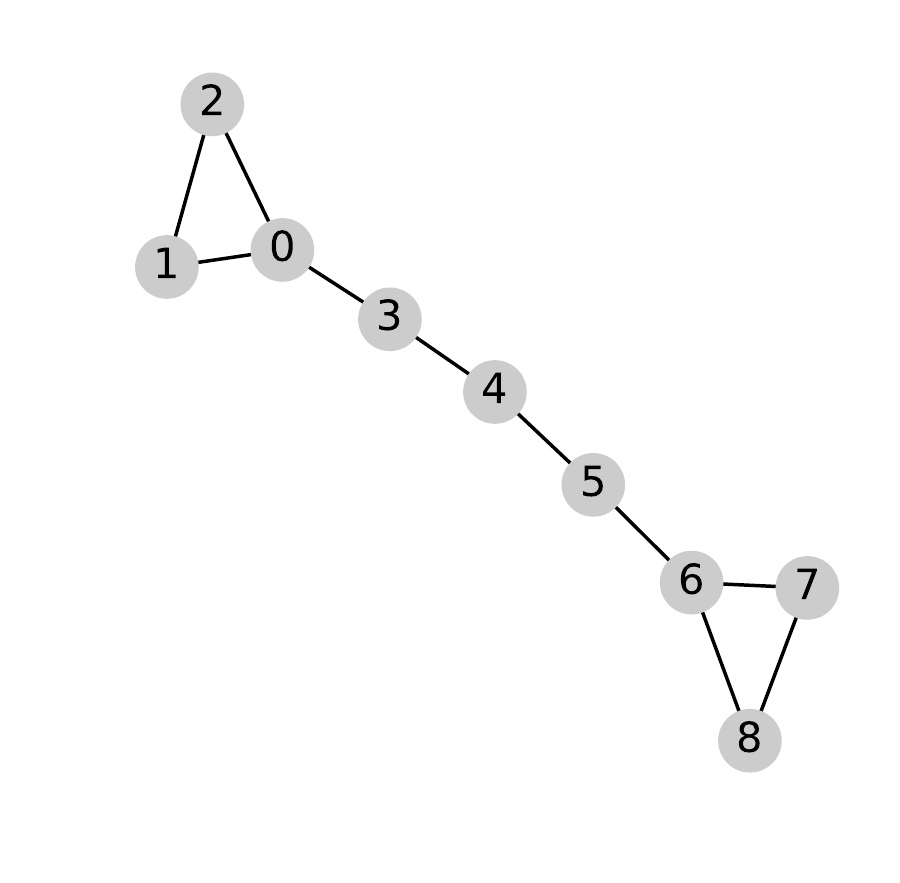} \hfill %
    \includegraphics[width=0.26\textwidth]{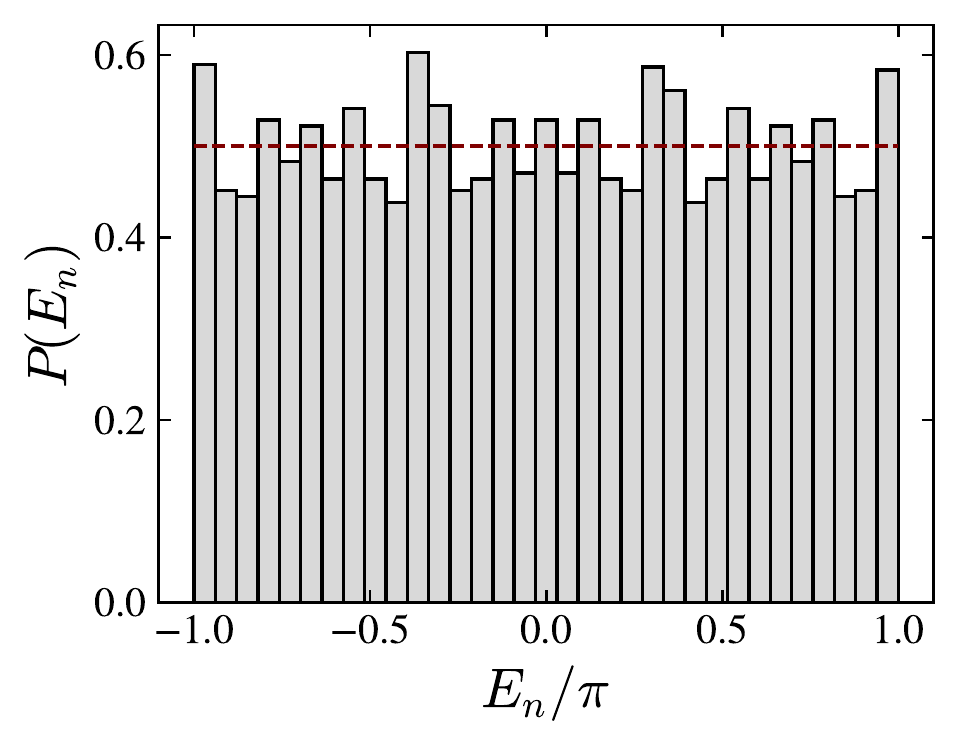} \hfill %
    \includegraphics[width=0.26\textwidth]{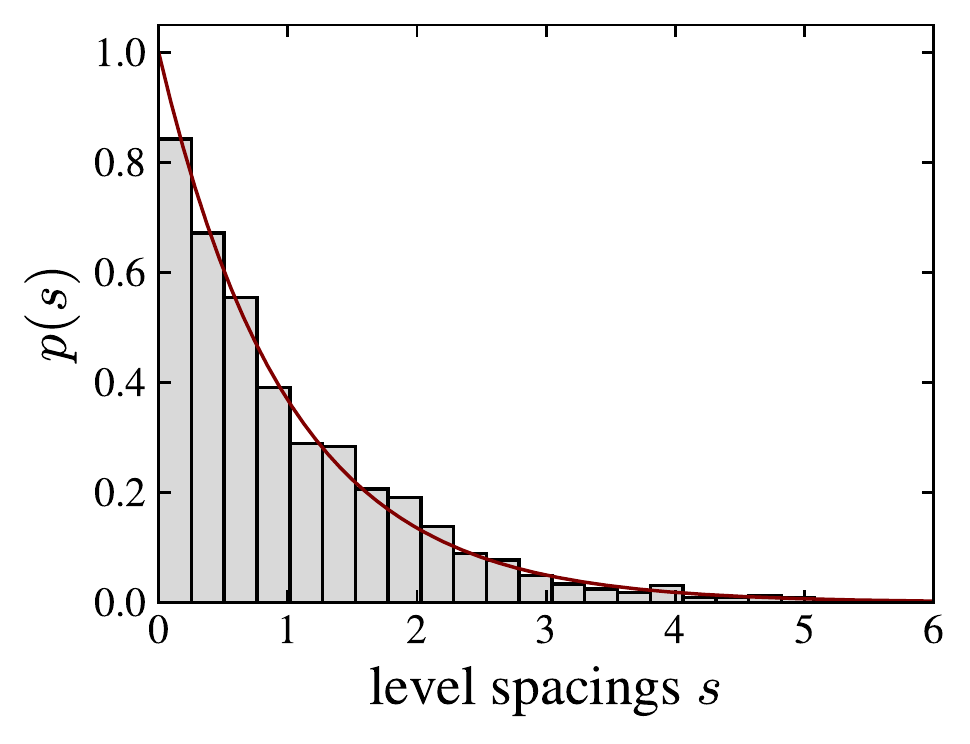} \hfill %
    \includegraphics[width=0.26\textwidth]{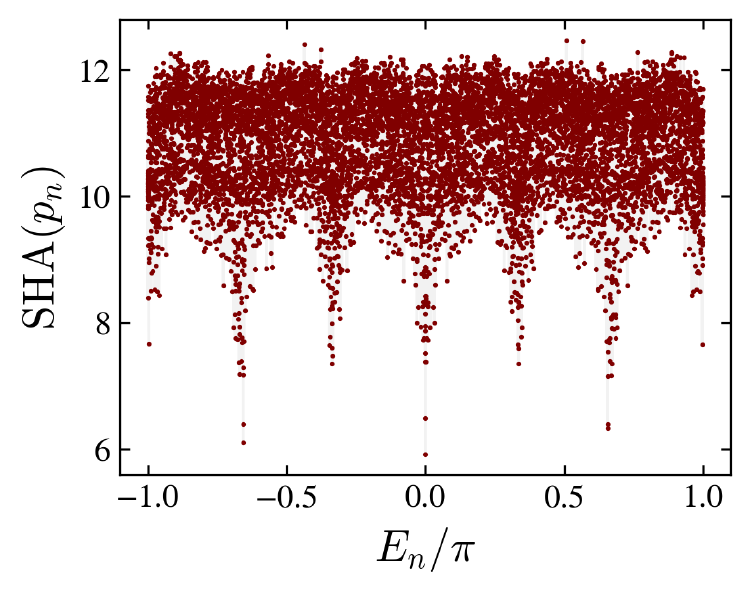} \\
    \includegraphics[width=0.16\textwidth]{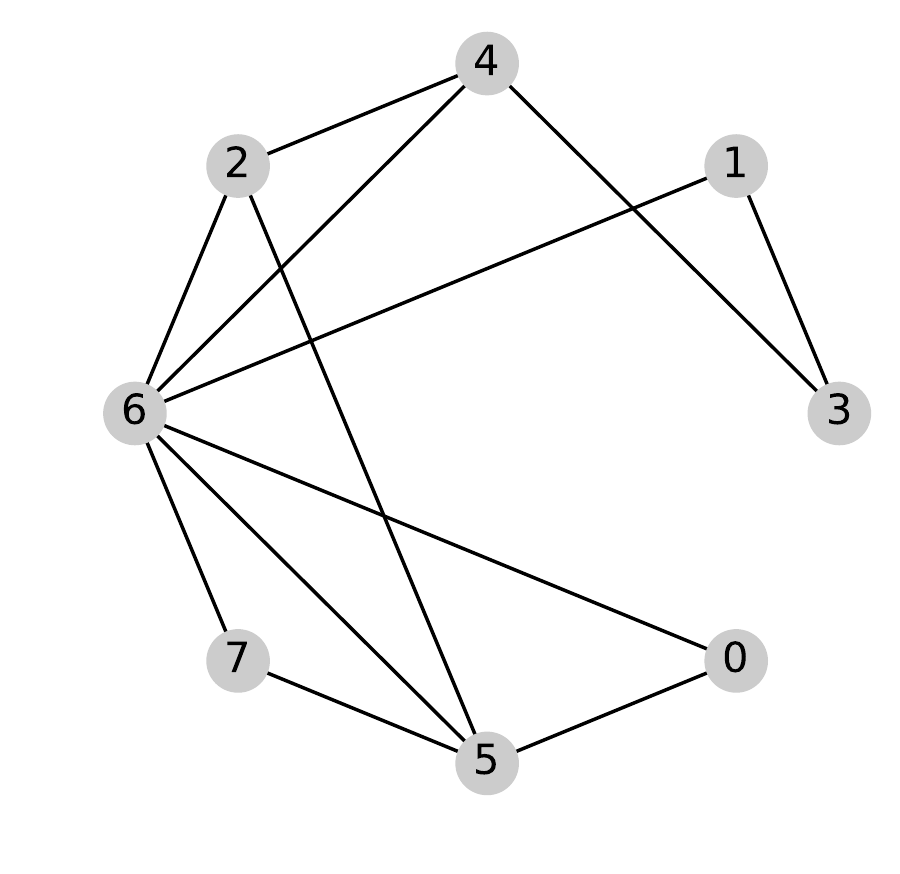} \hfill %
    \includegraphics[width=0.26\textwidth]{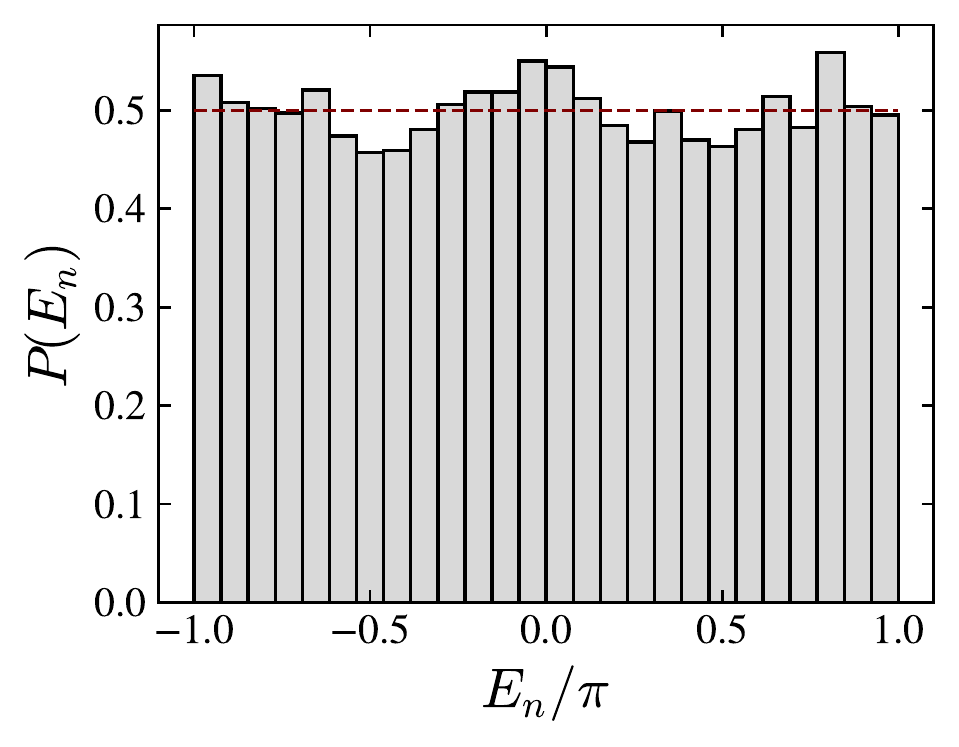} \hfill %
    \includegraphics[width=0.26\textwidth]{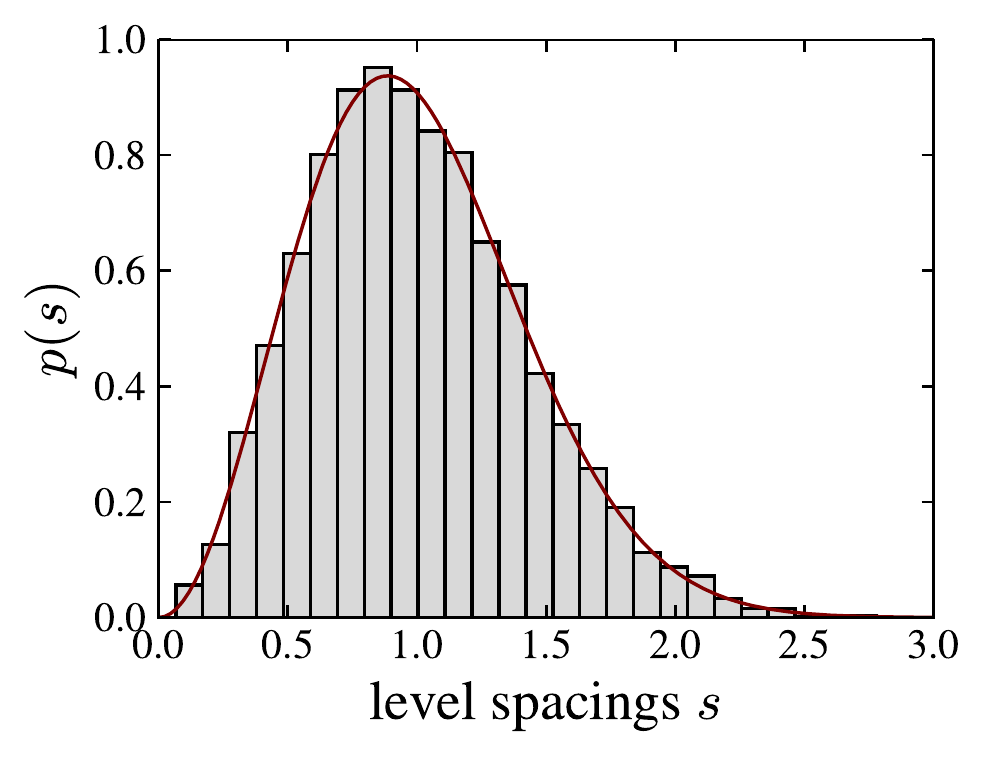} \hfill %
    \includegraphics[width=0.26\textwidth]{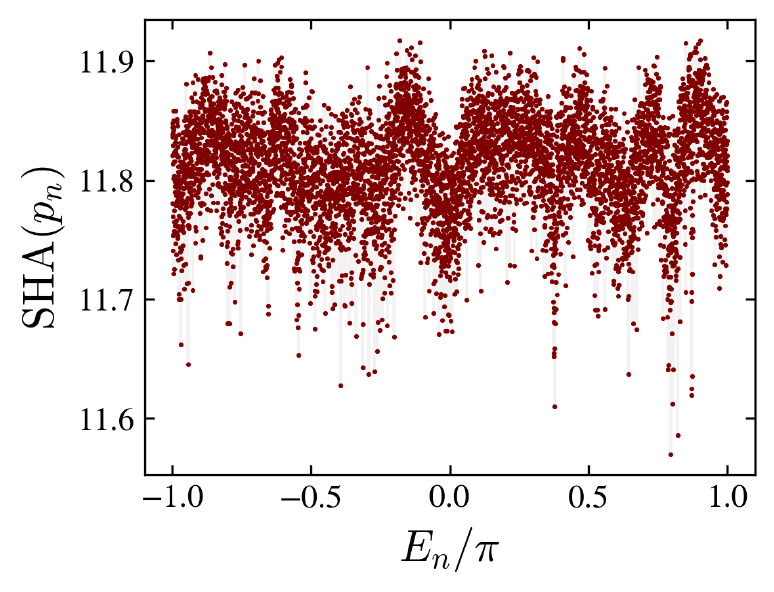}
    \caption{Evolution operator spectrum. Columns: 1) graph; 2) quasienergy histogram; 3) spacing distribution compared to the Gaussian unitary ensemble or Poisson distribution (solid line); Shannon entropy. Rows: 1) cube, 2) möbius, 3) chain (Grover coin), and 4) random graph $G = G(8,3)$ (Fourier coin). The dimensions of the Hilbert space are, for each graph: `c', `m' and `G', 6144; `X', 10240.
  \label{f:SP}}
\end{figure*}

\section{Spectrum}

To gain some insight into these phenomenological observations it is interesting to relate them with the spectrum of \(U\). Using exact diagonalization, we solve the eigenvalues equation,
\begin{equation}
  \label{e:Un}
  U \ket{n} = \E^{-\I E_n} \ket{n} 
\end{equation}
to find \(E_n\), the quasienergies in the band \(E_n\in[-\pi,\pi]\), and their corresponding eigenstates \(\ket{n}\). We denote \(v_n\) the complex eigenvector coordinates of level \(n\) in the \(|xcs\rangle\) representation:
\[v_n(xcs) = \braket{ xcs | n }\,.\] 
We define a Shannon entropy using \(p_n = |v_n|^2\) as the probability distribution, computed from the exact eigenvectors (a related definition is given by Borgonovi et al. in Ref.~\cite{Borgonovi-2016qe}). The Shannon entropy \(S\), 
\begin{equation}
  \label{e:SHA}
 S(n) = - \sum_{xcs} |v_n(xcs)|^2 \log|v_n(xcs)|^2 
\end{equation}
is a measure of the deviation of \(p_n\) to the uniform distribution, and hence it also quantifies the `localization' of the \(v_n\) eigenvectors in Hilbert space (at variance to localization in space); we will consider as ``typical'' an eigenvector \(v_\star\) whose Shannon entropy is about its maximum. Our system lacks a conserved local Hamiltonian from which we could define a typical energy, or associate an energy to the initial state; it is then convenient to define a ``thermal'' eigenvector as the one corresponding to maximum entropy, and verify a posteriori if expected values computed in this state are compatible with the microcanonical predictions. This method, which avoids the need of a semiclassical limit \cite{Srednicki-1996fq}, generalizes the method used in the case of system with a conserved Hamiltonian, used since the work of Rigol et al. \cite{Rigol-2008uq} on the thermalization of the boson Hubbard model. Therefore, to verify in our case the eigenvector thermalization hypothesis we may compare the actually observed results with the predictions of both the thermal and microcanonical states.

\begin{figure*}
  \centering
  ~\hfill \includegraphics[width=0.2\textwidth]{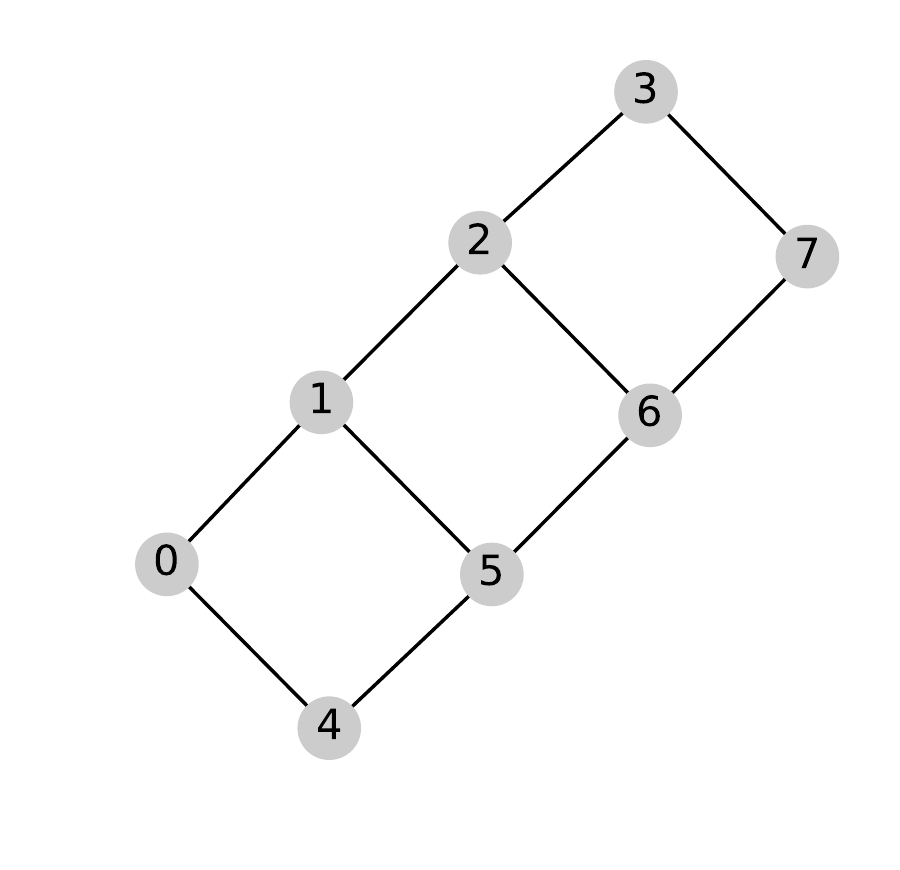}%
  \hfill \includegraphics[width=0.2\textwidth]{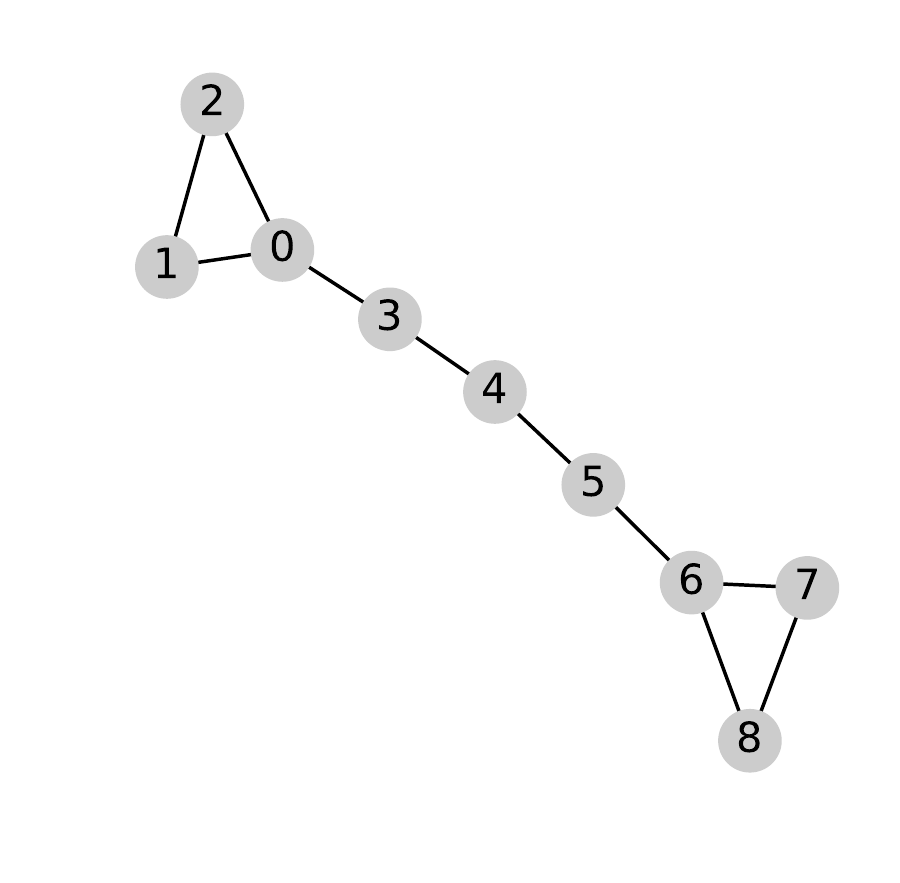}%
  \hfill \includegraphics[width=0.2\textwidth]{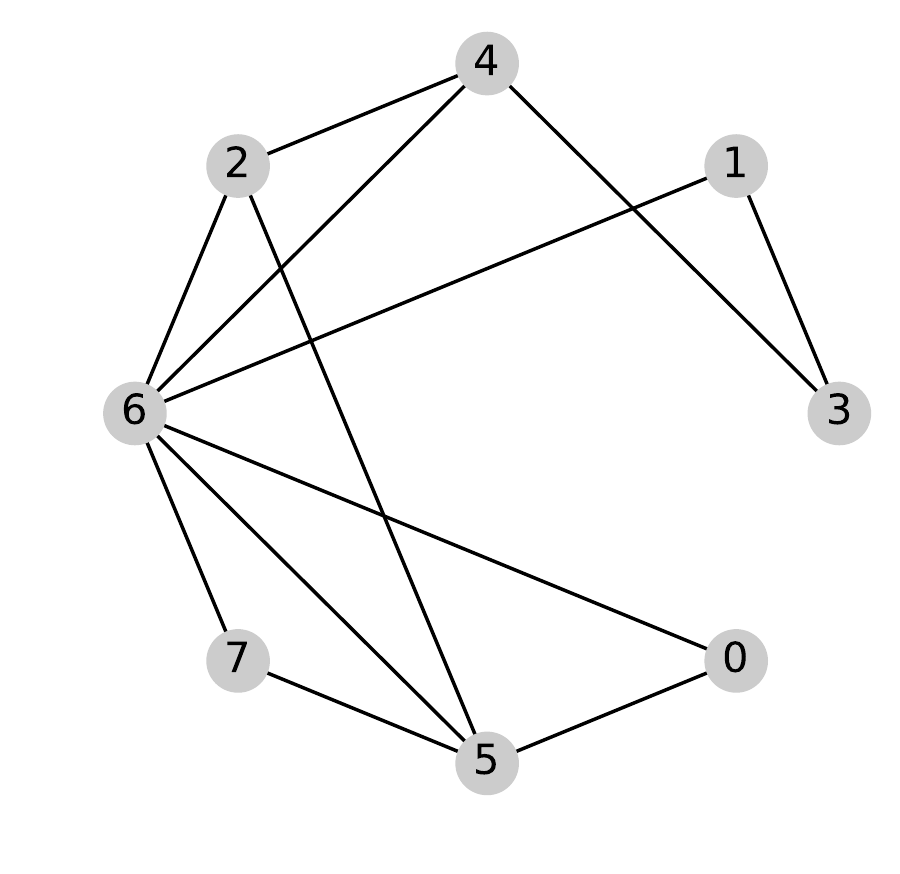} \hfill~\\
  \includegraphics[width=0.32\textwidth]{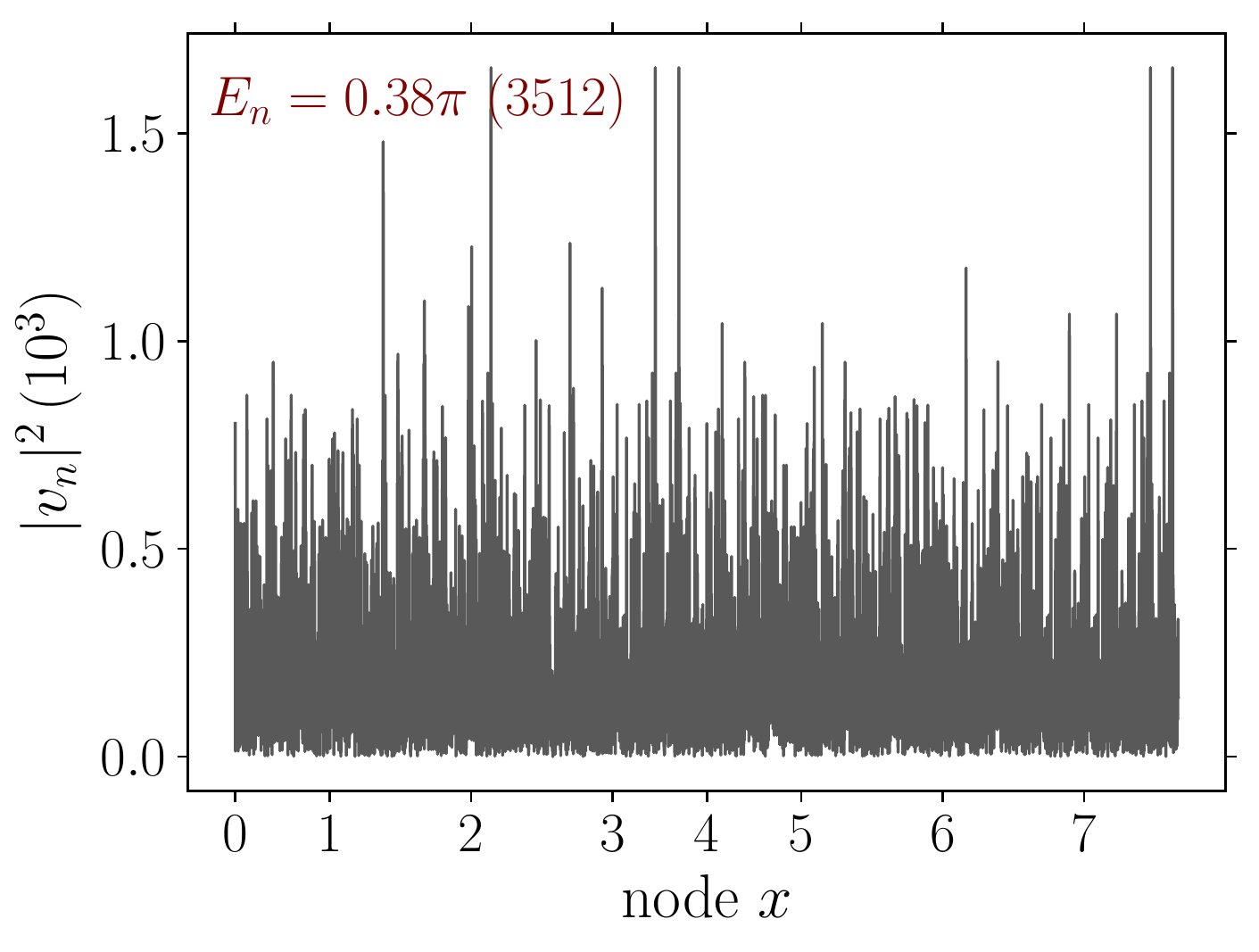}%
  \includegraphics[width=0.32\textwidth]{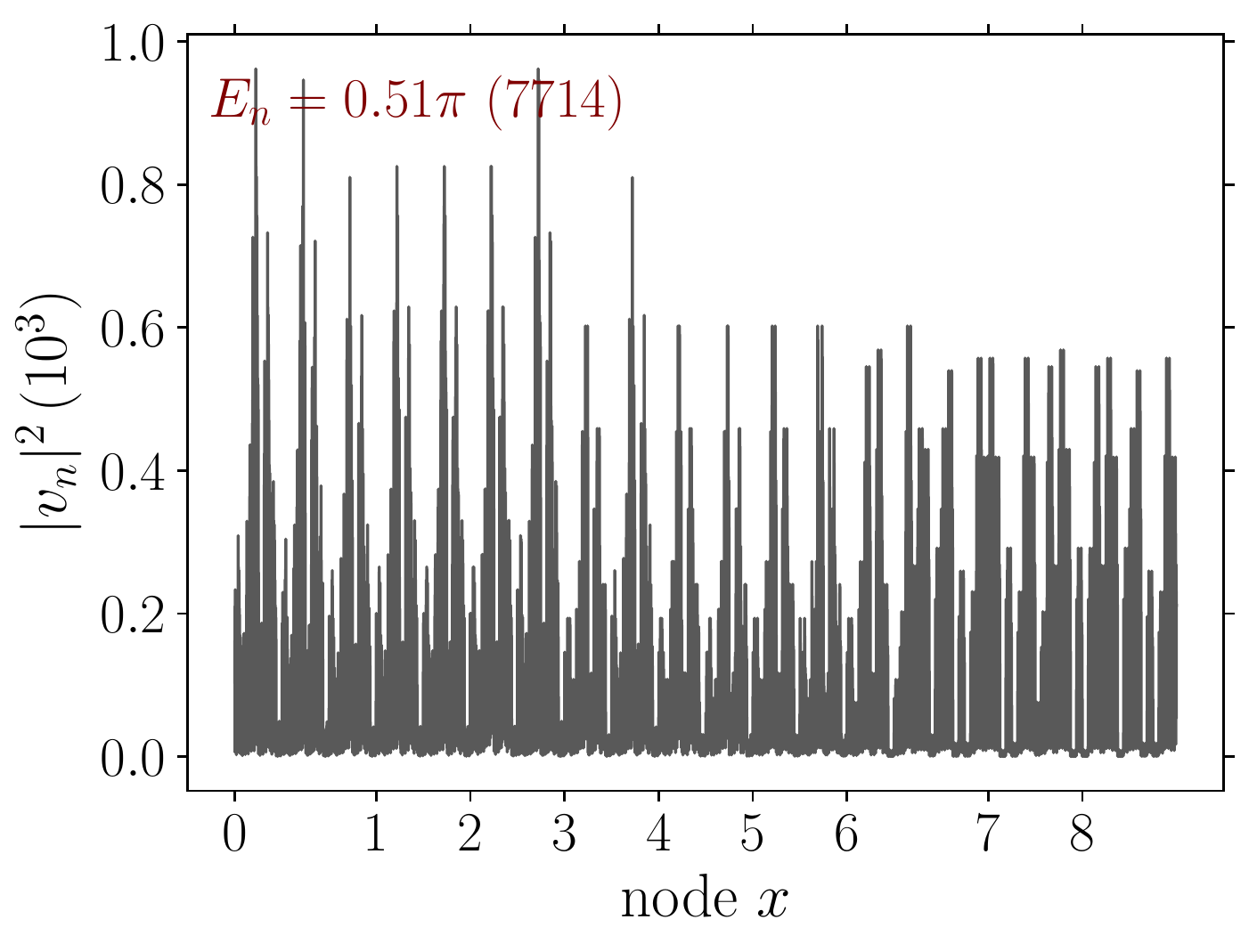}%
  \includegraphics[width=0.32\textwidth]{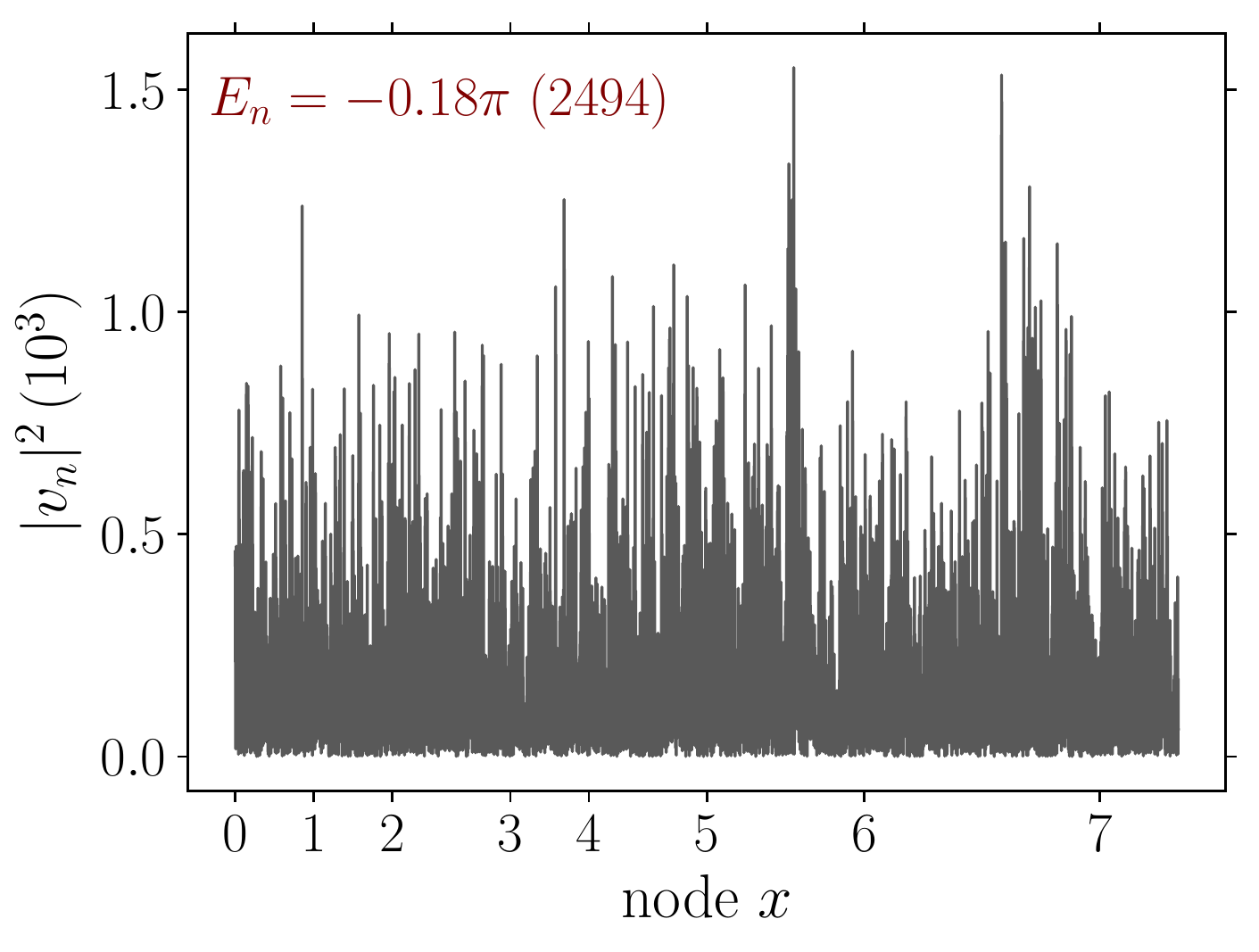}\\
  \includegraphics[width=0.32\textwidth]{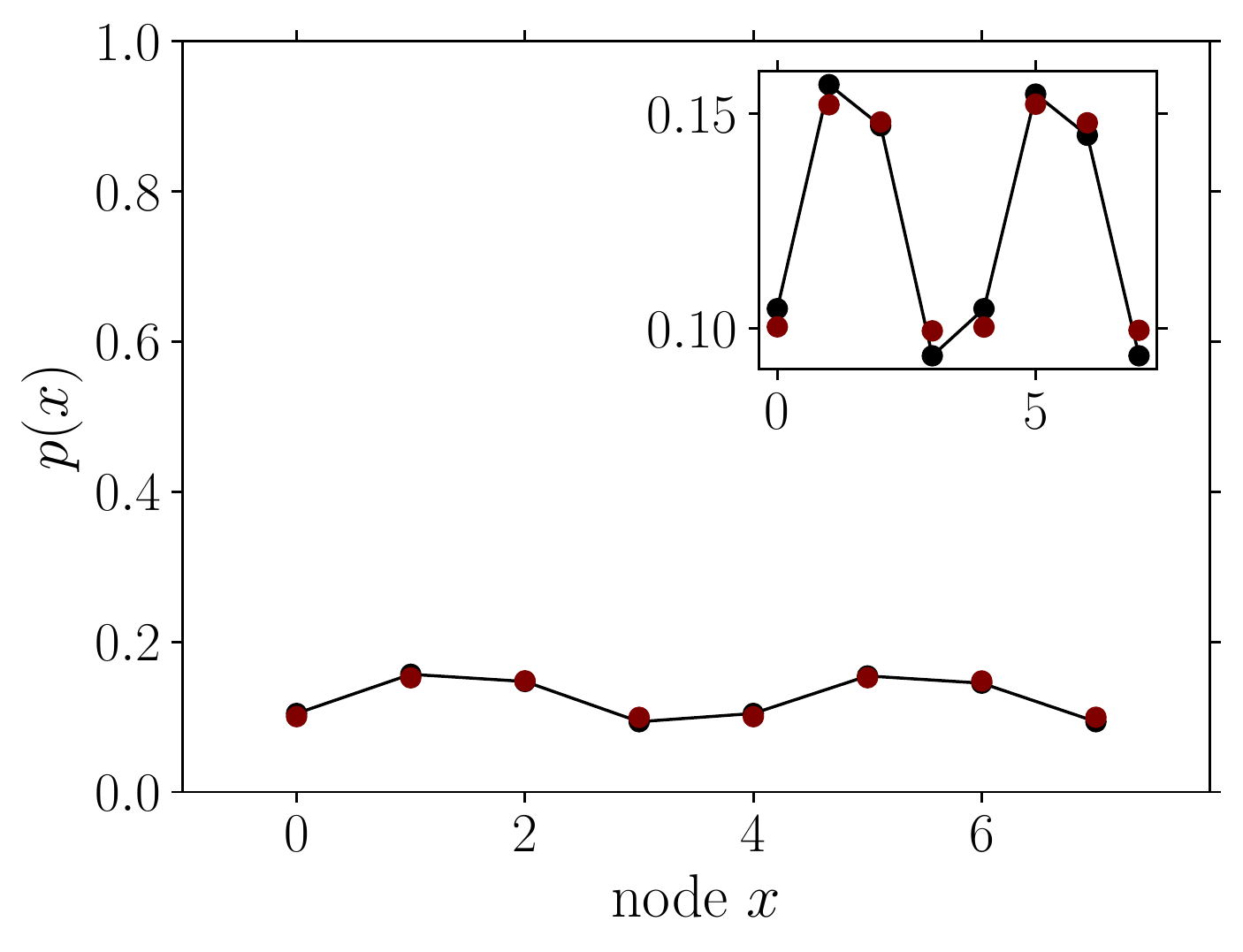}%
  \includegraphics[width=0.32\textwidth]{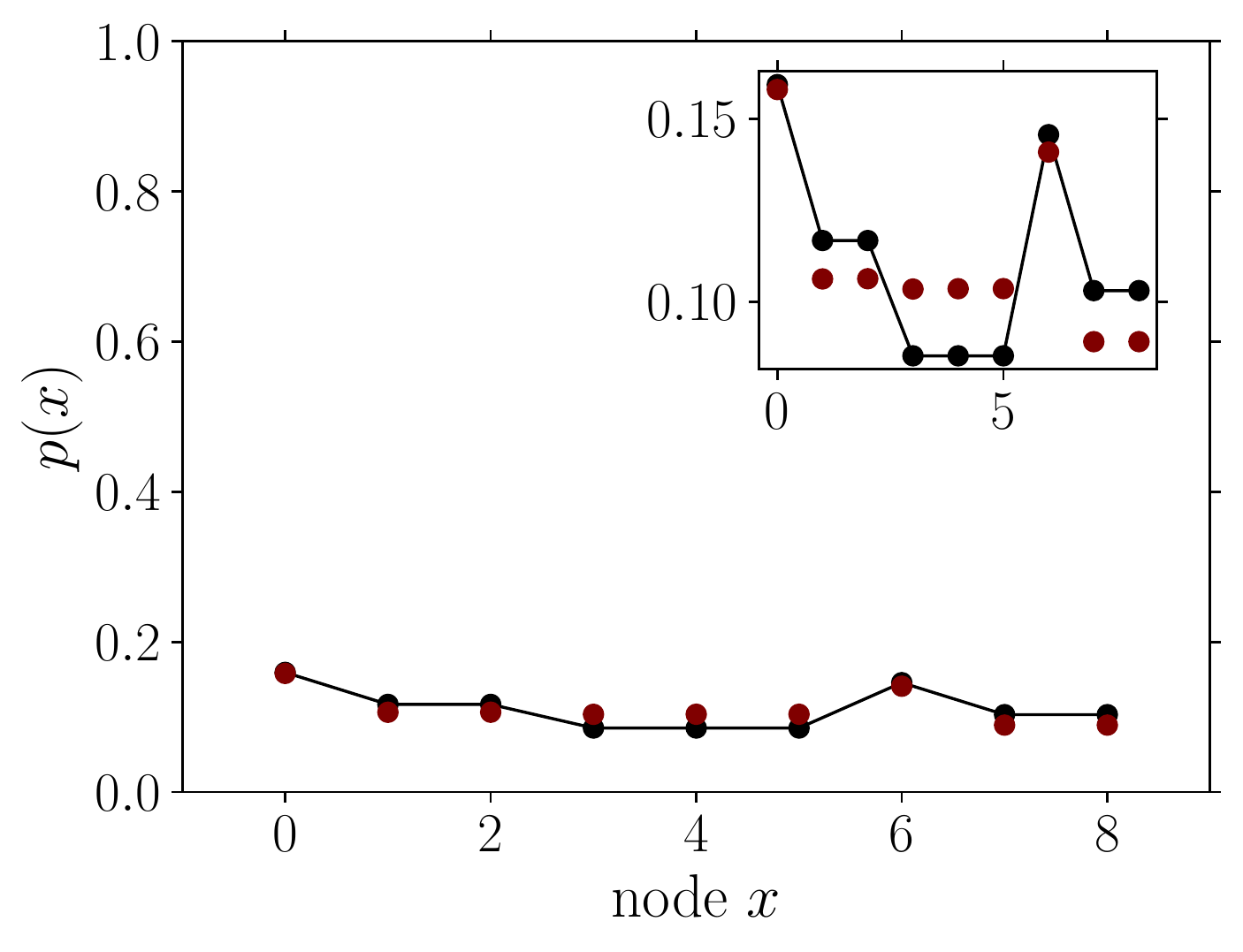}%
  \includegraphics[width=0.32\textwidth]{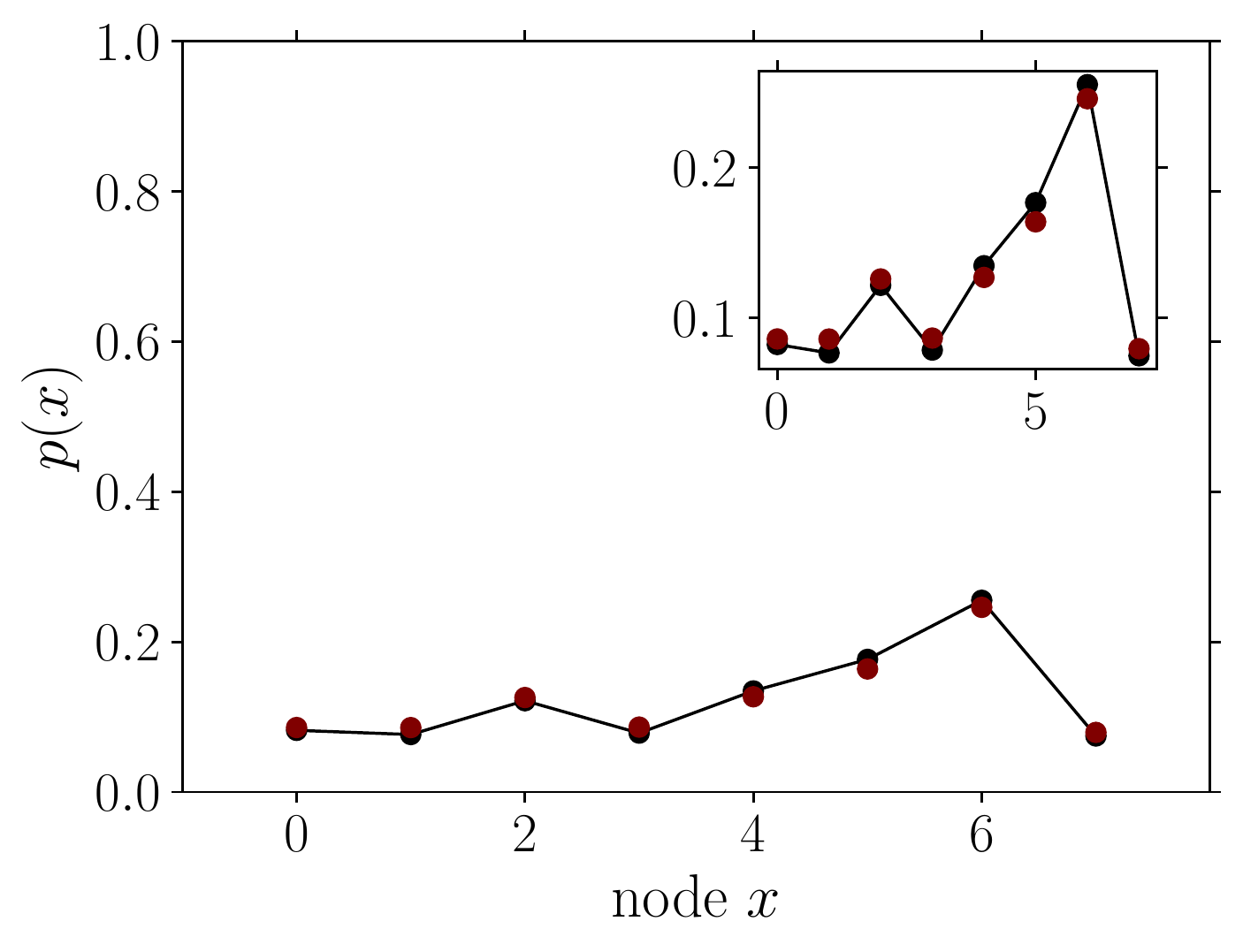}
  \caption{Eigenvector thermalization. Columns: ladder, chain and random $G(8,3)$ graphs. First row, maximum entropy eigenvectors, labeled by their quasienergy (between parenthesis the mode number $n$). Second row, position distribution $p(x)$ of the quantum walker, calculated from the exact state at step $t$, $\ket{\psi(t)}$ (red dots, exact numerical value, single marks), using the maximum Shannon entropy eigenvector (black line-dots, thermalization ansatz), and from the microcanonical expected value (yellow crosses, often superposed to other marks). The inset contains a zoom on the same data.
  \label{f:p}}
\end{figure*}

Central to the thermalization mechanism is the chaotic nature of the eigenvectors \cite{Santos-2010qr,Santos-2012ve,Lensky-2018rx}. We adopt the definition of quantum chaos based on the statistical behavior of the eigenvalues and eigenvectors of the unitary evolution operator, as being amenable to a description in terms of random matrices, generalizing the usual definition applied to the system's Hamiltonian \cite{Kampen-1985eu}. Figure~\ref{f:SP} summarizes the spectral properties of the cube `c', möbius `m', and chain `X' graphs (first three rows), for which \(U\) is defined with the Grover coin, and a random graphs `G', with the Fourier coin (last row). We computed the histogram of quasienergies (second column), the level spacing distribution (third column) and the Shannon entropy (fourth column). The general trend is that quasienergies are nearly uniformly distributed in the \((-\pi,\pi)\) band, with some pics related to degeneracy (especially around \(0\) and \(\pm\pi\)), and the level spacing \(s\) follows the Gaussian unitary ensemble statistics, well fitted by
\begin{equation}
  \label{e:Ws}
  p(s) = \frac{32 s^2}{\pi^2} \E^{-4 s^2/\pi}\,,
\end{equation}
the Wigner surmise \cite{Wigner-1967ve}; \(s\) is the normalized difference between two consecutive levels \(s = (E_{n+1} - E_n)/ \overline{\Delta E}\).

For the special case of the chain graph (row `X'), whose quasienergies are four times degenerated, we filtered out the degeneracy and obtained a Poisson distribution, characteristic of localized eigenvectors. For `c' and `m' cases, a small amount of degenerate levels is present. In the case of the Fourier coin, degeneracy tends to disappear, as in the case of the `G' graph, displayed in the last row; note the tight range of the Shannon entropy values in this case, compared withe the Grover coin cases. This points out an essential physical difference between the two coins, as already mentioned; the Fourier coin breaks the graphs structural symmetries and as a consequence, lift almost completely the degeneracy of the quasienergy levels. This effect allows us to explain the differences in the magnetization between the two cases: magnetization is favored by the presence of degeneracy. Even in the somewhat special case of `c' and `m', the residual magnetization observed in `m' can be attributed to the relative increase of degeneracy for this graph. Macroscopic degeneracy is attained in the extreme case of compete graphs, due to their very high symmetry, in which case we saw that the magnetization tends to its saturation value with the number of nodes (c.f. Fig.~\ref{f:K}).

The level spacing distribution \(p(s)\) is, in most cases, well described by the unitary Gaussian ensemble of random matrices \cite{Haake-2000ul}, corresponding to quantum chaotic systems. The almost one dimensional graph (`chain'), follows in contrast, a Poisson statistics, as in the case of quantum integrable systems \cite{Bohigas-1984gf}. The excellent fit shown in the figure with the Wigner surmise \eqref{e:Ws}, for the unitary ensemble, have no ajustable parameters. Note however, that at variance to the usual terminology, our classification do not refer to a classical model, which might be chaotic or integrable, and whose quantization would lead to our quantum system.

Besides the eigenvalues statistical properties it is also worth studying the eigenvectors. This can be done using the Shannon entropy we defined by the eigenvectors amplitudes: other statistical tools, like the inverse participation ratio, give essentially the same qualitative description. We find that the distribution of the Shannon entropy is almost uniform over the levels (Fig.~\ref{f:SP}, last column), even if for some eigenvectors it deviates from the average, and the width (variance) of the distribution may depend on the graph (compare `X' with the other graphs).

\begin{figure}
  \centering
  \includegraphics[width=0.4\textwidth]{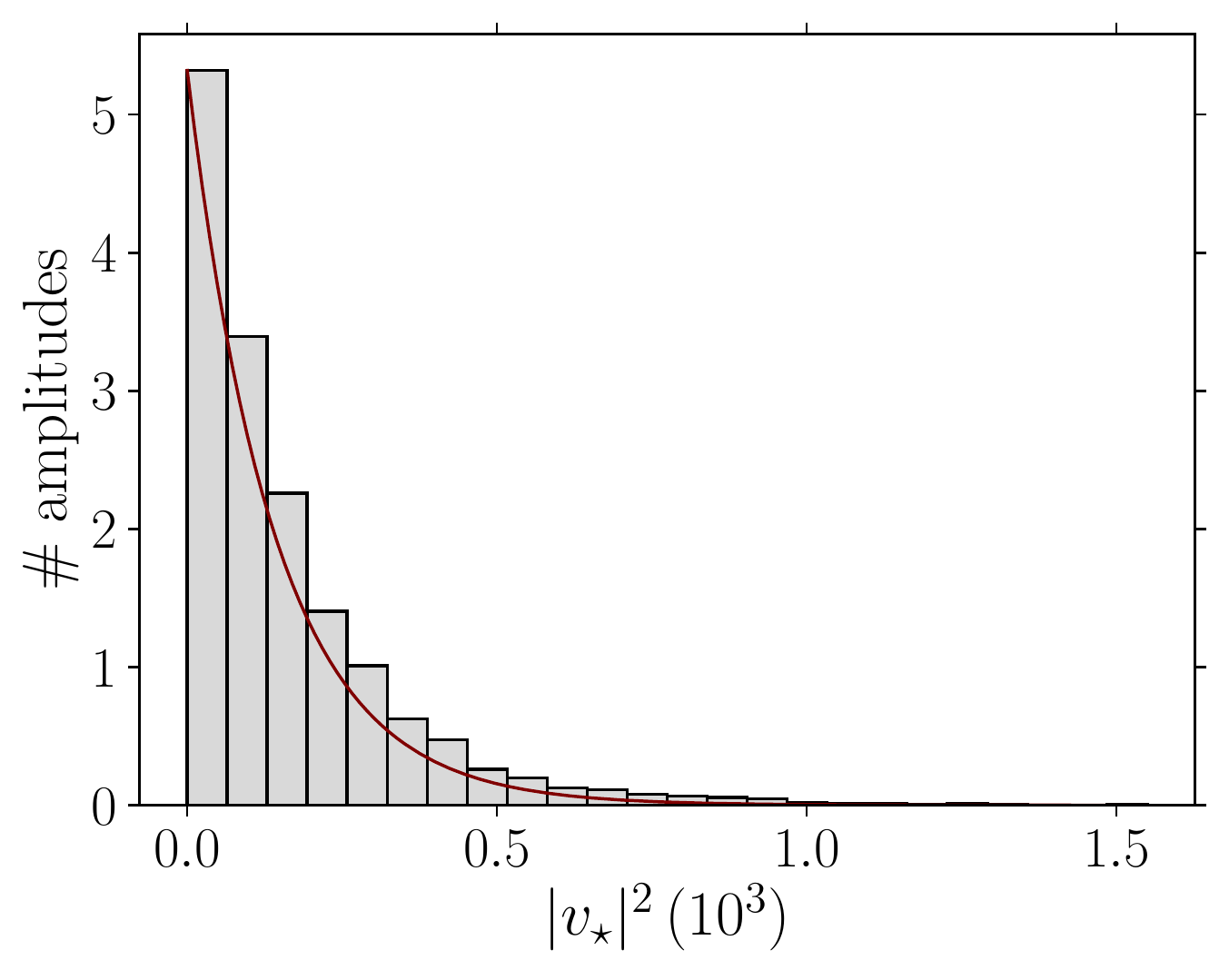}\\
  \includegraphics[width=0.4\textwidth]{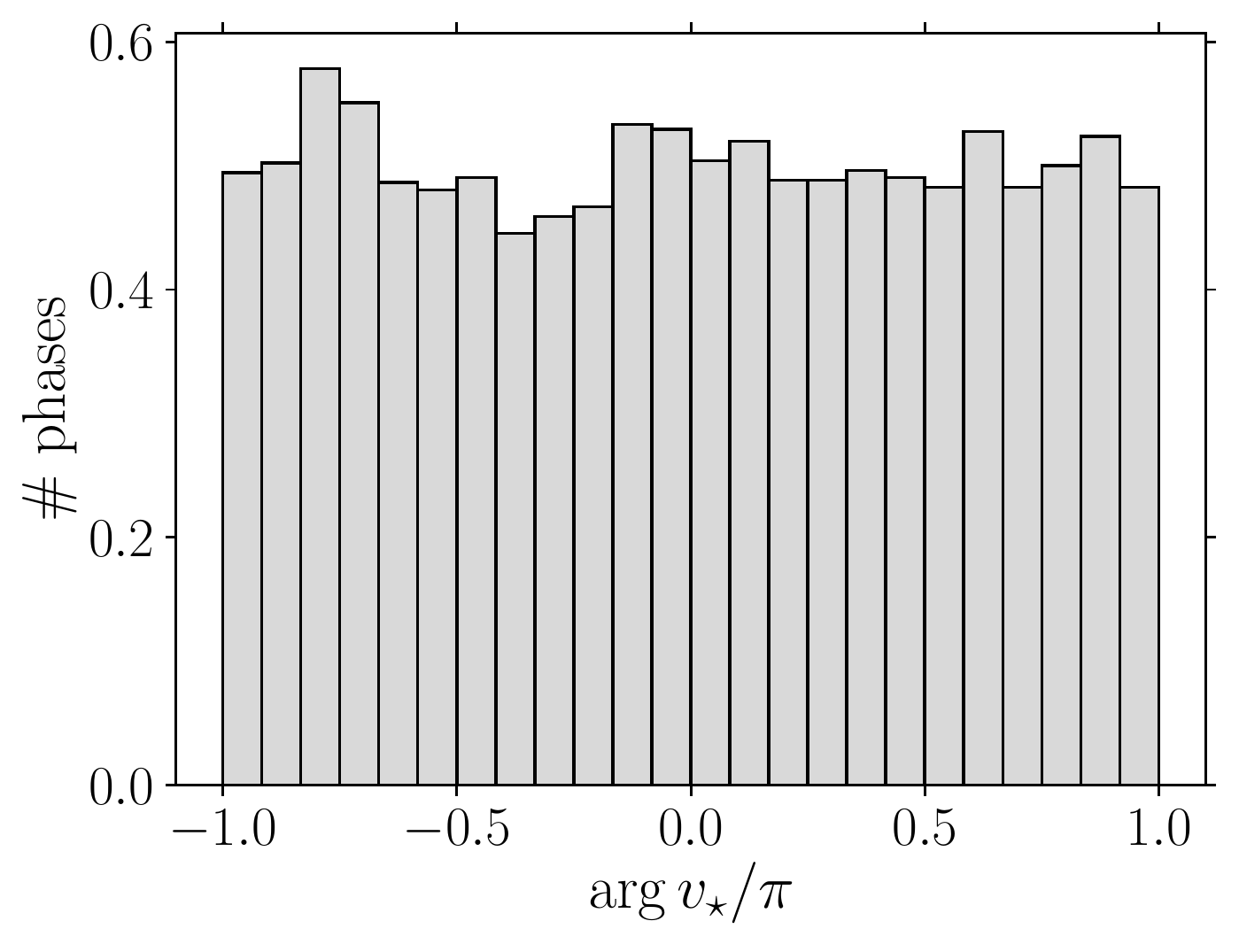}
  \caption{Amplitude and phase histograms of the random graph typical eigenvector $v_\star$ (mode $2494$) of Fig.~\protect\ref{f:p}. The fit corresponds to an exponential distribution.
  \label{f:amp}}
\end{figure}

We can also classify the eigenvectors according to their Shannon entropy: chaotic eigenvectors have near maximum entropy, degenerate eigenvectors have much smaller entropy. The chaotic systems exhibits a narrow distribution of the Shannon entropy, with deeps holes at the degenerate levels; the Shannon distribution for the integrable system is much boarder, and the degeneracy holes are much boarder than in the chaotic case. This classification in terms of the Shannon entropy properties of the quantum walks on graphs, is entirely compatible with the classification based on the spacing distribution and the corresponding random matrices ensembles. Moreover, differences with respect to the ideal behavior, such as the presence of degeneracy for certain graphs and coins, are physically significant, for example, because their are related to the presence of a finite magnetization (as in the case of the complete graphs).

Direct inspection of the eigenvectors can shed some light on their statistical properties, as we just described in terms of entropy. We plot in Fig.~\ref{f:p} the typical eigenvectors \(v_n = v_\star\), for the ladder, chain and random graphs, as a function of the node labels (basis vectors in node space \(\ket{xcs}\). We observe that the chain eigenvector (middle column) has much more structure than the other two eigenvectors, this structure is characteristic of almost independent modes (Hilbert space localization, reminiscent of the behavior of an integrable system). Here we use the word localization, not to mean that the quasienergy eigenvectors concentrate on a given node, but that they span only a subset of the available Hilbert space (as we mentioned above). The ladder eigenvector (left column), even if it is an essentially stochastic function, shows a set of modes with the same amplitude pertaining to different nodes (in this case the two neighboring nodes \(3\) and \(7\)). The eigenvector \(v_\star = v_{2494}\) of the random graph, with the Fourier coin, is featureless; we show in Fig.~\ref{f:amp} its amplitude histogram, well fitted by an exponential distribution, and its phase histogram, which is nearly uniform.

It is a remarkable fact that the simple interaction rules encoded in the unitary operator \(U\), which do not contain any random ingredient, lead to a Gaussian unitary ensemble statistics, and chaotic extended eigenvectors with maximum Shannon entropy. Taking one of these vectors and computing the expected value of the position in the graph, we obtain values comparable to the actually computed one and, more interestingly, to the microcanonical expectation of uniform distribution among the nodes:
\begin{equation}
  \label{e:pmc}
  p_M(x) = \frac{d_x}{\sum_x d_x}\,,
\end{equation}
explicitly,
\begin{equation}
  \label{e:eth}
  \sum_{c,s} |v_\star(xcs)|^2 \approx \sum_{c,s} |\psi_{xcs}(t)|^2 \approx p_M(x)\,.
\end{equation}
This confirms that the chaotic eigenvectors associated with the evolution operator, as the ones shown in Fig.~\ref{f:p}, are compatible with the thermalization hypothesis. At variance to this result, which applies to both coins and almost arbitrary graphs, graphs exhibiting localized eigenvectors and Poisson distribution of level spacing, do no thermalize; thermalization is also absent in highly degenerate graphs, like the complete graphs with Grover coin (which also exhibit large values of the magnetization).

For most graphs, and in particular for the Fourier coin, the thermal state is characterized by a vanishing magnetization; the spin expected value computed with the maximum entropy eigenvector, also vanishes in agreement with the eigenvector thermal hypothesis. However, in the presence of degeneracy, and in particular for highly connected graphs (for which locality does not apply and degeneracy is extensive), the finite magnetization implies some influence of the initial state. As a matter of fact, the value of the mean spin is well reproduced by its expected value in a state which is a superposition of the thermal eigenvector and the overlap vector with the initial state. Nevertheless, the magnetization at the stationary state, do not depend exclusively on the initial state, the same magnetization is found for a set of initial conditions with different total spin, but on the overlap with the quasienergy eigenvectors.

\section{Discussion and conclusion}

The traditional approach to chaotic quantum systems proceeds by analogy with classical systems, which is possible as long as the semiclassical approximation is valid (Berry \cite{Berry-1987lq}). Our starting point in this work reverses this traditional approach: we define, using the elementary principles of quantum mechanics, a system satisfying a set of simple unitary transformation rules, and ask whether it can exhibit some ordered collective behavior that can be attached to chaotic states.

\begin{figure}
  \centering
  \includegraphics[width=0.49\textwidth]{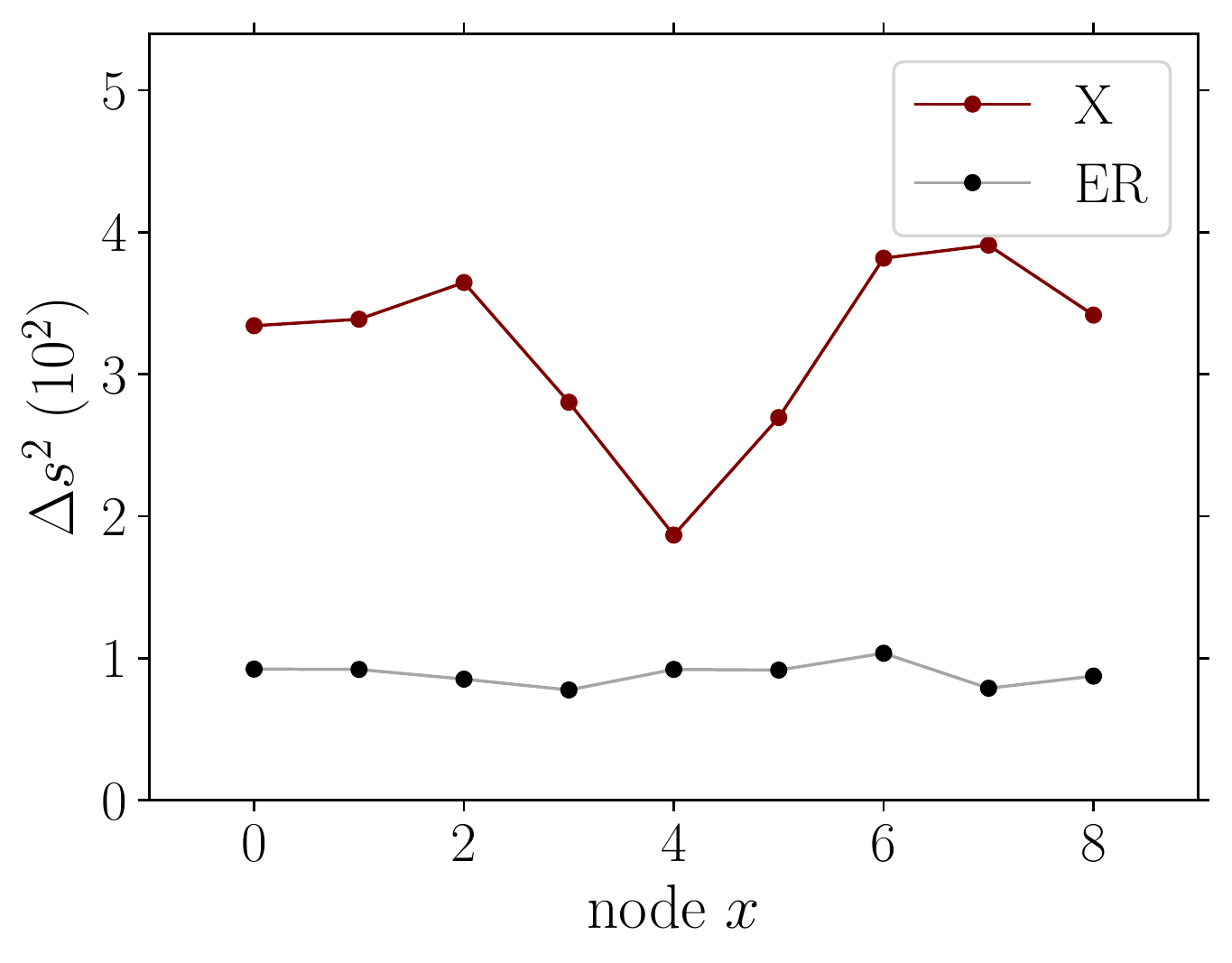}
  \caption{Spin fluctuations for the chain `X' and random Erdős-Rényi `ER' graphs with $9$ nodes (`X' upper line, `ER' lower line).
  \label{f:fl}}
\end{figure}

We observed that the interplay of interaction and entanglement, respectively local and nonlocal mechanisms, leads the system to a variety of states that can be characterized by well defined measurable properties: spatial distribution of the probability, magnetic order, thermalization. We initialized the walker at node zero, and the graph spins in an ordered ferromagnetic state. Each application of $U$ leads to the swapping of the position with neighboring nodes, according to the particle color, then introducing color-position entanglement; at the same time, the node color and spin are swapped, and the state of neighboring down spins acquires a phase; these interactions entangle the color and spin at a given node, and spins sharing an edge. As a result of the ballistic propagation of the walker, entanglement rapidly grows and, in most cases, saturates at a stationary value. The asymptotic stationary state is characterized by well defined macroscopic observables, as the position density and magnetization. Deviations to this phenomenological thermal behavior appear for instance, in highly symmetric graphs, like the complete or the quasi one dimensional graphs. In these cases significant fluctuations appear, as shown in Fig.~\ref{f:fl}, where we compare a typical graph (random Erdős-Rényi) with the chain graph. The spin standard deviation,
\begin{equation}
  \Delta s(x) = \braket{(s(x) - \bar{s})^2}^{1/2}
\end{equation}
as a function of the node \(x\), is much larger in the case of the chain graph than in the random graph, while both graphs have the same magnetization (\(\bar{s} = 0.2\). The entanglement entropy is larger for the random graph (\(S_s =4.4\)), than for the chain graph (\(S_s = 4.2\)), suggesting that the random graph reached a thermal equilibrium state. The symmetry of the graph has also another important consequence on the entanglement level. We exhibited cases where two graphs differing in their topology could encode a different number of qubits, depending on the existence of a swapping symmetry between two cycles.

\begin{figure*}
  \centering
  \includegraphics[width=0.35\textwidth]{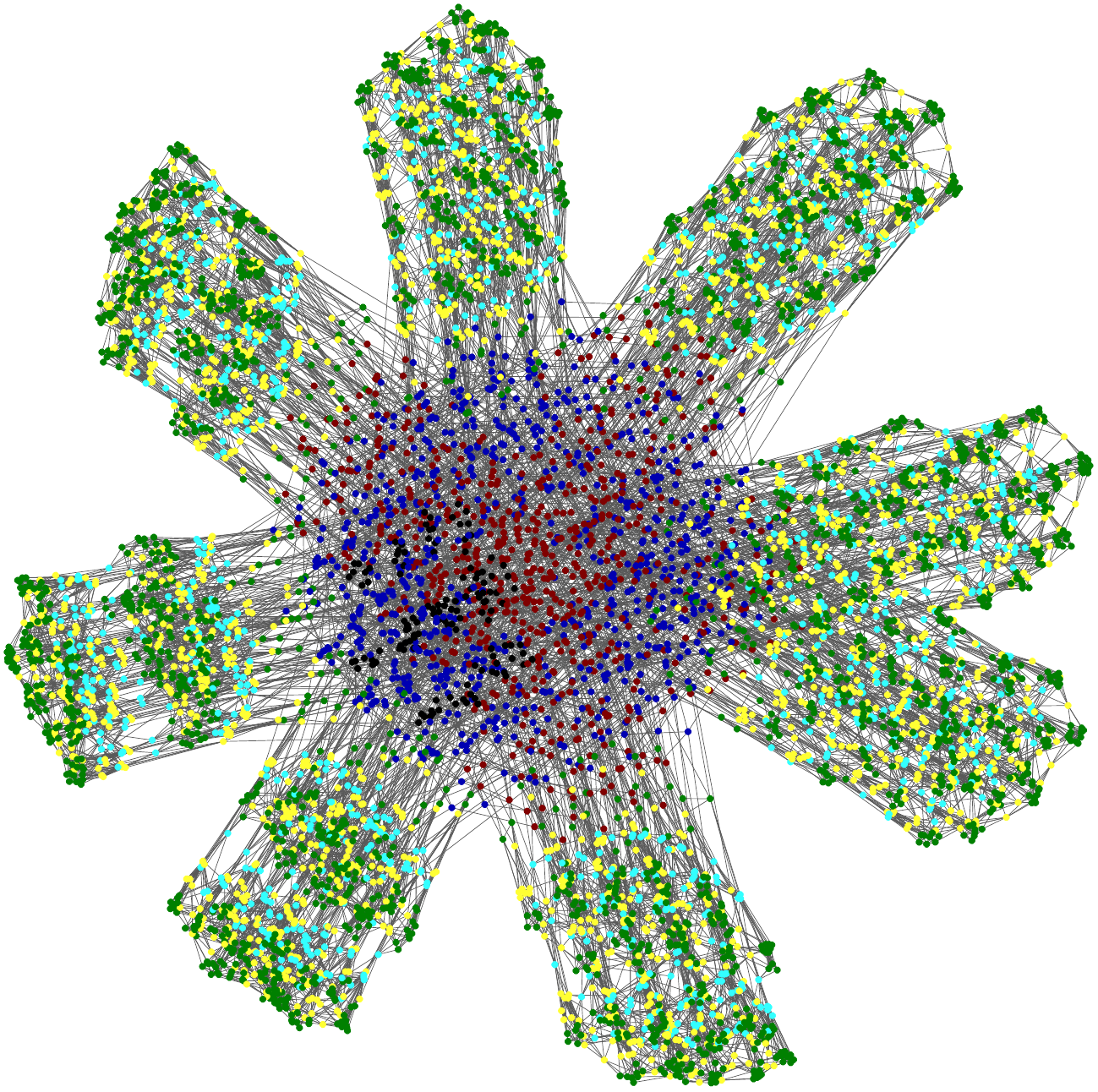}\hfill %
  \includegraphics[width=0.22\textwidth]{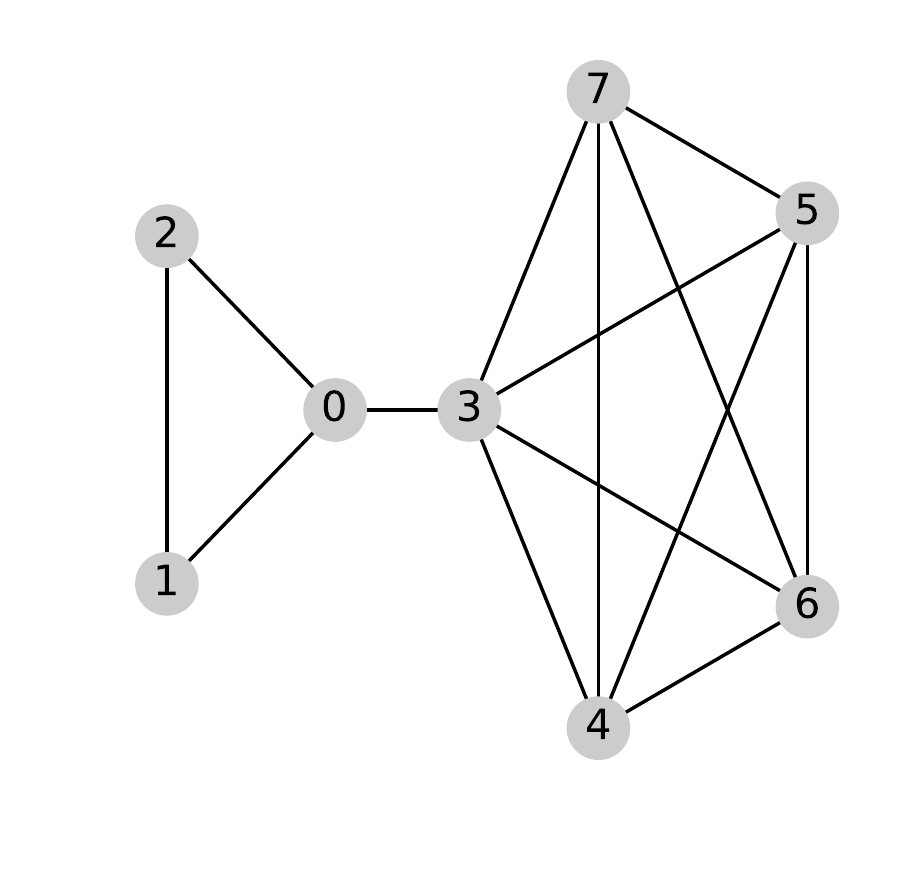}\hfill %
  \includegraphics[width=0.35\textwidth]{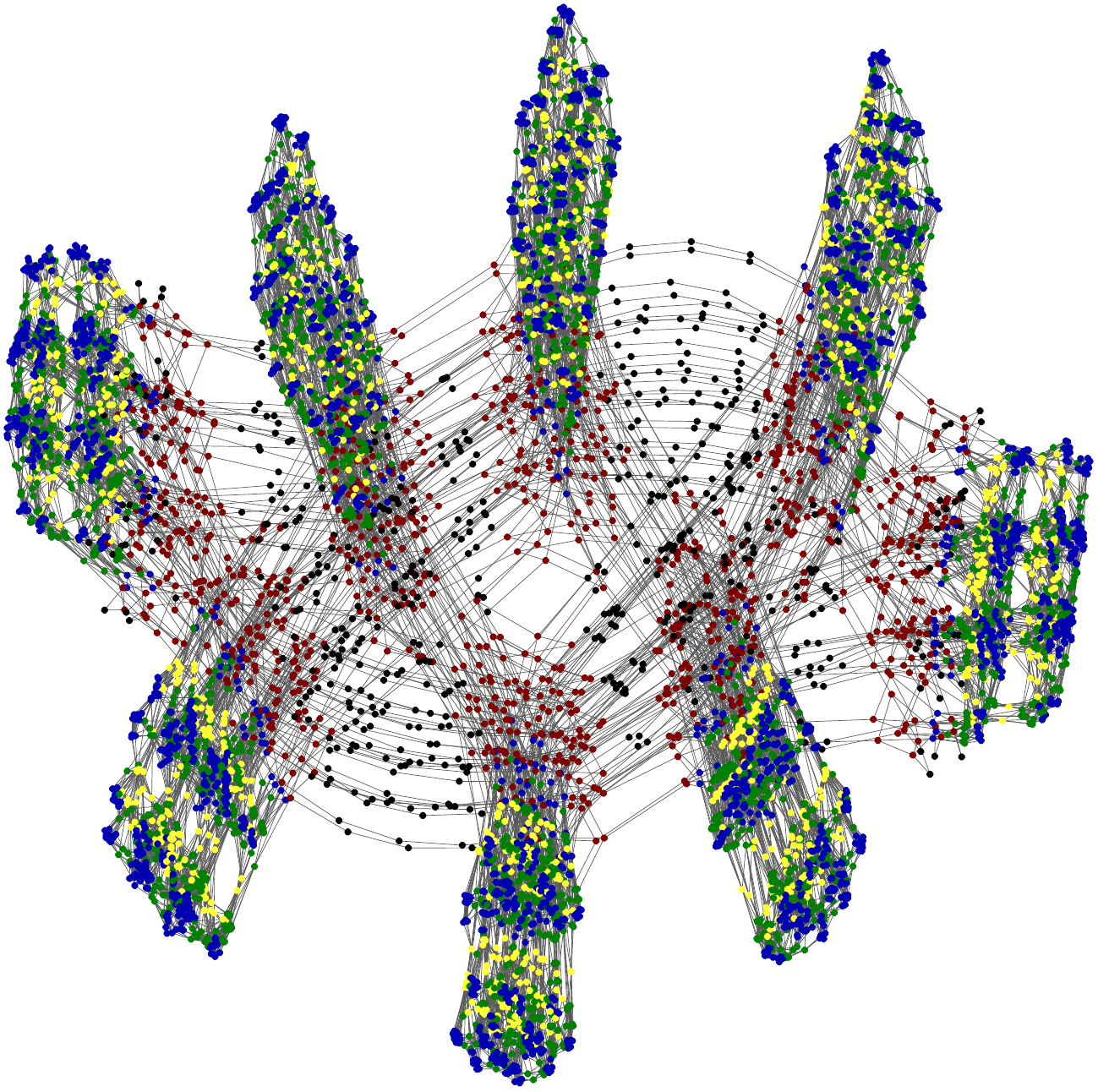}\\
  \includegraphics[width=0.35\textwidth]{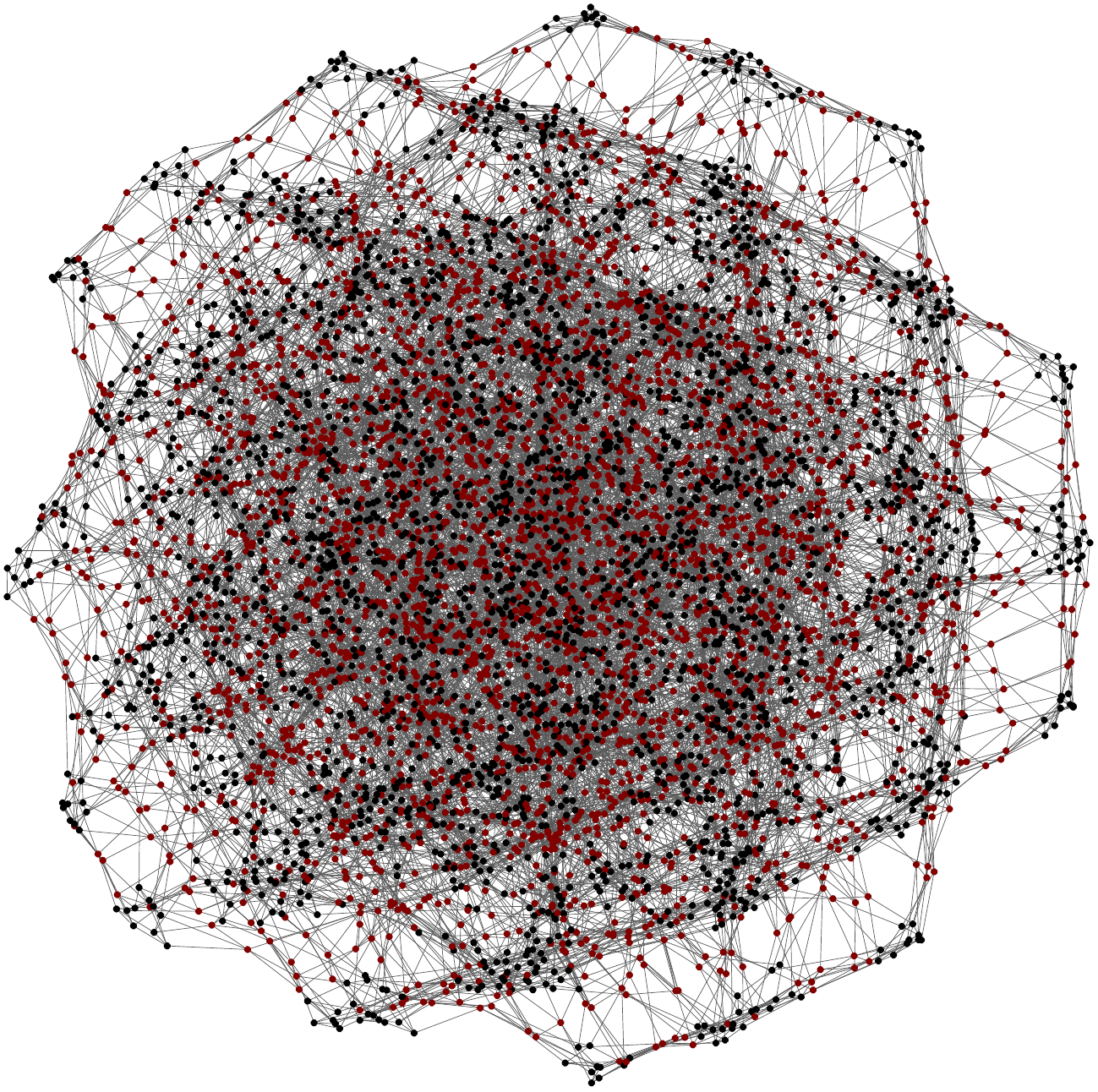}\hfill %
  \includegraphics[width=0.14\textwidth]{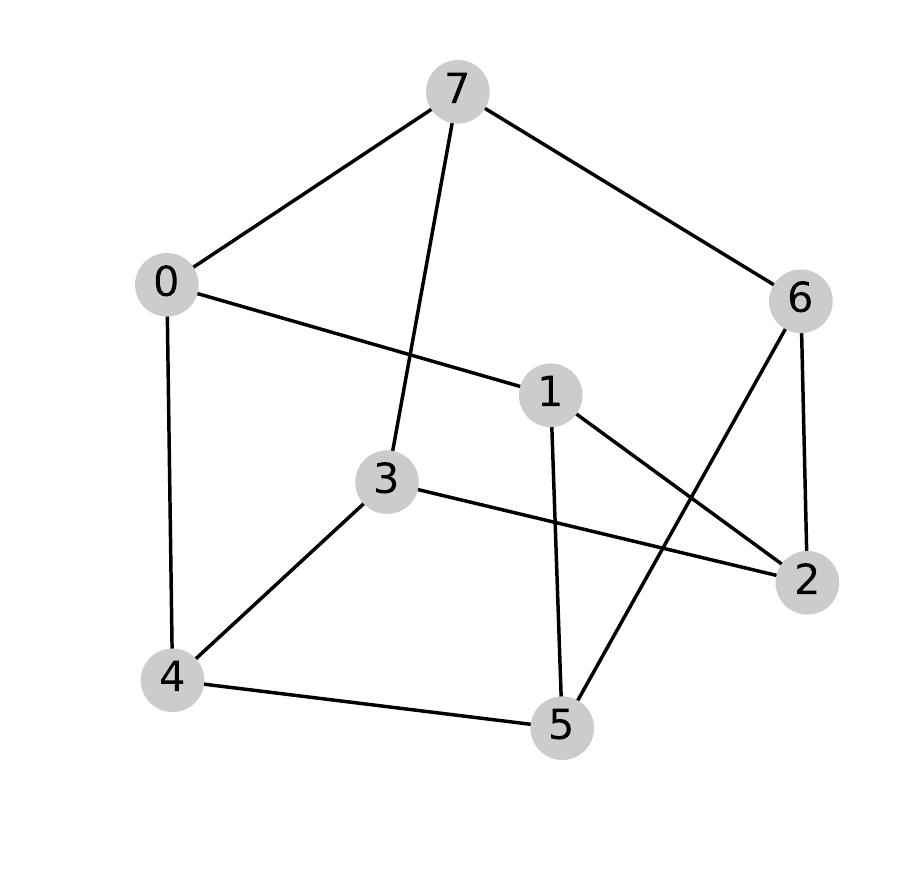}%
  \includegraphics[width=0.14\textwidth]{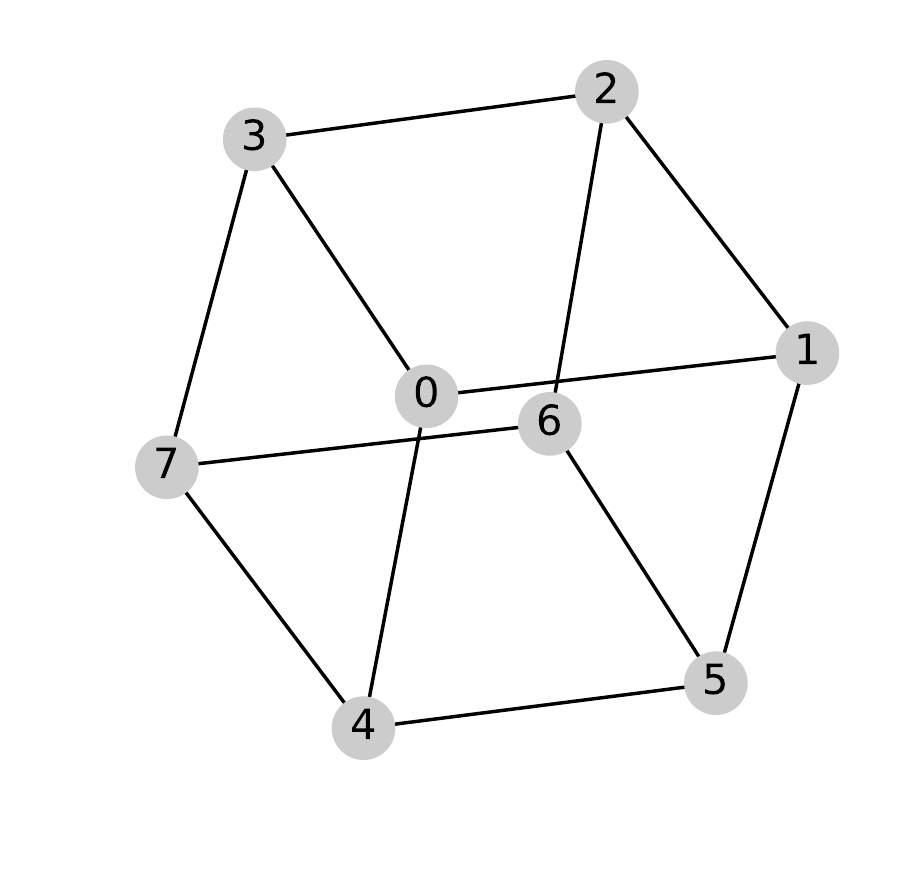}\hfill %
  \includegraphics[width=0.35\textwidth]{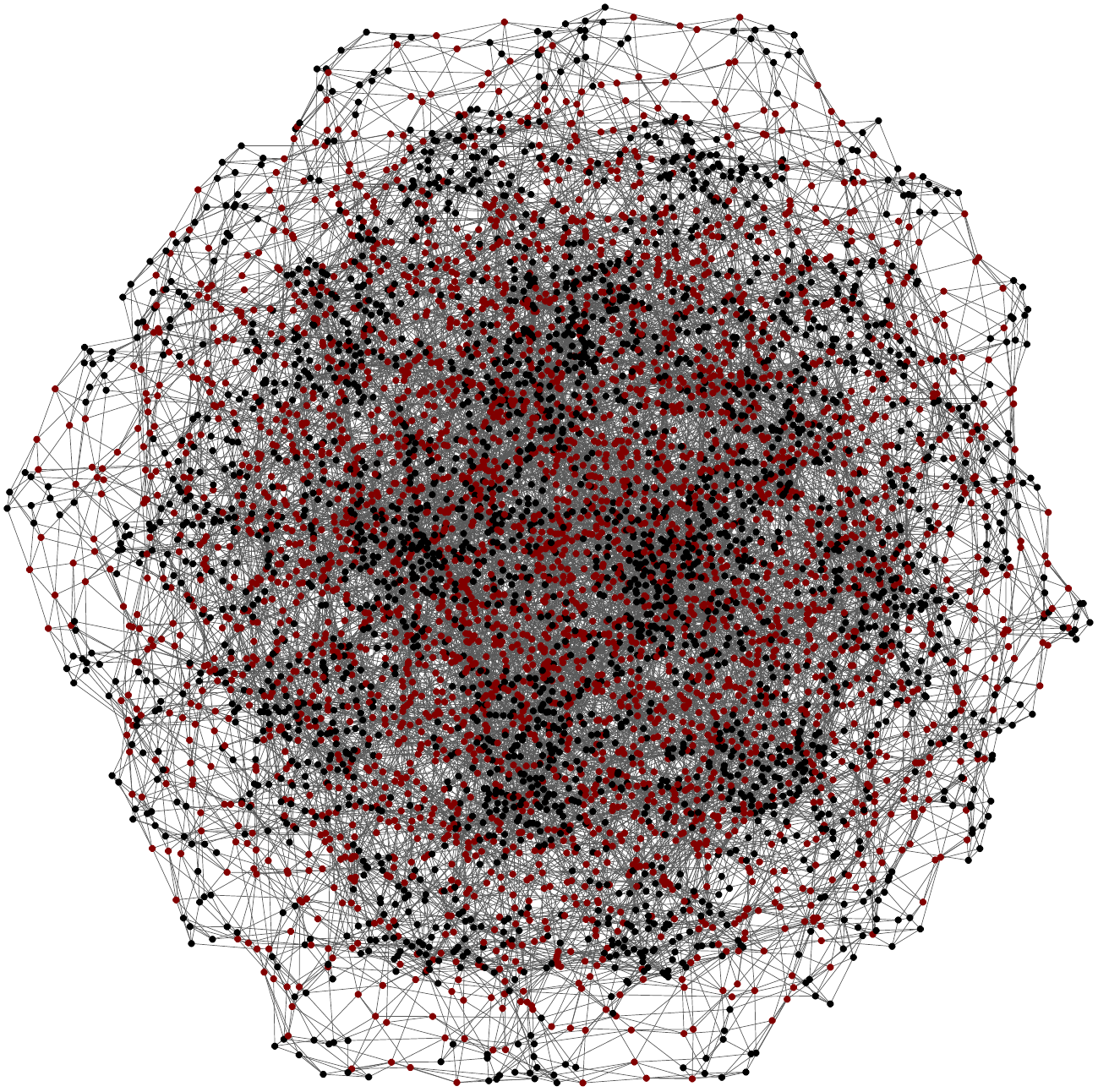}
  \caption{Quantum walk \(U\) network build from its adjacency matrix, for graphs `kite' (first row left, Fourier coin, and right, Grover coin), `möbius' (second row, left) and `cube' (right). Vertices are colored by degree: kite graph (10240 vertices) \(d=\{3,4,5,7,8,9\}\) for the Fourier coin, and \(d=\{2,4,7,8,9\}\) for the Grover coin (the order of colors is: black, red, blue, green, yellow, cyan, lower degree node tends to be in the inner part of the network, and higher degree nodes to the periphery); and `möbius' and `cube' graphs (6144 vertices), \(d = \{5,6\}\). The drawing uses the `sfdp' algorithm \protect\cite{Hu-2005pd}.
  \label{f:u}}
\end{figure*}

The phenomenological thermal properties of the relaxed stationary state, can be highlighted by an analysis of the spectral properties of the evolution operator. We may in particular relate these thermal properties with the chaotic behavior of the system, using the statistical properties of the \(U\) eigenvalues (quasienergies) and eigenvectors. We find that the quasienergy spectrum approaches a uniform distribution over the circle, and the spacing statistics corresponds to a random matrix unitary ensemble. These properties define a chaotic quantum system. Chaos is confirmed by the measure of the Shannon entropy and the statistical properties of the eigenvectors. Systems with high symmetry (graph and coin) display degenerate levels whose number may be comparable with the Hilbert space dimension in the case of complete graphs (degeneracy becomes, in this case, extensive). Magnetic properties are strongly correlated with the amount of degeneracy: breaking the symmetries by using the Fourier coin instead of the Grover one, leads to a non degenerated paramagnetic state. However, in the generic case degeneracy is marginal, and the system sets in a thermal state, sometimes with finite magnetization, amenable to a description in therms of its eingenvector properties. We find that the eigenvector thermalization hypothesis is well verified, allowing a comparison of the observables mean value with the expected value using a single high entropy eigenvalue, which is in turn, comparable with the microcanonical expectation (c.f.~\eqref{e:eth}). 

It is instructive to visualize the operator \(U\) as a network (Fig.~\ref{f:u}). We constructed an adjacency matrix from the nonzero off-diagonal entries of \(U\) and display it as a graph colored by degree. We represented the \(U\) network for the `kite' (union of two complete graphs of five and three nodes) and the `möbius' graph. The two images of the `kite' graph correspond to the two choices of the coin, Fourier (left) and Grover (center); they show a striking difference that manifests at the spectral level, by a difference in degeneracy. It is worth mentioning that the `kite' system, with Grover coin, slowly relaxes without reaching a thermal state, in contrast to the `möbius' system, which relaxes to a ferromagnetic state (with the Grover coin) and the `cube' which relaxes to a paramagnetic state. Note the difference in the symmetry of the `möbius' and `cube' graphs, which can be related to the crossing in `m', absent in `c'. These networks illustrate the complexity of the action of \(U\) on a state vector: each vertex corresponds to a basis vector and each link represents a superposition of two amplitudes. One may also associate a notion of locality of \(U\) associated to the sparsity of these networks, which is a consequence of the fact that only two-body interactions are allowed. 

More specifically, we may say that the quantum walk operator \(U\) is local in the sense that it is defined by the properties of a node and its associated edges (neighbors); the quantum state is instead nonlocal, which is simply a consequence of the quantum superposition principle. However, physical locality needs the extra condition that the degree of the graph is not commensurable with the number of vertices, \(d/N\) must vanishes for large \(N\), with \(d\sim O(1)\); without this property we cannot spatially separate two regions (in our model the propagation velocity is one). For the graphs studied this condition is difficult to satisfy because of the limited size amenable to exact computation. We avoided in some extent this limitation by considering a variety of graphs, in particular random sparse ones. The unitary step operator contains, in addition to the interactions of the walker and the spins, information about the specific geometric properties of the graph, its action \emph{shuffle amplitudes} to create an entangled state whose complexity increases without the need of external perturbations or quenched randomness, or even averaging over initial conditions. This complexity is naturally exhibited by the energy eigenvectors, which can be compared with a large random phase vector in Hilbert space \cite{Deutsch-1991vn} (c.f. Fig.~\ref{f:amp}).

The emerging general picture is that a system with local interactions coupling internal and translation degrees of freedom, thus giving rise to a complex structure of the evolution operator \(U\), develops correlations between these different degrees of freedom, to eventually create an entangled state carrying nonlocal information. These general considerations suggest that thermalization results from the interplay between the \emph{local} rules governing the interaction and evolution of the quantum state, and the \emph{nonlocal} entanglement these same rules create. The rules are strictly local, however, the quantum dynamics is global: the action of \(U\)) is over the full Hilbert space, allowing the spreading of correlations at each time step.

The model introduced here revealed so rich that many interesting perspectives arise. The most obvious direction is to generalize the system to take into account changes in the graph topology \cite{Hamma-2010fk,Arrighi-2017rm,Arrighi-2018qr}. Supposing the graph itself as a quantum object implies states with superpositions of different graphs, and therefore the possibility to answer questions like, for instance, the type of graphs that maximize, for a given degree (to impose locality), the entanglement entropy for some of the degrees of freedom \cite{Passerini-2008xy}. Within such framework a transition of the `cube' to the `möbius' graph can be interpreted as a transition between two magnetic orders, to study more general quantum transition is also of interest.
 
In summary, we investigated the behavior of a quantum walker in a graph of interacting spins. We found that the system naturally evolves towards a stationary state, that in most cases (for generic graphs) can be assimilated to a thermal state, well described by random matrix theory and chaotic eigenvectors. We also observed that, in the thermal state, the position of the walker can encode a maximum number of qubits, as measured by the partial trace of the von Neumann entropy, depending on the graph symmetries. Symmetry and degenerate levels introduce deviations with respect to the micorcanonical expectation value of the observables, lowering the entanglement level.

\begin{acknowledgments}
We greatly appreciated discussions with Giuseppe Di Molfetta and Pablo Arrighi.
\end{acknowledgments}

\appendix

\section{Numerical alghorithms}
\label{s:py}

We present in this appendix excerpts of the python codes used to compute the quantum walk. The code uses the standard scientific libraries \texttt{numpy} (arrays), \texttt{scipy} (in particular to compute eigenvalues and eigenvectors), and \texttt{networkx} (graphs definitions, and drawing with \texttt{graphviz}).

\subsection{Graph and state}

The graph is defined using a list \texttt{G}, of a list of neighbors for each node, ordered in incrasing node number. For the bull graph of Fig.~\ref{f:b} we have,
\begin{lstlisting}[language=Python]
G = [[1,3,4], [0], [4], [0,4], [0,2,3]]
E = [[0,1], [0,3], [0,4], [2,4], [3,4]]
D = [3, 1, 1, 2, 3]
S = [[0, 0, 0], [0], [1], [1, 2], [2, 0, 1]]
\end{lstlisting}
where \texttt{E} is the list of edges, \texttt{D} the degrees per node, and \texttt{S} the list of subnodes, whose labels are the coin colors.

The quantum state \texttt{psi} \eqref{e:psi}, is defined as a complex array of shape \((N,d, 2^N)\).

\subsection{Operators}

It is easier to understand the action of the coin, motion, and interaction operators when applied to the relevant degrees of freedom. The Grover \eqref{e:GR} and Fourier \eqref{e:FT} coins applied to a node \(x\) with \(d_x=4\) neighbors, are represented by the matrices:
\begin{equation}
  \label{a:grft}
  \GR = \frac{1}{2} \begin{pmatrix}
     -1 & 1 & 1 & 1 \\
     1 & -1 & 1 & 1 \\
     1 & 1 & -1 & 1 \\
     1 & 1 & 1 & -1
  \end{pmatrix},
\end{equation}
and
\begin{equation}
  \FT = \frac{1}{2} \begin{pmatrix}
    1 & 1 & 1 & 1 \\
    1 & \E^{\I \pi/4} & \E^{\I  \pi/2} & \E^{\I 3 \pi/4} \\ 
    1 & \E^{\I \pi/2} & \E^{\I \pi} & \E^{\I 3 \pi/2} \\ 
    1 & \E^{\I 3\pi/4} & \E^{\I 3\pi/2} & \E^{\I 9 \pi/4}
  \end{pmatrix}.
\end{equation}
If \(d_x < d\) the matrices are completed with a unity block. The coin operator is build using the degree list \texttt{D}, by the comprehension
\begin{lstlisting}[language=Python]
[coin(d_x, d) for d_x in D]
\end{lstlisting}
where \texttt{coin} is \(GR\) or \(\FT\), resulting in an array of shape \((N,d,d)\). The three index tensor \texttt{coin} is multiplied by the three index state \texttt{psi} using the \texttt{einsum} routine: \texttt{einsum('...ik,...kl', coin, psi)}, avoiding the use of huge matrices. For example, the dimensions of \texttt{coin} and \texttt{psi} in the case of the ladder graph `l', are \(8\times 3 \times 3\) and \(8 \times 3 \times 2^8\), respectively.

The quantum walk \(\MV\) operator \eqref{e:MV}, is implemented by the code:
\begin{lstlisting}[language=Python]
def move_op(psi):
  """swap position amplitudes between neighbors"""
  psi_1 = zeros( shape(psi), dtype = complex )
  for x in range(N):    
    psi_1[ix_(G[x]),ix_(S[x])] = psi[x,:len(G[x])]
return psi_1
\end{lstlisting}
using the adjacency list \texttt{G} and the subnode list \texttt{S}, to exchange the amplitudes labeled by \(c_y\) by \(c_x\). It is diagonal in the spin index (third array index).

The interaction between two neighboring spins \eqref{e:CZ}, can be represented by the two qubits phase gate:
\begin{equation}
  \label{a:CZ}
  \CZ = \begin{pmatrix}
    1 & 0 & 0 & 0 \\
    0 & 1 & 0 & 0 \\
    0 & 0 & 1 & 0 \\
    0 & 0 & 0 & -1
  \end{pmatrix}
\end{equation}
The python code consists in a loop over the edges \texttt{E}, and a search of neighbors having down spins \(\texttt{sx}=1\) and \(\texttt{sy}=1\). The value of the spin at a given node \(x\) is obtained from the binary representation of the spin label \texttt{xs}:
\begin{lstlisting}[language=Python]
def CZ_sp(psi):
  """controlled phase Z interaction between neighboring spins"""
  psi_1 = psi.copy()
  for x, y in E: # loop on undirected edges
    xcs = nonzero(psi[x]) # list of pairs (c,s)
    xc = xcs[0] # array of coin values
    xs = xcs[1] # array of spin values
    s_x = mod(xs//2**x, 2) # list of 0 and 1
    s_y = mod(xs//2**y, 2) # list of 0 and 1
    # indices of sx = sy = 1
    s = array(nonzero(s_x*s_y==1)) 
    for i in s:
      psi_1[x,xc[i],xs[i]] = - psi[x,xc[i],xs[i]]
return psi_1
\end{lstlisting}

The coin-spin interaction, that corresponds to the swap two qubits gate,
\begin{equation}
  \label{a:SW}
  \SW = \begin{pmatrix}
    1 & 0 & 0 & 0 \\
    0 & 0 & 1 & 0 \\
    0 & 1 & 0 & 0 \\
    0 & 0 & 0 & 1
  \end{pmatrix}
\end{equation}
is implemented in a similar form, but using the first color and the local spin:
\begin{lstlisting}[language=Python]
#
# SW_p_sp particle spin interaction
......
s_x = mod(xs//2**x, 2) # list of 0 and 1
for i in range(len(s_x)):
  c = xc[i]
  s = s_x[i]
  if c == 0 and s == 1:
    psi_1[x,1,xs[i] - 2**x], psi_1[x,0,xs[i]] = \
        psi[x,0,xs[i]], psi[x,1,xs[i] - 2**x]
  if c == 1 and s == 0:
    psi_1[x,0,xs[i] + 2**x], psi_1[x,1,xs[i]] = \
        psi[x,1,xs[i]], psi[x,0,xs[i] + 2**x]
......
\end{lstlisting}

The main loop of the quantum walk, takes an initial state array with equal amplitudes over a list of nodes, usually the base ket \(\ket{000}\), and applies \(T\) times the operator \eqref{e:U}
\begin{lstlisting}
# initial state
for x, c, s in nodes: # nodes = [[0,0,0],...]
  psi[x, c, s] = 1
psi = psi/norm(psi)
# coin: Grover or Fourier
coin = coin_C(D)
# time loop
for t in range(1,T+1):
  psi = CZ_sp(SW_p_sp(move_op(coin_op(coin, psi))))
\end{lstlisting}

The \(U\) operator can be straightforwardly build by applying the above operators to the set of basis vectors; however, a more efficient implementation uses a swept over the nodes and edges of the graph \texttt{gx} (a \texttt{networkx} object). In addition to the lists defining the graph we need a list of the starting indices of each node of basis vectors \texttt{sD}, which is the cumulative sum of the array \texttt{D}, such that \texttt{sD[x]*NS} gives the index of the ket \(\ket{x00}\) (with \texttt{NS=2**N}). 

The coin unitary matrix is simply given by,
\begin{lstlisting}[language=Python]
def coin_U(gx, coin_d):
  ......
  c = zeros((dim,dim), dtype = complex)
  i = 0
  for d in D:
    c[i:i+d, i:i+d] = coin_d(d)
    i += d
  return kron(c, eye(NS))
\end{lstlisting}
where the object \texttt{gx} contains the graph information, \texttt{dim} is the maximum degree dimension and \texttt{d} the node degree.

The particle motion operator matrix is:
\begin{lstlisting}[language=Python]
def move_U(gx):
  ......
  sD = roll(cumsum(D),1)
  sD[0] = 0
  a = zeros((dim,dim))
  ix = 0
  for x in range(N):
    iy = 0
    for y in G[x]:
      a[ix,sD[y]+S[x][iy]] = 1
      ix += 1; iy += 1
  return kron(a, eye(NS))
\end{lstlisting}
where \texttt{x} run over the nodes and \texttt{y} over the corresponding neighbors; the \texttt{sD+S} combination allows to swept for each node \texttt{x} the coin values corresponding to each neighbor \texttt{y}.

In a similar way we build the particle-spin interaction unitary matrix:
\begin{lstlisting}[language=Python]
def exchange_U(gx):
  ......
  up = eye( dim*NS )
  for x in arange(N)[D>1]: # exclude degree 1 nodes
    for s in range(NS):
      if mod(s//2**x, 2) == 1:
        up[sD[x]*NS+s, sD[x]*NS+NS+s-2**x] = 1
        up[sD[x]*NS+s, sD[x]*NS+s] = 0
      else:
        up[sD[x]*NS+NS+s, sD[x]*NS+s+2**x] = 1
        up[sD[x]*NS+NS+s, sD[x]*NS+NS+s] = 0
  return up
\end{lstlisting}
which implements the swap \eqref{a:SW} between the coin and the spin (non diagonal terms equal to one, and diagonal terms equal to zero).

Finally, the spin spin interaction code is,
\begin{lstlisting}[language=Python]
def spin_U(gx):
  ......
  us = eye( dim*NS )
  for x, y in E:
    for sx in range(NS):
      if mod(sx//2**x, 2)*mod(sx//2**y, 2) == 1:
        for c in range(D[x]):
          us[sD[x]*NS+c*NS+sx,sD[x]*NS+c*NS+sx] = -1
  return us
\end{lstlisting}
where the first loop is over the list of edges, the second one over the spin (most rapidly varying index in the basis list) and the over the color \texttt{c}, only if the two interacting spins are down.

The product of the matrices gives the \(U\) operator \eqref{e:U},
\begin{lstlisting}[language=Python]
U = dot(spin_U(gx), dot(exchange_U(gx),
        dot(move_U(gx), coin_U(gx, coin_d)))) 
\end{lstlisting}
that can be diagonalized using the routine \texttt{eig} of \texttt{scipy}.

%
%
%
\end{document}